\pdfoutput=1
\documentclass[atlasdraft=false, cernpreprint, UKenglish, texmf, orcidlogo]{atlasdoc}
 
\usepackage{tablefootnote}

\usepackage[subfigure]{atlaspackage}
\usepackage{atlasbiblatex}
 
\usepackage{atlasphysics}
\usepackage{atlasjetetmiss}
 
\usepackage{todonotes}
\usepackage{xpatch}
\usepackage{multirow}
\makeatletter
\xpretocmd{\todo}{\@bsphack}{}{}
\xapptocmd{\todo}{\@esphack}{}{}
\makeatother
 
\graphicspath{{logos/}{figures/}}
 
\usepackage{ANA-STDM-2018-55-PAPER-defs}

\AtlasTitle{
Differential cross-sections for events with missing transverse momentum and jets measured with the ATLAS detector in 13~TeV proton--proton collisions}
 
\AtlasVersion{3.0}
\AtlasAbstract{
Measurements of inclusive, differential cross-sections for the production
of events with missing transverse momentum in association with jets in proton--proton collisions at $\sqrt{s}=13$ TeV are presented.
The measurements are made with the ATLAS detector using an integrated luminosity of
140 fb$^{-1}$ and include measurements of dijet distributions in a region in which vector-boson fusion processes are enhanced.
They are unfolded to correct for detector resolution and efficiency within
the fiducial acceptance, and are designed to allow robust comparisons with a wide range of theoretical predictions.
A measurement of differential cross sections for the $Z~\to \nu\nu$ process is made.
The measurements are generally well-described by Standard Model predictions except for the dijet invariant mass distribution.
Auxiliary measurements of the hadronic system recoiling against isolated leptons, and photons, are also made in the same phase space.
Ratios between the measured distributions are then derived, to take advantage of cancellations in modelling effects and some of the major systematic uncertainties.
These measurements are sensitive to new phenomena, and provide a mechanism to easily set constraints on phenomenological models.
To illustrate the robustness of the approach, these ratios are compared with two common Dark Matter models,
where the constraints derived from the measurement are comparable to those set by dedicated detector-level searches.
}
 
\author{The ATLAS Collaboration}
 
\AtlasRefCode{STDM-2018-55}

\AtlasJournal{JHEP}
 
\AtlasJournalRef{JHEP 08 (2024) 223}
\AtlasDOI{10.1007/JHEP08(2024)223}

\PreprintIdNumber{CERN-EP-2024-034}

\hypersetup{pdftitle={ATLAS document},pdfauthor={The ATLAS Collaboration}}
 
\begin{document}
 
\maketitle
 
\tableofcontents

\newpage
\section{Introduction}
\label{sec:intro}
 
A high priority goal of experiments at the Large Hadron Collider (LHC) is to
establish to what extent the Standard Model (SM) remains valid at the accessible energies
above the electroweak symmetry-breaking scale.
If the measured data agree with the SM, it is important to both quantify that agreement and to
interpret it in terms of limits on physics beyond the SM (BSM).
If the data are inconsistent with SM predictions, this could constitute evidence for BSM physics.
 
One key reason to suspect BSM physics exists is the astrophysical and cosmological evidence
for the existence of Dark Matter (DM)~\cite{Bertone:2004pz,Trimble:1987ee,Feng:2010gw}. Many BSM theories postulate a DM particle that
may be produced at the LHC, giving rise to missing transverse
momentum (\ptmiss) in proton--proton (\pp) collision events, over and above that expected from SM processes producing neutrinos.
Searches have exploited this signature to set limits on DM models~\cite{ATLAS:2021kxv,CMS:2021far}. In addition, the cross-section for the
principal SM process producing large \ptmiss, in which a \Z\ boson decaying to neutrinos recoils against jets, has recently been
measured by the CMS Collaboration~\cite{CMS-SMP-18-003}. The ATLAS Collaboration has also recently produced a measurement of the
$Z$-boson invisible width which exploits this signature~\cite{ATLAS:2023ynf}.
 
The main purpose of this analysis is to make precise, detector-corrected measurements of \ptmiss produced in association with jets, inclusively and
with as little model dependence as possible. This is the first such measurement made using the full 140~fb$^{-1}$ of integrated luminosity collected by the ATLAS
detector in Run 2 of the LHC\@. The results are presented alongside auxiliary measurements, made in the same phase space,
of the transverse momentum of the hadronic system, \ptrecoil, recoiling against isolated leptons, and photons.
This allows modelling effects and major uncertainties to cancel when a ratio of cross-sections is taken.
This complements and extends the approach of taking ratios presented in a previous study~\cite{EXOT-2016-03}.
The results are compared quantitatively to state-of-the-art SM predictions.
 
The measurements also serve another purpose. Contributions from DM production would typically \textit{not} cancel out in the cross-section ratios,
making them sensitive to DM and other BSM signatures.
A secondary objective of the paper is therefore to demonstrate that the measurements can be used for searches and setting constraints,
with only a minor penalty in sensitivity, and without the need to repeat complex and time-consuming detector simulation. This means they can be readily
reinterpreted to gain information about models, and model parameter points, beyond those considered here.
 
Cross-sections differential in \ptmiss and \ptrecoil, and in several jet observables, are defined in fiducial phase spaces designed to
probe different aspects of the SM\@.
The dominant SM contribution to the \ptmiss-plus-jets final state comes from \Zboson bosons produced in association with jets and decaying into neutrinos,
\Znunujets; other contributions
come from leptonic \Wboson boson decays where the lepton does not enter the fiducial phase space. Diboson and triboson production can also
provide small contributions.
All relevant kinematic selections
are included in the fiducial phase space definition,  and detector effects, including instrumental sources of \ptmiss,  are corrected for using an
unfolding procedure.
Motivated by ease of comparison to SM predictions, and to validate the consistency of the approach, a measurement of \Znunu production
is also made, where the contributions from other SM processes are treated as backgrounds and subtracted before unfolding.
 
For the BSM interpretation, two example models are chosen to illustrate the constraints that can be extracted from the measurements.
First, a common simplified DM model~\cite{Abdallah:2015ter}, which was searched for previously in this final state by ATLAS~\cite{ATLAS:2021kxv} 
using the same data sample as the current analysis, and by CMS~\cite{CMS:2021far}.
Second, a more complicated model that introduces an additional Higgs doublet and a pseudoscalar
that couples to DM~\cite{Bauer:2017ota,LHCDarkMatterWorkingGroup:2018ufk} and has also been searched for previously~\cite{CMS:2020ulv,ATLAS:2023rvb} is considered.
 
The paper is structured as follows.  After a brief description of the experimental apparatus in Section~\ref{sec:atlas-detector},
the cross-sections and observables to be measured are defined in Section~\ref{sec:fiducial}.
The theoretical predictions, Monte Carlo event generation,
and detector simulation are discussed in Section~\ref{sec:sim}.
The details of the event selection and object reconstruction are given in Section~\ref{sec:events},  and
the treatment of backgrounds is described in Section~\ref{sec:backgrounds}.
The correction for detection effects, and the associated systematic uncertainties, are described in Section~\ref{sec:correction}.
Results are reported in Section~\ref{sec:results}, and interpreted in terms of SM and BSM calculations
in Section~\ref{sec:interpretation}. Finally, the conclusions are summarised.


\section{ATLAS detector}
\label{sec:atlas-detector}
 
\newcommand{\AtlasCoordFootnote}{
ATLAS uses a right-handed coordinate system with its origin at the nominal interaction point (IP)
in the centre of the detector and the \(z\)-axis along the beam pipe.
The \(x\)-axis points from the IP to the centre of the LHC ring,
and the \(y\)-axis points upwards.
Polar coordinates \((r,\phi)\) are used in the transverse plane,
\(\phi\) being the azimuthal angle around the \(z\)-axis.
The pseudorapidity is defined in terms of the polar angle \(\theta\) as \(\eta = -\ln \tan(\theta/2)\).
Angular distance is measured in units of \(\Delta R \equiv \sqrt{(\Delta\eta)^{2} + (\Delta\phi)^{2}}\).}

The ATLAS experiment~\cite{PERF-2007-01} at the LHC is a multipurpose particle detector
with a forward--backward symmetric cylindrical geometry and a near \(4\pi\) coverage in
solid angle.\footnote{\AtlasCoordFootnote}
It consists of an inner tracking detector surrounded by a thin superconducting solenoid
providing a \qty{2}{\tesla} axial magnetic field, electromagnetic and hadronic calorimeters, and a muon spectrometer.
The inner tracking detector covers the pseudorapidity range \(|\eta| < 2.5\).
It consists of silicon pixel, silicon microstrip, and transition radiation tracking detectors.
Lead/liquid-argon (LAr) sampling calorimeters provide electromagnetic (EM) energy measurements
with high granularity within the region \(|\eta|< 3.2\).
A steel/scintillator-tile hadronic calorimeter covers the central pseudorapidity range (\(|\eta| < 1.7\)).
The endcap and forward regions are instrumented with LAr calorimeters
for EM and hadronic energy measurements up to \(|\eta| = 4.9\).
The muon spectrometer surrounds the calorimeters and is based on
three large superconducting air-core toroidal magnets with eight coils each.
The field integral of the toroids ranges between \num{2.0} and \qty{6.0}{\tesla\metre}
across most of the detector.
The muon spectrometer includes a system of precision tracking chambers up to \(|\eta| = 2.7\) and fast detectors for triggering up to \(|\eta| = 2.4\).
The luminosity is measured mainly by the LUCID--2~\cite{LUCID2} detector, which is located close to the beampipe.
A two-level trigger system is used to select events~\cite{TRIG-2016-01}.
The first-level trigger is implemented in hardware and uses a subset of the detector information
to accept events at a rate below \qty{100}{\kHz}.
This is followed by a software-based trigger that
reduces the accepted event rate to \qty{1}{\kHz} on average
depending on the data-taking conditions.
A software suite~\cite{ATL-SOFT-PUB-2021-001} is used in data simulation, in the reconstruction
and analysis of real and simulated data, in detector operations, and in the trigger and data acquisition
systems of the experiment.


\section{Measured observables and fiducial phase spaces}
\label{sec:fiducial}
 
The differential cross-sections to be measured are defined within a fiducial phase space, specified in terms of requirements applied
to final state particles.
These requirements are chosen to closely reflect the acceptance of the detector, thus reducing the need for theory-based
extrapolations. For the dilepton regions, the leptons are required have the same flavour and opposite charge.
 
\subsection{Particle-level objects \label{sec:particlelevelobjects}}
 
At particle-level, the following objects are defined.
Charged leptons (electrons or muons) are required to be prompt, in that they do not originate from the decay of a hadron.
Leptons from the decay of prompt $\tau$-leptons are allowed. The four-momenta of prompt photons within a cone of \(\Delta R = 0.1\) is
added to the four-momentum of the lepton to produce a `dressed' lepton.
 
Photons are required to be prompt and isolated. The photon isolation is chosen such that it mimics the
isolation requirement at the detector level, requiring that the transverse energy in a cone of \(\Delta R = 0.4\) around
the photon be less than ($2.45~\GeV + 0.044 \times \pt$) where \pt is the transverse momentum of the photon in \GeV.
 
For the inclusive \ptmiss measurement, the particle-level \ptmiss is defined as the magnitude of a vector,
which is the negative two-momentum ($x,y$ components) sum of all visible
final-state particles with $|\eta|<5$, excluding muons with $|\eta| > 2.5$ or $\pT < 7\,\GeV$.
The \ptrecoil observable is defined in a similar way, but the identified charged dressed leptons, and isolated photons,
are excluded from the sum. Thus, for the inclusive \ptmiss measurements, $\pTrecoil \equiv \ptmiss$.
For the measurement of \Znunu, the particle-level \ptmiss{} is defined as the summed
\pt{} of the neutrinos from the decayed boson.
 
Jets are defined using the \antikt{} jet algorithm~\cite{Cacciari:2008gp,Cacciari:2011ma} with a radius parameter of 0.4.
All stable final-state particles are used as input to the jet
algorithm, except that for the inclusive \ptmiss measurement neutrinos and other invisible particles as well as muons
are excluded, while for the \Znunu measurement invisible particles as well as the boson decay products are excluded.
Any jets that contain a hadron coming from the decay of a prompt $\tau$-lepton are
classified as hadronically decaying $\tau$-leptons.
 
Jets are removed if the jet momentum direction is closer than \(\Delta R < 0.2\) to any lepton.
Next, all leptons that are within \(\Delta R = 0.4\) of a jet are discarded.
If no leptons remain, jets are removed if the jet momentum direction is closer than \(\Delta R < 0.2\) to any photon.
These conditions mirror closely the overlap removal of reconstructed objects described in Section~\ref{sec:recosel}.
 
\subsection{Phase-space regions}
\label{sec:particle-level-regions}
 
Measurements are made in six regions defined in terms of the number and
flavour of leptons or the presence of a photon: \ptmissjets, \oneeljets, \twoeljets{}, \onemujets, \twomujets and \onegjets.
The first of these is the primary measurement, while the others are auxiliary measurements with similar topologies to the primary,
which constrain the uncertainties through correlations across the regions.
The similarity in topologies is ensured by using the same event selection,
and by the fact that in the main \ptmiss measurement region, $\ptrecoil \equiv \ptmiss$.
Table~\ref{tab:leptons-phasespace} summarises the selections that define
these regions. The differences between the pseudorapidity requirements between electrons and muons arise from their different experimental acceptance, and the desire to minimise extrapolation during the unfolding procedure.
 
For each of these regions, two sub-regions are defined by further selection on the jet content of the hadronic
recoil system, the \onejet and vector boson fusion (\vbf{}) regions. These are designed to enhance the sensitivity
to particular classes of BSM physics involving DM, such as those that are studied in Section~\ref{sec:interpretation}.
Table~\ref{tab:jets-phasespace} summarises the selections that define these sub-regions.
 
\begin{center}
\begin{table}[htb]
\centering
\caption{Requirements
defining the six principal phase-space regions of the measurement. For the inclusive \ptmiss measurement, $\ptmiss \equiv \ptrecoil$. In the \Znunu measurement, it corresponds to the \pt of the \Zboson\ boson. Transverse mass, \mT{}, is defined as $\sqrt{2\pT\ptrecoil (1 - \cos(\phi))}$
where \pT is the lepton transverse momentum and $\phi$ is the azimuthal angle between the lepton and \ptrecoil. For the dilepton regions, the leptons are required have the smae flavour and opposite charge.
\label{tab:leptons-phasespace}
}
\begin{tabular}{ l||c|c|c|c|c|c}
\toprule
Attribute              &  \ptmissjets & \oneeljets & \twoeljets & \onemujets & \twomujets & \onegjets \\
\hline
Lepton or photon        &      \multirow{2}{*}{--} &  \multicolumn{2}{|c|}{\(|y|\le 1.37\)  or } & \multicolumn{2}{c|}{\multirow{2}{*}{\(|y|\le 2.5\)} } & \(|y|\le 1.37\)  or \\
rapidity              &            &  \multicolumn{2}{|c|}{ \(1.52\le |y| \le 2.47\)  }                                & \multicolumn{2}{c|}{          }  &   \(1.52 \le |y| \le 2.47\) \\  \hline
Leading lepton or       &   \multirow{2}{*}{--}       & \multirow{2}{*}{  $ > 30 $ }  & \multirow{2}{*}{ $ > 80 $}     &\multirow{2}{*}{  $ > 7 $ }    & \multirow{2}{*}{ $ > 80$ }    &  \multirow{2}{*}{ $ > 160$ } \\
photon \pt[GeV]       &            &            &            &            &            &    \\ \hline
Sub-leading             &   \multirow{2}{*}{--}       &   \multirow{2}{*}{--}        & \multirow{2}{*}{ $ > 7$ }     & \multirow{2}{*}{--}          &  \multirow{2}{*}{ $ > 7 $}    &  \multirow{2}{*}{--} \\
lepton \pt[GeV]       &            &            &           &            &            &    \\ \hline
Dilepton mass,         &   \multirow{2}{*}{--}       &   \multirow{2}{*}{--}        & $\mll \in$ & \multirow{2}{*}{--} &  $\mll \in$ & \multirow{2}{*}{--} \\
\mll[GeV]           &            &            &  $  (66,116) $       &            &  $  (66,116) $       &    \\ \hline
(Additional) muons & \multicolumn{6}{c}{None with \pt  $ > 7 $~\GeV, \(|\eta| <  2.5\)   } \\                       \hline
(Additional) electrons & \multicolumn{6}{c}{None with \pt  $ > 7$~\GeV, \(|\eta| < 1.37\) or \(1.52 <|\eta| < 2.47\)}  \\ \hline
\mT{} [GeV]            &   \multirow{2}{*}{--}       & $ \mT \in$ & \multirow{2}{*}{--} & \multirow{2}{*}{--} & \multirow{2}{*}{--} & \multirow{2}{*}{--} \\
&     &   $(30,100)$         &            &            &            &    \\ \hline
\ptmiss{} [GeV]   &    $ > 200 $ &  $ > 60 $ & -- & -- & -- & -- \\ \hline
\ptrec{} [GeV]   &    $ > 200  $ &  $ > 200 $&  $ > 200 $ &  $ > 200 $ &  $ > 200 $ &  $ > 200 $\\
\bottomrule
\end{tabular}
\end{table}
\end{center}

\begin{table}[htb]
\centering
\caption{A summary of the fiducial selections
applied to the hadronic recoil system to define the subregions
of the measurement.
The veto on `in-gap jets' is applied to jets with a rapidity lying between the rapidities of the leading and the
sub-leading jets. \label{tab:jets-phasespace} }
\begin{tabular}{c || c | c}
\toprule
Attribute & \;\;\onejet \;\; & \;\; \vbf \;\; \\
\hline
\DPhiJM{}          & \multicolumn{2}{c}{  $ > 0.4 $ for four leading \pt{} jets} \\ \hline
Hadronic $\tau$-lepton & \multicolumn{2}{c}{None with \pt  $ > 20$~\GeV,  } \\
&\multicolumn{2}{c}{\(|\eta| < 1.37\) or \(1.52 <|\eta| < 2.47\)}
\\ \hline
Leading jet \pt[GeV] &  $ > 120 $
&  $ > 80 $ \\  \hline
Sub-leading jet \pt[GeV] & --
&  { $ > 50 $ }\\ \hline
Leading jet $|y|$ &  $ < 2.4 $ &   $ < 4.4 $ \\ \hline
Sub-leading jet $|y|$ & -- &  $ < 4.4 $\\ \hline
Dijet invariant mass \mjj[GeV]  & -- &   $ > 200 $ \\ \hline
$|\dyjj|$  & -- &   $ > 1 $\\ \hline
In-gap jets & -- &  None with  $\pt>  30~\GeV$ \\
\bottomrule
\end{tabular}
\end{table}

\subsection{Measured observables}
\label{sec:particle-level-variables}
 
Differential cross-sections as a function of several observables are measured in the regions defined in Section~\ref{sec:particle-level-regions}.
The distribution of \ptrec, defined in Section~\ref{sec:particlelevelobjects}, is measured for all selections in all regions.
It is sensitive
both to the SM processes involving neutrinos (predominantly \Znunu) and to potential contributions
from BSM invisible particles.
In addition, in the \vbf phase-space region the \mjj and \dphijj distributions are also measured, where \mjj is the invariant mass of the two leading jets and
$\dphijj{} = \phi_{1} - \phi_{2}$ is the signed difference in azimuthal angle between the jets ordered in their rapidities such that $y_1 > y_2$.
This observable probes the CP structure of the \vbf{} interaction~\cite{Bernlochner:2018opw}.
The \mjj distribution is sensitive to the presence of potential new particles decaying into jets.
All these distributions are available from HEPData~\cite{Maguire:2017ypu} and implemented in Rivet~\cite{Bierlich:2019rhm}.


\section{Theoretical predictions and simulation}
\label{sec:sim}
 
Monte Carlo (MC) event generators capable of simulating the complete final state of collision events are used as input to a detailed
\GEANT~\cite{Agostinelli:2002hh,Allison:2006ve} simulation of the
ATLAS detector~\cite{SOFT-2010-01}. The output from this is passed through the same reconstruction and analysis chain as the data,
to evaluate efficiencies and the
migration matrix used to unfold for detector effects, and to make measurements at particle level. Given the inclusive
nature of the measurement, several important processes contribute, requiring a wide range of sophisticated configurations
for the event generators. The generated samples are also reweighted as appropriate to improve their modelling of the data (with a negligible effect on the unfolded results).
 
Event generator predictions are also used for comparison with the final particle-level results.
Since the data are corrected for detector effects, new predictions can be used directly for the comparisons, without the full detector simulation.
For this reason, the predictions used for the final comparison are in some cases improved versions that embody the most accurate and precise predictions available at the time of publication.
The samples employed for each use case are described in turn below.
 
\subsection{Fully simulated Standard Model samples}
 
Events containing a single $W$ or $Z/\gamma^\ast$ boson in association with jets ($V+$jets),
as well as prompt single-photon production, were simulated
with the \SHERPA[2.2.1]\cite{Gleisberg:2008ta} event generator. In this set-up,
the \OPENLOOPS~\cite{Cascioli:2011va,Denner:2016kdg} and
\Comix~\cite{Gleisberg:2008fv} libraries provided matrix elements
with next-to-leading-order (NLO) virtual quantum chromodynamics (QCD) corrections for up to two jets,
and matrix elements accurate to leading-order (LO) for up to four jets.
The default \SHERPA parton shower~\cite{Schumann:2007mg} based on
Catani--Seymour dipoles and the cluster hadronisation model~\cite{Winter:2003tt}
was used. This used the parameters developed by the
\SHERPA authors for this version based on the \NNPDF[3.0nnlo] parton distribution function (PDF) set~\cite{Ball:2014uwa}.
The NLO matrix elements of a given jet multiplicity were matched to the parton
shower using a colour-exact variant of the \MCatNLO
algorithm~\cite{Hoeche:2011fd}. Different jet multiplicities were then merged
into an inclusive sample using an improved CKKW matching
procedure~\cite{Catani:2001cc,Hoeche:2009rj} that was extended to NLO
accuracy using the \MEPSatNLO prescription~\cite{Hoeche:2011fd,Hoeche:2012yf,Catani:2001cc,Hoeche:2009rj}.
The merging scale
was set to $\Qcut=20\,\GeV$.
For single-photon production a dynamic merging scale~\cite{Siegert:2016bre} of 20~\GeV was used,
and photons were required to be isolated according to a smooth-cone isolation criterion~\cite{Frixione:1998jh}.
In all cases, matrix elements were matched
with the \SHERPA parton shower~\cite{Schumann:2007mg} using the \MEPSatNLO
prescription.
 
Electroweak (EW) production of two forward jets in association with a $W$ or $Z/\gamma^\ast$ boson
and up to one additional parton emission at LO accuracy was simulated using \SHERPA[2.2.11]~\cite{Bothmann:2019yzt}.
These predictions are labelled `EWK' in the comparisons to data.
The Catani--Seymour dipole-based parton shower was used, matched with the matrix element using the
\MEPSatLO prescription, and a cluster hadronisation model was employed.
Diagrams arising from semileptonic diboson production, with one boson decaying hadronically, were removed
in a gauge-invariant manner by requiring a colour-singlet exchange in the $t$-channel,
also known as the `VBF approximation'.
 
Samples of leptonically decaying dibosons were simulated with
\SHERPA[2.2.2]~\cite{Gleisberg:2008ta}, with a similar set-up to the \vjets samples\cite{ATL-PHYS-PUB-2017-005}.
The QCD corrections to matrix elements at NLO accuracy were
provided by the \OPENLOOPS\ library~\cite{Cascioli:2011va,Denner:2016kdg}.
The parameters and PDFs were the same as for the \vjets samples.
Triboson production was simulated with the same set-up as the fully leptonically decaying diboson samples.
Semileptonically decaying diboson samples were simulated with almost the identical set-up to the fully leptonic ones,
except that the \SHERPA[2.2.1]~\cite{Gleisberg:2008ta} generator was used.
These predictions together are labelled `diboson' in the comparisons to data.
 
The production of on-shell \ttbar events was modelled using the
\POWHEGBOX~\cite{Frixione:2007nw,Nason:2004rx,Frixione:2007vw,Alioli:2010xd}~v2
generator at NLO with the \NNPDF[3.0nlo]~\cite{Ball:2014uwa} PDF set
and the \hdamp parameter\footnote{The \hdamp\ parameter controls the transverse momentum \pt of the first additional emission
beyond the leading-order Feynman diagram in the parton shower and therefore regulates the  high-\pt emission against which the
\ttbar system recoils.} set to 1.5~\mtop~\cite{ATL-PHYS-PUB-2016-020}.
The events were interfaced to \PYTHIA~8.230~\cite{Sjostrand:2014zea}
using the A14 set of tuned parameters (tune)~\cite{ATL-PHYS-PUB-2014-021} and the \NNPDF[2.3lo] PDF set~\cite{Ball:2012cx}.
The NLO \ttbar\ inclusive production cross-section was corrected to the theory prediction at next-to-next-to-leading order (NNLO)
in QCD including the resummation of next-to-next-to-leading logarithmic (NNLL) soft-gluon terms calculated using
\textsc{Top++2.0}~\cite{Beneke:2011mq,Cacciari:2011hy,Baernreuther:2012ws,Czakon:2012zr,Czakon:2012pz,Czakon:2013goa,Czakon:2011xx}.
 
Single-top $tW$ associated production was modelled using the \POWHEGBOX~\cite{Re:2010bp,Nason:2004rx,Frixione:2007vw,Alioli:2010xd}~v2
generator at NLO in QCD in the five flavour scheme with the \NNPDF[3.0nlo]~\cite{Ball:2014uwa} PDF set.
The diagram removal scheme~\cite{Frixione:2008yi} was used to treat the overlap with top-quark pair
production~\cite{ATL-PHYS-PUB-2016-020}.
Single-top $t$-channel and $s$-channel production were modelled using the
\POWHEGBOX~\cite{Frederix:2012dh,Nason:2004rx,Frixione:2007vw,Alioli:2009je,Alioli:2010xd}~v2
generator at NLO in QCD in the four- and five-flavour schemes with the corresponding \NNPDF[3.0nlo]~\cite{Ball:2014uwa} PDF sets respectively.
The matrix element generators were interfaced to \PYTHIA~8.230~\cite{Sjostrand:2014zea} using the A14 tune~\cite{ATL-PHYS-PUB-2014-021}
and the \NNPDF[2.3lo] PDF set. The inclusive cross-sections were corrected to the theory prediction calculated at
NLO in QCD with \HATHOR v2.1~\cite{Aliev:2010zk,Kant:2014oha}.
 
Additional pile-up collisions were overlaid, based on soft QCD processes simulated with \PYTHIA~8.186
using the  \NNPDF[2.3lo] PDF set and the A3 tune~\cite{ATL-PHYS-PUB-2016-017}
over the original hard-scattering events. Additional weighting factors are applied to the fully simulated samples to improve the modelling,
including a factor to reproduce the distribution of the average number of interactions per bunch crossing observed in the data.
 
\subsection{Standard Model predictions}
 
For the particle-level predictions, no data-driven scale factors are applied, and pile-up events are not added.
The settings described above are also valid for the SM predictions used for comparison with the final result,
with the following exceptions.
 
A particle-level prediction for top-quark pair production was produced with
\SHERPA[2.2.11], using NLO-accurate
matrix elements for up to one additional parton, and LO-accurate
matrix elements for up to four additional partons calculated with the
Comix~\cite{Gleisberg:2008fv} and \OPENLOOPS[2]~\cite{Buccioni:2019sur,Cascioli:2011va,Buccioni:2017yxi,Denner:2016kdg}
libraries. They were matched with the \SHERPA parton shower~\cite{Schumann:2007mg} using
the \MEPSatNLO prescription~\cite{Hoeche:2011fd,Hoeche:2012yf,Catani:2001cc,Hoeche:2009rj}.
 
The diboson and single EW boson MC samples were also replaced with calculations produced
with \SHERPA[2.2.11] and \OPENLOOPS[2]~\cite{Buccioni:2019sur,Cascioli:2011va,Buccioni:2017yxi,Denner:2016kdg}.
In the case of \vjets production, the matrix-element-level description of additional emissions was extended
up to five jets at LO\@. The \PDFforLHC PDF set~\cite{Butterworth:2015oua} was used,
supplemented with quantum electrodynamics (QED) effects from \LUXQED~\cite{Bertone:2017bme}.
To improve the description of $W+$jets and $Z+$jets processes, these MC predictions were then
reweighted to account for higher-order EW corrections. The reweighting procedure was based on parton-level
predictions for $W/Z+$jets production from Ref.~\cite{Lindert:2017olm},
and included NLO EW corrections~\cite{Denner:2011vu,Denner:2012ts,Kallweit:2015dum,Denner:2009gj}
supplemented by Sudakov logarithms at two loops~\cite{Kuhn:2004em,Kuhn:2007qc,Kuhn:2005az,Kuhn:2007cv}.
 
In the \onejet region, an alternative prediction was obtained by extending the reweighting procedure
to include NNLO QCD corrections~\cite{Gehrmann-DeRidder:2015wbt,Gehrmann-DeRidder:2016cdi,Boughezal:2016isb,Boughezal:2016dtm}
from Ref.~\cite{Lindert:2017olm}.
These corrections were provided separately for $W+$jets, \Zlljets (where $\ell=e  \text{ or } \mu$) and
\Znunujets processes, as a function of the vector-boson \pT,
to improve the description of the measured $Z$ boson \pT distribution~\cite{STDM-2016-01}.
The reweighting procedure took into account the difference between the intrinsic perturbative accuracy
of the generated MC samples and the provided parton-level calculations.
In addition, the reweighting was extended in the \vbf region to include NLO EW corrections~\cite{Lindert:2022ejn}
for $Vjj$ production, which were provided separately for each decay channel as a function of
dijet invariant mass and azimuthal difference between the tagging jets.
 
In the \vbf region, an alternative prediction for $V+$jets was obtained using the high energy jets (\HEJ)
framework~\cite{Andersen:2009nu,Andersen:2011hs}. \HEJ calculates the tower of leading logarithmic QCD corrections
in the ratio of the partonic centre-of-mass energy and transverse momentum squared, $s/\pt^2$, to all orders in the strong coupling \alphas,
for all relevant SM processes.
These corrections are relevant in regions of phase space where jets span a large range of rapidity or where pairs
of jets have a large invariant mass. The framework includes the matching of these corrections to both tree-level high-multiplicity
matrix elements point-by-point in phase space, and to NLO corrections for distributions.
The framework was implemented in a partonic event generator~\cite{Andersen:2023kuj}. These predictions are only
available for the leptonic measurements.
 
\subsection{Standard Model theory uncertainties}
 
The uncertainties in the SM predictions are estimated following the prescription developed in Ref.~\cite{Lindert:2017olm}.
Uncertainties on pure QCD higher-order corrections in the SM processes are estimated by varying the renormalisation and
factorisation scales by factors of 0.5 and 2. The difference between the nominal prediction
and the envelope of the seven possible versions (the cases where both the renormalisation and
factorisation scales vary upwards or downward at the same time are excluded) is assigned as an uncertainty,
denoted by $\delta^{(1)}K_\mathrm{(N)NLO}$. To account for possible differences in the shape
of the $\pT^V$ and \mjj spectra between low and high scales in the \vjj{} channels,
additional uncertainties $\delta^{(2)}K_\mathrm{(N)NLO}$ and $\delta^{(4)}K_\mathrm{(N)NLO}$
are constructed from conservative shape distortions of the nominal scale-uncertainty band
as a function of $\pT^V$ and $m_{jj}$, respectively.
The distortion is given by $(x^2 - x_0^2) / (x^2 + x_0^2)$ where $x_0$ is the midpoint of the
observable of interest in logarithmic space, namely 650\,\GeV and 1300\,\GeV
for $\pT^V$ and \mjj, respectively. As discussed in Ref.~\cite{Lindert:2017olm}, this
shape uncertainty increases the scale uncertainties by a factor of up to $\sqrt{2}$.
 
All pure-QCD systematic uncertainties are taken as correlated between observable bins
as well as weak bosons and their decay channels,
but uncorrelated between different processes, including EW \vjj{} production.
Following the approach in~\cite{Lindert:2017olm}, the residual level of decorrelation between
the decay modes is estimated from the difference between the differential higher-order $K$-factors
for \wjets production and \Zlljets and \Zvvjets production relative to their average
and assigned as an additional uncertainty, denoted by $\delta^{(3)}K_\mathrm{(N)NLO}$.
A similar uncertainty is estimated for EW $Vjj$ production using the
higher-order $K$-factors calculated in Ref.~\cite{Lindert:2022ejn}.
The uncertainty band for the \HEJ predictions in the \vbf region is estimated from the envelope
of the seven variations of the renormalisation and factorisation scales by factors of 0.5 and 2.
 
Pure EW uncertainties in \vjets and $Vjj$ production arise predominantly from
unknown high-\pT EW effects due to truncation of the perturbative series.
This uncertainty is estimated through naive Sudakov exponentiation,
denoted $\delta^{(1)}\kappa_\mathrm{nNLO\,EW}$, which is taken to be correlated between
the weak boson decay channels.
In the case of \vjets production, an additional conservative uncertainty
$\delta^{(2)}\kappa_\mathrm{nNLO\,EW}$ is assigned, given by 5\,\% of the absolute
full NLO EW correction, which is taken to be uncorrelated between the weak boson decay channels.
Moreover, an uncertainty in the Sudakov approximation at two-loop level in \vjets production is estimated by assigning
an additional uncertainty $\delta^{(3)}\kappa_\mathrm{nNLO\,EW}$, given by the difference between
the next-to-leading logarithmic Sudakov approximation and the naive exponentiation of the full
NLO EW correction, which is also taken to be uncorrelated between the weak boson decay channels.
All pure EW systematic uncertainties are taken to be correlated across bins in a given observable.
 
An uncertainty in unknown non-factorising mixed QCD and EW effects is estimated
from the difference between the additive and multiplicative combination of QCD and
EW higher-order corrections, denoted $\delta K_\mathrm{mix}$.
This systematic uncertainty is taken to be correlated between bins in a given observable
and between weak boson decay channels.
 
PDF uncertainties are estimated by using the sum in quadrature of the set of independent
\PDFforLHC$+$\LUXQED Hessian eigenvectors. The \alphas uncertainty is estimated from
$\pm0.001$ shifts around the nominal value of 0.118 in the PDF sets.
 
To evaluate systematic uncertainties, prompt single-photon production was also simulated
using the \PYTHIA~8.186~\cite{Sjostrand:2007gs} generator.
Events were simulated using tree-level matrix elements for \onegjet final
states and LO QCD dijet events, with the inclusion of initial-
and final-state parton showers.
 
The \PYTHIA simulation includes LO \onegjet events from both
the direct processes (the `hard' $qg \to q\gamma$ and $q\bar{q}\to g\gamma$ component)
and the photon bremsstrahlung in LO QCD dijet events.
The bremsstrahlung component was modelled by final-state QED radiation arising from calculations
of all $2 \to 2$ QCD processes. The \NNPDF[2.3lo] PDF
set was used in the matrix element calculation, the parton shower, and
the simulation of the multi-parton interactions. The samples
include a simulation of the underlying event with parameters set
according to the A14 tune~\cite{ATL-PHYS-PUB-2014-021}. The Lund
string model~\cite{Andersson:1983ia,Sjostrand:1984ic} was used for the
description of the fragmentation into hadrons.
 
Finally, the uncertainty in the interference between top-quark pair production and $tW$ production
is estimated by taking the difference relative to the prediction where the nominal $tW$ sample
is replaced with a version employing an alternative diagram subtraction scheme~\cite{Frixione:2008yi}.


\section{Event selection and reconstruction}
\label{sec:events}

The data used in this analysis were collected with the ATLAS detector in \pp\ collisions  at \sqs~=~13~\TeV{} during the \RunTwo{}
data-taking from 2015 to 2018.
After applying necessary selections to ensure good detector operation conditions, a total integrated luminosity of 140~\ifb{} is available.
The uncertainty in the combined \RunTwo{} integrated luminosity is 0.83\%~\cite{DAPR-2021-01},
obtained using the LUCID-2 detector~\cite{LUCID2} for the primary luminosity measurements.
The average number of inelastic \pp{} collisions per bunch crossing
is 33.7 in the data sample considered. 
Most of  these \pp{} collisions have an interaction vertex that is consistent with the beam-spot envelope.
 
\subsection{Trigger selection}
 
Events for the primary \ptmiss measurement were selected by the two-stage trigger system~\cite{TRIG-2016-01} using the transverse momentum
imbalance within the calorimeter system~\cite{TRIG-2019-01} and requiring hadronic jets in the final state.
Due to  increasing  number of simultaneous \pp interactions in different years of the \RunTwo data-taking the
minimum \ptmiss threshold of the triggers used increased from 70~\GeV to 120~\GeV over the data-taking period
to suppress the impact of the energy contributed by \pileup{} collisions on the rate of accepted events. The algorithm
used to calculate this \ptmiss also varied for the same reason.
All \ptmiss  triggers used in the analysis  were  fully efficient in events for which the offline \( \ptmiss > 200~\GeV\).
 
Since muons deposit very little energy in the calorimeters, the calorimeter-based \ptmiss triggers also selected events with high-\pt
muons in the final-state.
These events are  used for the  single-muon and double-muon auxiliary measurements.
For events with $\ptrecoil>200~\GeV$, this trigger selection 
was 100\% efficient for the subset of those events with a muon with \( \pt \ge 30~\GeV\).
 
A combination of low- and high-\pt single-electron triggers was used to select events for the single-electron and double-electron auxiliary
measurements.
Two  single-electron triggers, with a minimum \pt threshold of 24~(26)~\GeV{} in 2015--2016 (2017--2018)
and \tightID electron identification criteria~\cite{TRIG-2018-05}, selected events in the low-\pt{} region.
In the high-\pt region, where the rate of single electron triggers is low compared to that which can be accommodated by the trigger system bandwidth,
several triggers with less restrictive electron identification were employed to increase the trigger efficiency.
Events satisfying the low- or high-\pt threshold trigger were retained, with an efficiency of around 97\% for electrons with \(\pt\ge80~\GeV\).
The details of the electron trigger combination procedure are summarised in Ref.~\cite{TRIG-2018-05}.
The simulation reproduces the single-electron trigger efficiency measured in data to within 5\% for electrons with \(\pt< 60~\GeV\),
and to better than 1\% in the high-\pt region.
The residual mismodelling is corrected for by reweighting simulated events using data-driven scale factors.
Both statistical and systematic uncertainties in the derived trigger scale factors are propagated to the measured observables.
 
Events selected for the single photon auxiliary measurement are required to have a photon candidate with a minimum \pt of 120~(140)~\GeV{}
at the trigger-level satisfying the \looseID photon identification criteria~\cite{TRIG-2018-05} in 2015~(2016--2018).
In the \pt range above 200~(300)~\GeV a trigger with only a \pt selection was used in addition
(logical `OR') to improve the efficiency of the trigger selection during
2015~(2016--2018) data-taking.
The photon triggers were fully efficient in the single photon auxiliary measurement region phase space.
 
\subsection{Reconstruction and offline selection}
\label{sec:recosel}

Events selected by the trigger system undergo a number of offline reconstruction and calibration steps before they can be used for the analysis.
 
Candidate interaction vertices are reconstructed by associating at least two reconstructed tracks with $\pT > 500$ \MeV to a common origin along
the \pp{} collision axis~\cite{ATL-PHYS-PUB-2015-026}.
Events with at least one such vertex are selected.
In the case of multiple candidate vertices in an event, the primary vertex is defined to be one with the highest sum of squared transverse momenta of associated tracks.
 
Reconstructed tracks in the inner detector (ID) and clusters of energy deposits in the EM calorimeter are  used as inputs to the reconstruction
of electrons and photons.
The  electron and photon reconstruction~\cite{EGAM-2018-01} uses three-dimensional clusters of energy depositions (\topos)
built from topologically connected EM and hadronic calorimeter cells~\cite{PERF-2014-07} to restore energy from bremsstrahlung photons or
from electrons from photon conversions.
The  transition region between the barrel and endcaps of the EM calorimeter,  \(1.37<|\eta|<1.52\), is excluded.
The electron candidates are reconstructed from \topos{} matched to ID tracks.
These tracks are refitted to account for energy losses due to bremsstrahlung.
\Topos{} not matched to any track or matched to conversion vertices are reconstructed as unconverted or converted photon candidates, respectively.
The conversion vertices are formed from one or two tracks that are consistent with a massless   particle decaying within the ID volume.
Electron candidates in the \oneeljets (\twoeljets) region are required to satisfy the \tightID( \mediumID) identification  working point~(WP)~\cite{PERF-2017-03}.
The efficiency to select \tightID(\mediumID) electron candidates reaches a plateau of 88\%~(93\%) for \(\pt>80~\GeV\) electrons.
To reject electrons from  heavy-flavour decays the \customID{HighPtCaloOnly} isolation selection~\cite{EGAM-2018-01}, with 92\%--98\% efficiency
depending on the electron \pt, is applied.
 
Photon candidates with shower shape variables corresponding to the \tightID{} identification working point and satisfying \tightID
isolation criteria are accepted for the single-photon auxiliary measurement~\cite{EGAM-2018-01}.
This combination of identification and isolation  requirements provides a good rejection of photons from non-prompt backgrounds
while maintaining high efficiency for prompt photon selection.
The electron or photon candidate energy is calibrated using energy depositions in the calorimeters and track
measurements in the ID~\cite{EGAM-2018-01}.
The precision of the energy calibration of electrons~(photons) is better than 0.2\%~(0.5\%), verified \insitu using  \Zll and \Zllg events.
 
The muon reconstruction uses track segments in the ID and muon spectrometer (MS), as well as calorimeter information.
Muon candidates are formed by matching the MS and ID tracks and performing a combined fit that makes use of corresponding MS and ID hits,
and accounts for the energy depositions in the calorimeter cells along the muon candidate trajectory.
Identification requirements for muons are formed using selections on track quality and the compatibility between the ID and MS tracks.
Muon candidates in single-muon and two-muon auxiliary measurements are required to satisfy the \mediumID~\cite{MUON-2018-03} identification WP\@.
The  efficiency for identifying \mediumID muons exceeds 98\% for the selection criteria applied in this analysis.
To reject muons produced in semileptonic decays of hadrons, the \customID{FixedCutLoose}~\cite{PERF-2015-10}
requirement is imposed on the activity around muon candidates in the muon auxiliary measurement.
The  \customID{FixedCutLoose} efficiency for selecting a prompt muon ranges from 93\% in the low-\pt{} region to 100\% for
muons with \(\pt>50~\GeV\).
The muon momentum scale is calibrated  using \JPmm and \Zmumu events.  The precision of the muon momentum measurements
changes from 0.05\% for muons within $|\eta|=1$ to 0.15\% for muons in the $|\eta|\sim 2.5$ forward region.
 
Events with no \looseID electrons or muons~\cite{EGAM-2018-01,MUON-2018-03} (regardless of their isolation conditions) are selected for the primary \ptmiss  measurement.
These criteria ensure a very high purity of the signal event sample, since the \looseID identification WPs select at least 92\% of
prompt fiducial electrons or photons and more than 99\% of prompt fiducial muons.
These requirements also reject events with electrons or muons coming from $\tau$-lepton decays.
Hadronically decaying $\tau$-leptons are reconstructed using jets identified by the \antikt jet algorithm, with the radius parameter \(R=0.4\), as a seed, which is then
associated to tracks consistent with $\tau$-lepton production at the interaction vertex~\cite{ATL-PHYS-PUB-2015-045}.
They are then identified as $\tau$-leptons by a recurrent neural network (RNN) algorithm~\cite{ATL-PHYS-PUB-2019-033}.
The \looseID identification WP provides between 87\% and 79\% identification efficiency for $\tau$-leptons while providing a multijet background rejection factor of 21 to 90 respectively depending on the number of associated tracks.
Events with at least one hadronically decaying $\tau$-lepton satisfying the \looseID selection are removed.
 
Jets are reconstructed using the \antikt jet algorithm with the radius parameter \(R=0.4\) 
using an algorithmic combination of the calorimeter energy depositions and the charged-particle tracks.
First, calorimeter cells are grouped into \topos{} using a nearest-neighbour
algorithm~\cite{PERF-2014-07} that exploits the significance of the cell energy compared to the noise expected in the \pileup{} environment for each year of running.
The direction of each \topo{} receives an {origin  correction} to account for the primary vertex position that is different in every event.
The jet measurements are further improved using the particle flow (\customID{PFlow}) algorithm~\cite{PERF-2015-09}, which replaces the charged particle
calorimeter energy deposits by the momenta of the tracks measured in the ID that are associated with the \topos. 
The \customID{PFlow} jets have better energy and angular resolution, as well as reduced sensitivity to \pileup{}, compared to jets reconstructed
from calorimeter information only.
To suppress signals arising from calorimeter noise and other non-collision backgrounds, reconstructed jets are required to satisfy
a \looseID identification selection~\cite{ATLAS-CONF-2015-029}. This selection has a better than 99.5\% efficiency for keeping jets from \pp collisions.
Due to the large instantaneous luminosity, the jets reconstructed in one bunch crossing could originate from different \pp collisions.
To suppress jets arising from vertices other than the primary collision vertex a jet-vertex tagging algorithm~(JVT)~\cite{PERF-2014-03,ATL-PHYS-PUB-2015-034}, 
based on a combination of track-based variables, is used.
Jets with  \(\pt<60~\GeV\) in the central \(|\eta|<2.5\) region are accepted only if the \tightID{} JVT selection 
is satisfied.
In addition, a \tightID requirement from the forward jet vertex-tagging algorithm (fJVT)
is used to reject \pileup jets in the forward region~\(|\eta|\ge2.5\).
 
The jet four-momentum measurement is calibrated using information from both simulation and data~\cite{JETM-2018-05}.
First, the jet energy is corrected for \pileup{} contamination.
An MC-based absolute jet energy correction is used to restore the energy and direction of the jet to that at the particle-level.
Next, the {global sequential calibration} is employed to remove the dependence of the jet response on the energy distribution inside the jet,
and the fluctuations of the shower development in the calorimeter.
Finally a residual  \insitu{} correction, determined from \Z+jets, \gammajet{} and multijet events,
is applied to recover the remaining differences between data and simulation.
Over the rapidity range considered, the jet energy is measured with 1\%--3.5\% accuracy depending on transverse momentum.
 
The missing transverse momentum vector \ptmiss (\ptrecoil in the events with prompt leptons or photons)
is calculated as the magnitude of the vector sum of the transverse momenta of all particles produced in the event~\cite{PERF-2014-04}.
Detector signals associated with identified physics objects constitute a hard term,
while the signals that are not part of these objects form a soft term.
The \ptmiss reconstruction uses energy deposits from the calorimeter, muons reconstructed in the MS, and tracks from the ID.
 
The \ptmiss is then given  by $\ptmiss = \sqrt{(p_{x}^{\mathrm{miss}})^2 + (p_{y}^{\mathrm{miss}})^2}$, where $p_{x(y)}^{\mathrm{miss}}$ are calculated as follows:
\begin{equation}\label{METcalculation}
p_{x(y)}^{\mathrm{miss}} = p_{x(y)}^{\mathrm{miss},e} + p_{x(y)}^{\mathrm{miss},\gamma} + p_{x(y)}^{\mathrm{miss},\tau} + p_{x(y)}^{\mathrm{miss},\mu} + p_{x(y)}^{\mathrm{miss,jets}} + p_{x(y)}^{\mathrm{miss,soft}}
\end{equation}
where each term is calculated as the negative sum of the calibrated reconstructed objects, projected onto the $x$ and $y$ directions.
The soft term, $p_{x(y)}^{\mathrm{miss,soft}}$, is calculated from tracks associated with the primary vertex but not to any of the
high-\pt objects.
 
For the calculation of \ptrecoil in the auxiliary measurements, the same expression is used, but the identified prompt leptons or photons are excluded.
 
Events involving $\tau$-leptons can enter the signal region if the $\tau$-lepton is not reconstructed. Conversely, in the auxiliary measurement regions, it is possible for events where a jet is misreconstructed as a $\tau$-lepton to affect the $p_{x(y)}^{\mathrm{miss}}$ calculation.
In principle, if this effect is not accounted for  it could lead to biases when correlating the regions.
However, $\tau$-lepton reconstruction is found to be well modelled across all the measured regions.
 
Since physics objects are reconstructed independently of each other, there is a possibility that the same
detector signals are used to build multiple jets, photons or leptons.
To avoid double-counting of particle level physics objects, the following procedure is employed.
First, leptonically decaying $\tau$-leptons closer than \(\Delta R=0.2\) to an electron or muon are discarded.
Second, electrons that share the same ID track with a muon are rejected.
Third, jets are removed if the jet momentum direction is closer than \(\Delta R < 0.2\) to any electron candidate.
In turn, all electrons that are within \(\Delta R = 0.4\) of a jet are discarded.
Similarly, jets are discarded if they are within \(\Delta R = 0.2\) of a muon and have less than three associated tracks,
while muons within \(\Delta R = 0.4\) of a jet are rejected.
Finally, jets that are within \(\Delta R = 0.2\) of a hadronically decaying $\tau$-lepton are removed.
If a lepton is removed by this procedure, the event is still considered for other measurement regions with fewer or no leptons.


\section{Background estimation}
\label{sec:backgrounds}
 
Two categories of background contribute to all the measurements made: non-collision backgrounds, produced by beam--gas or cosmic rays events or calorimeter noise,
and reducible backgrounds, which arise when a miscalibration or a misidentification of physics objects leads to an artificially large missing transverse momentum or
a spurious particle candidate.
In addition, in the \Znunu measurement, contributions from other SM processes that satisfy the true event selection and therefore
cannot be distinguished from \Znunu events within the fiducial region are treated as an irreducible background.
In this section the background estimation methods are briefly described.
 
\subsection{Non-collision background\label{sec:noncol}}
 
Muons produced away from  the proton--proton collision but in-time with it can create significant energy depositions in the calorimeter that can be reconstructed
as hadronic jets, and thus lead to events with a single jet and a large \ptmiss signature.
Such muons can be created by the cosmic-ray showers, or by interactions upstream of the ATLAS detector between the beam and the LHC collimators, or residual gas in the beam pipe.
The azimuthal angle distribution of the fake jets they produce has a characteristic shape, with pronounced peaks at \(\phi=0\)  and \(\phi=\pi\).
Moreover, these muons enter the calorimeter earlier than jets from the interaction point, and so have very different timing properties.
The non-collision background contribution is strongly reduced by the identification requirements applied to the leading jet in event.
The residual contribution from this background source is evaluated using a data-driven approach that exploits the differences in time between the signal jets
produced in the collision vertex and the non-collision background jets, and is subtracted. This amounts to a few thousand events over the course of the data-taking period.
The difference between the
subtracted and non-subtracted sample is taken as the uncertainty and propagated through to the final results.
 
\subsection{Multijet background in the \ptmissjets selection}
\label{sec:multijet}
 
Jet production processes containing no prompt \ptmiss can contribute to the event yield in the primary measurements
when jets are mis-reconstructed or mis-calibrated, giving rise to fake \ptmiss.
In addition, decays of heavy flavour hadrons among the jet constituents may produce neutrinos that can generate the \ptmiss.
In such cases the \ptmiss vector will typically be aligned with the direction of the jet, and the \(\ \DPhiJM>0.4\) requirement removes most of this type of
background. However, the \ptmissjets selection will contain a residual multijet background, since in such cases the \ptmiss can receive contributions from
several jets and the resulting \ptmiss direction may not align with any one of them.
The probability to reconstruct a large \ptmiss in any given multijet event is rather low,  but the jet production cross-section is large. This implies that
a simulation-based approach would require a very large simulated event sample, with an extremely accurate modelling of hadron production and calorimeter performance,
especially in the tails of the distributions. These considerations mandate a data-driven method.
 
Reference~\cite{SUSY-2011-20} contains a detailed description of the jet smearing method used to estimate the multijet background contribution in the primary measurements.
A high-statistics sample of low-\ptmiss events with well-measured hadron jets is collected using a set of inclusive jet triggers with different \ptjet thresholds.
A set of `pseudodata' events  is created by fluctuating the jet energies in these events using a function, constrained using data,
that models the detector response to jets. Each fluctuation is considered as a separate event;
the altered four-momenta of the jets are stored and the \ptmiss vector is recalculated.
This approach produces pseudodata events with fake \ptmiss populating a range up to about 2~\TeV.
 
The distribution of the multijet background in each differential cross-section measured is taken from the pseudodata distributions after
applying the \onejet or \vbf event selection as appropriate,
while the normalisations are obtained from the fit to data in a dedicated multijet background-enriched control region, defined using the \ptmissjets event selection
as given in Table~\ref{tab:jets-phasespace}, but with the requirement on the azimuthal angle between the jet direction and \ptmiss vectors inverted i.e., \(\DPhiJM\leq 0.3\).
 
Events with  \(0.3< \DPhiJM < 0.4\) satisfying the \ptmissjets requirements but with the \ptmiss selection relaxed to 130~\gev (as low as possible considering trigger thresholds)  are used to verify the multijet background estimation procedure.
In this region, where the multijet background contributes approximately four times more events than the other SM processes, very good agreement is observed between the data and predictions.
A similar region without the relaxed \ptmiss requirement, which is closer to the kinematic regime probed by this analysis, is also tested.
Figure~\ref{fig:bkg_multijet_ptmiss_vr} shows the distribution of events in this validation region  as a function of \ptmiss. Although the multijet background is not the dominant contributor to the yield in this validation region, it represents a significant fraction of the first few bins, where its inclusion leads to a good agreement between prediction and data. The multijet component in these validation regions is approximately an order of magnitude more important than in the signal region. There is a very good agreement between the prediction and data.
 
\begin{figure}
\centering
\subfigure[]{\includegraphics[width=0.49\linewidth]{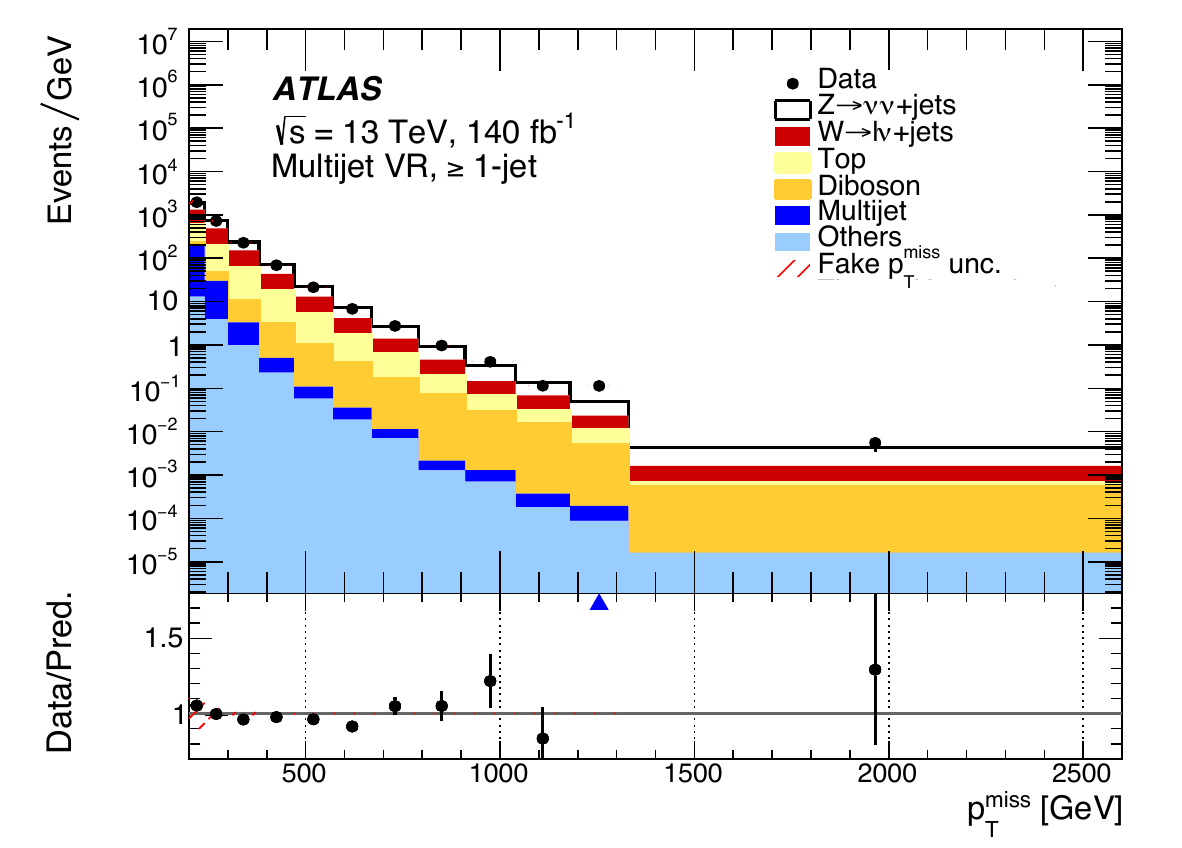}}
\subfigure[]{\includegraphics[width=0.49\linewidth]{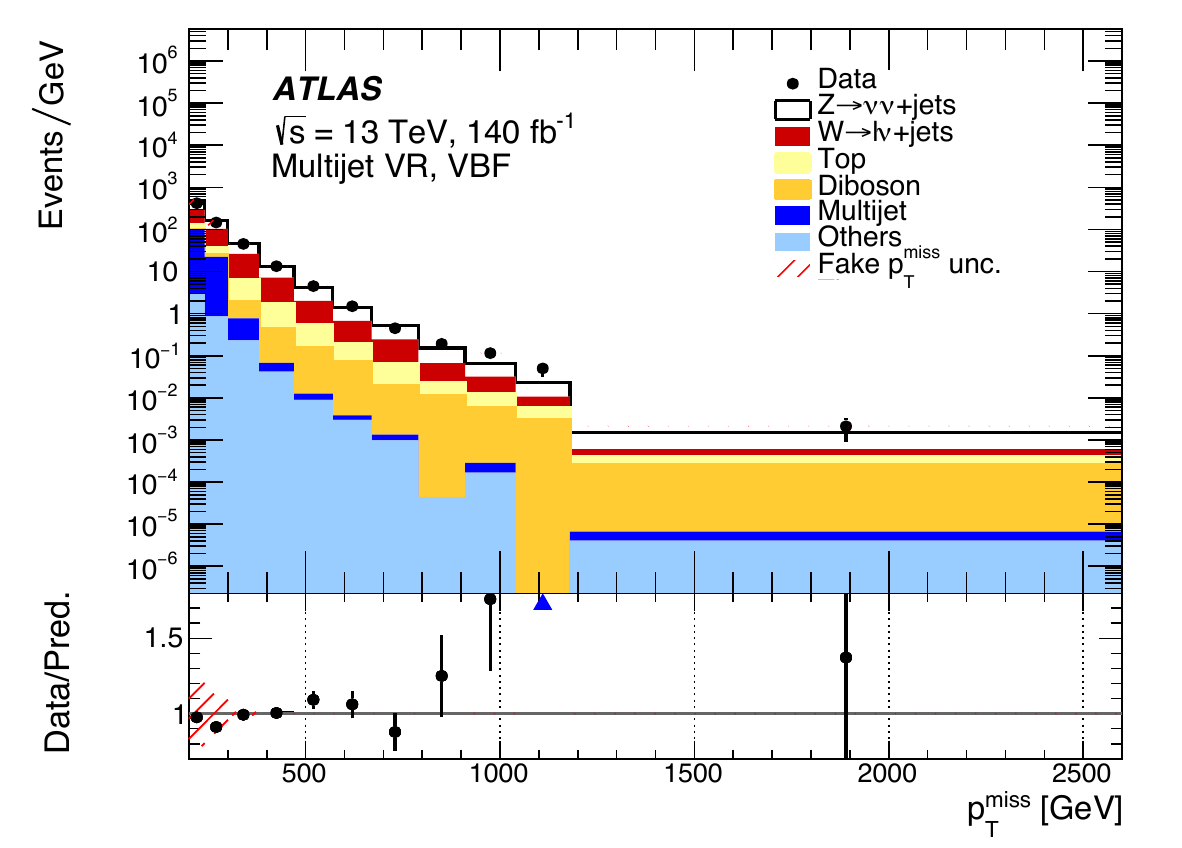}}\\
\caption{
The event yield in the multijet background validation region as a function of \ptmiss for the (a) \onejet{} selection and (b) \vbf{} selection.
Points denote the data, and different SM backgrounds are shown as histograms.
The hatched band shows the full uncertainty assigned to the estimate of the fake \ptmiss contribution from multijet events, which, since it represents a very small contribution to the final event sample and is subject to statistical variations due to the nature of the evaluation, is conservatively taken to be 100\% of the estimated yield. The vertical lines represent the statistical uncertainty. The bottom panels show the ratios of the data to the predictions; the blue triangle
indicate values which are out of the display range.
}
\label{fig:bkg_multijet_ptmiss_vr}
\end{figure}
 
The multijet background contribution in the \ptmissjets event selection in the \onejet region is smaller than 1\%  for
\(\ptrec\leq 300~\GeV\) and falls steeply as \ptmiss increases.
In the \vbf region the multijet background contribution is around 1\%--2\% in the \(\mjj<2~\TeV\) region, and is negligible in the high-\mjj{} range.
 
\subsection{Background from misidentified photons and leptons in the auxiliary measurements}
\label{sec:fakebkg}

The main source of background in the electron and photon auxiliary measurements comes from jets that are misidentified as leptons or photons.
This can occur due to fluctuations in jet formation, or in the development of hadronic showers in the calorimeter, leading to an apparently
large amount of EM energy. In conjunction with inefficiencies in the inner tracker, this can lead to energy depositions that
are reconstructed as a photon or an electron.
Jets containing heavy flavour hadrons that decay into final states including a muon are the main source of the non-prompt muon background.
The yields of such `fake' photons and leptons are heavily suppressed by the reconstruction algorithms, while the remaining contributions are removed
using a set of data-driven techniques described below.
 
\subsubsection{Jet-photon misidentification contribution to the \onegjets selection}
\label{sec:fakegamma}
 
Multijet production is the dominant source of background in the \onegjets auxiliary measurement.
The \customID{Tight} photon identification WP~\cite{EGAM-2018-01} together with the \customID{Tight}
requirement on the photon isolation energy strongly suppress the jet-photon misidentification.
A data-driven two-dimensional side-band method~\cite{STDM-2014-09, STDM-2017-01} is used to determine the shape and the normalisation of the residual
photon misidentification background.
 
For this, four samples (A, B, C and D) are selected by splitting the \customID{Tight} (A, B)  and non-\customID{Tight} (C, D) photon samples
into \customID{isolated} (A, C) and non-\customID{isolated} (B, D) samples.
Background-enriched samples of non-\customID{Tight} photons are collected by requiring the photon to satisfy the \customID{LoosePrime4}
and to simultaneously fail to satisfy the \customID{Tight} identification criteria.
The \customID{LoosePrime4} WP has a relaxed selection on the photon shower-shape properties that is not correlated with the photon isolation conditions, meaning
the background in the four sub-regions can be treated as being uncorrelated.
The yield $N_\text{A}$ of background events in the \onegjets region is determined by interpolating the measured event yields in the
other three control regions, \( N_\text{A} = N_\text{B} \times \frac{N_\text{C}}{N_\text{D}} \).
The effect of leakage of signal photons into the control regions is in the range from 1-8\% for
the C and D regions and from 12-20\% for the B region; the contribution from other processes is negligible. These contributions are
evaluated from the simulation and accounted for in the systematic uncertainties.
 
In most parts of the phase space the contribution of the jet-photon background is around 2\%--3\%.
It increases to 4\% in the \(\ptmiss\le 300\)~\GeV{} region, while in the \(\mjj\le 400~\GeV\) range up to  5\% of events are  from the photon misidentification.
 
\subsubsection{Misidentified electrons in the \oneeljets selection}
\label{sec:fakeel}
 
Multijet events may satisfy the electron identification and isolation requirements because of the presence of semileptonic heavy-flavour decays, photon conversions, or
hadrons inside jets being misidentified as electrons.
This background is evaluated using a data-driven  matrix method~\cite{EGAM-2019-01} that exploits the fact that prompt electrons (P) are better isolated in
comparison to background electrons (B).
 
The number of electrons that satisfy the \customID{Tight} selection,  \(N_\textrm{T}\), and which satisfy the \customID{Loose} but fail to satisfy the  \customID{Tight} selection,
\(N_\textrm{!T}\), can be expressed in terms of the efficiency \( \epsilon_\textrm{P} \left( \epsilon_\textrm{B}\right) \) with which prompt~(background)
electrons that satisfy the \customID{Loose} selection also satisfy the \customID{Tight} selection as:
 
\begin{equation} \label{eqn:mmstart}
\begin{pmatrix}
N_\textrm{T} \\N_\textrm{!T}
\end{pmatrix}
=
\begin{pmatrix}
\epsilon_\textrm{P}  & \epsilon_\textrm{B} \\
1- \epsilon_\textrm{P}   & 1-\epsilon_\textrm{B} \\
\end{pmatrix}
\begin{pmatrix}
N_\textrm{P} \\N_\textrm{B}
\end{pmatrix},
\end{equation}
 
where  \(N_\textrm{P}\) (\(N_\textrm{B}\))  is the number of prompt (background) electrons.
Solving this matrix equation, the number of background electrons in data that satisfy the \customID{Tight} requirements
can be obtained using event yields  measured in data as:
\begin{equation} \label{eqn:mmfinal}
\epsilon_\textrm{B} N_\textrm{B} = N_\textrm{T}^\textrm{Bkg} = \frac{\epsilon_\textrm{B}}{\epsilon_\textrm{P} -\epsilon_\textrm{B}}
\left(  \left(\epsilon_\textrm{P}-1 \right)  N_\textrm{T}  + \epsilon_\textrm{P}N_\textrm{!T}   \right).
\end{equation}
 
The  \( \epsilon_\textrm{P}  \) efficiency is estimated from simulation binned in electron \pt and \(\eta\).
It changes from 85\% to 95\% as the electron \pt increases from 100~\GeV to 600~\GeV, where it reaches a plateau.
 
The  \( \epsilon_\textrm{B}  \)  efficiency is determined from data in a dedicated control region with the nominal \oneeljets
requirements on the \mT{} and \ptmiss inverted to enhance the background electron contribution, and the \ptrec requirement removed.
The events with genuine prompt electrons in this region are subtracted by using the simulation.
The resulting  \( \epsilon_\textrm{B} \) efficiency is measured in bins of electron \pt and \(\eta\).
It decreases rapidly from 15\% for electrons with \pt around 50~\GeV to about 1\% in the \(\pt \geq 500~\GeV\) region.
Combining the measured \( \epsilon_\textrm{P} \) and \( \epsilon_\textrm{B} \) efficiencies, the multijet background in the \oneeljets auxiliary measurement
is found to be less than 5\% in the \(\ptrec<500~\GeV\) region, while it increases up to almost 20\% in the \(\ptrec>1500~\GeV\) range.
 
\begin{figure}
\centering
\subfigure[]{\includegraphics[width=0.49\linewidth]{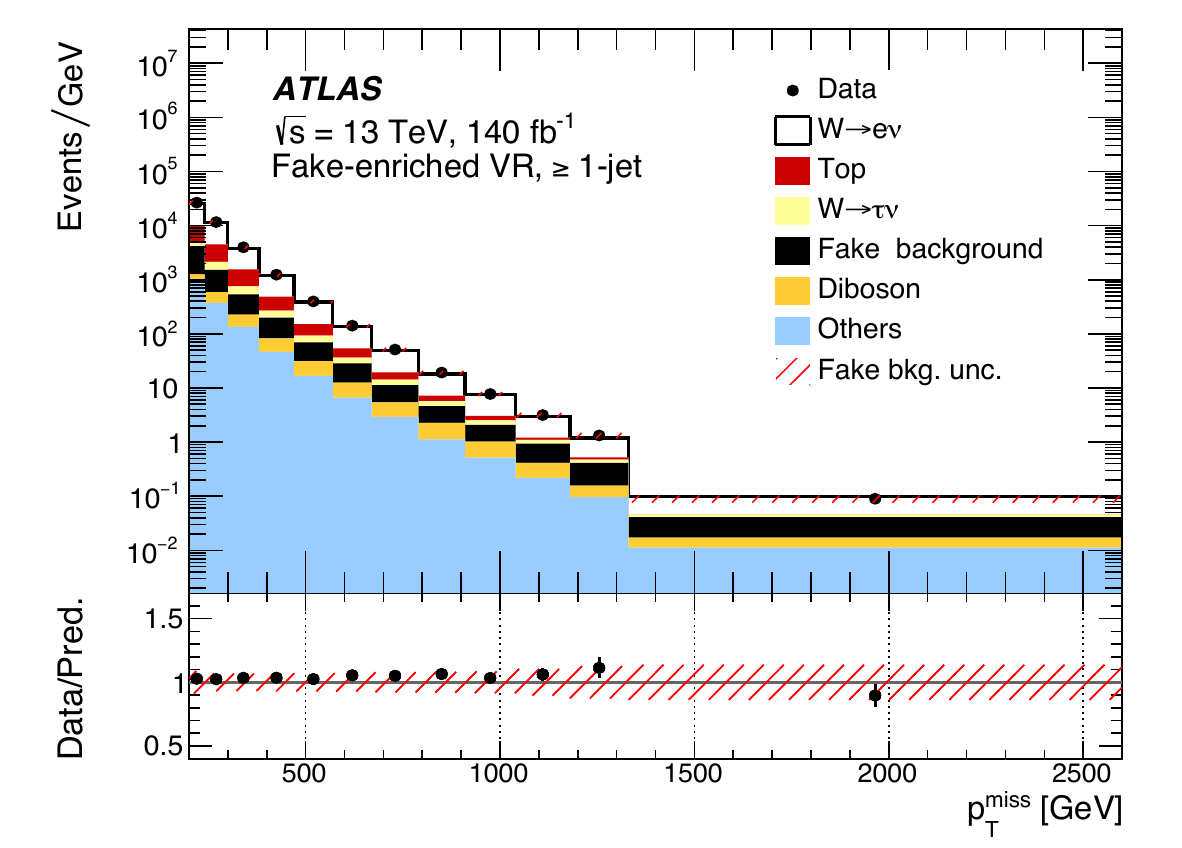}}
\subfigure[]{\includegraphics[width=0.49\linewidth]{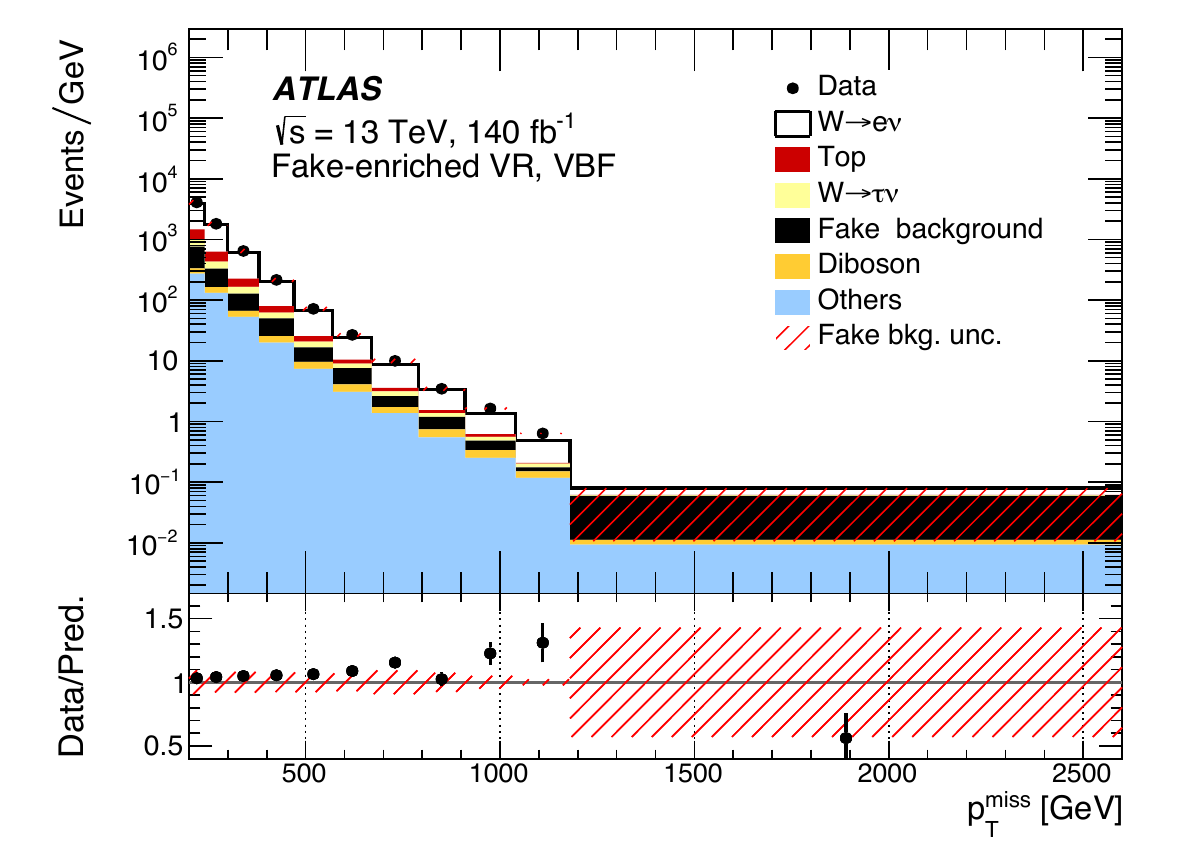}}\\
\caption{
The event yield in the \oneeljets background validation region as a function of \ptmiss for the (a) \onejet{} selection and (b) \vbf{} selection.
Points denote data, and estimated backgrounds are shown as histograms.
The hatched band shows the full uncertainty in the `fake background' arising from misidentified leptons (described in Sec.~\ref{sec:uncertainties})
and the vertical lines represent the statistical uncertainty. The bottom panels show the ratios of the data to the predictions.
}
\label{fig:bkg_multijet_wenu_vr}
\end{figure}
 
This estimate of the background contribution from misidentified electrons is validated using the \oneeljets event selection
but without the requirement
on \mT{} or \ptmiss. Figure~\ref{fig:bkg_multijet_wenu_vr} shows the distributions of events passing this modified selection as a function of \ptmiss.
The misidentified electron background contributes a significantly larger proportion of the total yield in these validation regions compared with the auxiliary measurement regions.
Very good agreement between data and prediction is seen after accounting for the misindentified electron background.

\subsubsection{Non-prompt lepton background in the \onemujets{}, \twomujets{} and  \twoeljets{}  auxiliary measurements  \label{sec:fakemu}}
 
Heavy-flavour hadron decays with a muon in the final state are the dominant source of muon misidentification in multijet events.
The relatively low branching ratio for muon-channel hadronic decays, together with a high efficiency for identifying prompt muons,
leads to a very high purity in the  single-muon and dimuon event samples, allowing the use of the `fake-factor' method.
A similar argument (high purity of the dielectron sample) implies that the same method can also be used in the \twoeljets{} auxiliary measurement.
 
The fake-factor method is a simplification of Eq.~(\ref{eqn:mmstart}).  It relies on the quality of the simulation in describing the detector response to a prompt lepton.
Accordingly,  the number of non-prompt leptons satisfying the  \customID{Tight} lepton identification, \(N_\textrm{T}^\textrm{Bkg}\),  can be estimated as:
\begin{equation} \label{eqn:fffinal}
N_\textrm{T}^\textrm{Bkg} = \frac{\epsilon_\textrm{B}}{1 -\epsilon_\textrm{B}}
\left(   N_\textrm{!T}  - N_\textrm{!T}^\textrm{P} \right),
\end{equation}
where \(N_\textrm{!T}  \) is the yield of \customID{Loose} leptons that fail to satisfy the \customID{Tight} identification WP observed in data and  \(N_\textrm{!T}^\textrm{P} \)
is the number of prompt leptons satisfying \customID{Loose} but failing the  \customID{Tight} lepton identification determined in the simulation.
 
The fraction of fake leptons satisfying the selection \(\epsilon_\textrm{B}\) is measured in data using a
dedicated sample of events containing a pair of leptons with different flavour but the same sign charge.
A small fraction of events in this sample can be attributed to true prompt lepton production in the two-boson or \ttbar{} processes,
while the rest are fakes.
The prompt lepton contribution is accounted for using the simulation.
 
As a function of lepton \pt{}, \(\epsilon_\textrm{B}\) for muons is flat at 5\% up to around 30~\GeV, after which it increases up to 55\% at 1~\TeV.
The  same \(\epsilon_\textrm{B}\)  is used in both single-muon and dimuon  auxiliary measurements.
The fake muon background in the \onemujets(\twomujets) event selection is around 5\% ($<$1\%).
Similarly, \(\epsilon_\textrm{B}\) for the electron definition in the \twoeljets{} region is measured as a function of electron \pt{}.
It is found to be 14\% for 7~\GeV\ electrons and steadily increases up to 31\% for 110~\GeV\ electrons.
Fake electrons are predominantly at low \pt{}, and due to the analysis selection these are generally
accompanied by a high \pt{} electron, where the fake rate is low.
Therefore the estimate of non-prompt electron background for \twoeljets{} events is found to be relatively small, and does not exceed 1\%.

\subsection{Contributions from other SM processes \label{sec:smbkg}}
 
SM processes with charged leptons in the final-state, involving single or double electroweak bosons as well as single top-quark
and \ttbar production, can contribute to the
measured observables. This can even occur in the zero-lepton measurements, for example when one or two final-state leptons are produced outside
of the analysis \pt or \(\eta \) acceptance, or are not reconstructed because of
detector inefficiencies or a poor lepton isolation due to underlying event or \pileup activity around the final-state lepton.
 
In the inclusive \ptmiss and \ptrec measurements, events from these sources that satisfy the fiducial phase-space selection are treated as signal, and instrumental effects are
corrected for later. In the \Znunu measurement, they are treated as irreducible backgrounds, and are accounted for with a semi-data-driven approach:
the shapes of the distributions are taken from the SM predictions, while the normalisations are extracted from fits to data using a set of four control regions defined by
adding to the \ptmiss and other selection criteria the requirement that one or more charged leptons with $\pt > 30$ \GeV be present.
A summary of the SM contributions and their relative importance in various signal regions of the analysis is given in Table~\ref{tab:process_composition}.
The level of contribution varies strongly with measurement region. The photon auxiliary region contains \textgreater\,99\% \onegjets{} events, and the two-lepton auxiliary measurements are expected to consist of around 95\%  \Zboson\ boson production events.  Due to this very high purity,
both the shape and normalisation of the top-quark and multi-boson contributions in these regions are taken directly from the corresponding
simulation.
 
\begin{table}[htb]
\centering
\label{tab:process_composition}
\caption{The relative contributions of SM processes to the \ptmissjets{} and auxiliary measurements event selections.
The contributions are calculated using the MC simulation.}
\begin{tabular}{c  r r r r r r}
\toprule
& \multicolumn{6}{c}{Final-state event selection}\\ \cmidrule{2-7}
Production process &  \ptmissjets  &  \twoeljets  & \twomujets  & \oneeljets  & \onemujets  & \onegjets \\
\midrule
\Znunu + jets  & 55\%  &  --  &  --   &  -- &  --  &   --\\
\Zee   + jets     &  --   & 94\% &  --   &  -- &  --  &   --\\
\Zmumu + jets &  --  &  --  & 95\% &  --  &  2\%  &   --\\
\Wenu  + jets   &  6\%   &  --  &  --  & 68\%  &  --  &   --\\
\Wmunu + jets &  9\%   &  --  &  -- &  --    & 67\% &   --\\
\Wtaunu+ jets  & 20\% &  --  &   --  &  5\%  &  7\%   &   --\\
\Gamma + jets &  --  &  --  &  --  &  --  &  --   & $>$99\%\\
Top                     &  7\%  &  3\%  &  2\%  & 25\%  & 21\%  &   --\\
Multi-boson     &  3\%  &  3\%  &  3\%  &  2\%  &  3\%   &  $<$1\%\\
\bottomrule
\end{tabular}
\end{table}
 
For the charged lepton measurements, the normalisations of the  \Wtaunu and top-quark background distributions are determined
in a combined fit to the \Wboson\ and top-background control regions.
The fit is performed separately in the electron and muon channels for each  \onejet and \vbf event selection.
The scale-factors for the top-quark background are found to be around 0.70--0.78,  while the  \Wtaunu{} distributions are rescaled by factors of 1.10--1.12.
The background normalisation factors in different phase-space regions and \(\tau\)-lepton  decay channels agree with each other within statistical and systematic
errors in the background estimation procedure.
 
The normalisations of the \Wenu,\Wmunu, \Wtaunu and top-quark  contributions to the \ptmissjets event selection are extracted in a simultaneous
fits to data in control regions for the \onejet and \vbf event selections separately.
As a result, the \Wboson production contributions in the \ptmissjets event selection are rescaled by factors ranging from 1.04 to 1.13,
depending of the phase-space region and the \Wboson\ boson leptonic decay channel.
The top-quark background distributions scale-factors are 0.97--0.98 in both the \onejet{} and \vbf{} regions.
 
For the inclusive measurement, rescaled contributions are used to construct the simulated sample for unfolding. For the \Znunu measurement,
they are subtracted from the data before unfolding.


\section{Detector correction and systematic uncertainties}
 
\label{sec:correction}
 
The data are corrected for detector effects so that they are presented in terms of particle-level objects, as
defined in Section~\ref{sec:particlelevelobjects}.
 
\subsection{Unfolding procedure}
 
An efficiency correction and an iterative Bayesian unfolding technique are used to correct the data and obtain particle-level differential
cross-sections.
A prior truth distribution is used as input, and biases from the prior are accounted for by iterating the unfolding after reweighting the input
distributions to the corrected data from the previous iteration.
The optimal number of iterations is determined by balancing the fact that fewer iterations results in a stronger bias from the input prior
(which is treated as a systematic uncertainty), whilst increasing the number of iterations increases the statistical uncertainty.
Two iterations proved to be an optimal number for all measured distributions.
 
The inputs to the unfolding procedure are:
 
\begin{itemize}
\item \textbf{Migration matrix}. Events in a specific bin at particle level can migrate to a different bin in
the reconstructed distribution, due to finite detector resolution. The migration matrix maps the true distribution onto the
reconstructed distribution using events satisfying both the particle- and detector-level
selections.
 
\item \textbf{Reconstruction matching efficiency}. Due to the efficiency and acceptance of the detector, only a fraction of particle-level
events are reconstructed within the target phase space. The reconstruction efficiency accounts for this
and is defined as the ratio of simulated events that satisfy both the particle- and detector-level selections to all events satisfying the
particle-level selection, as a function of the particle-level value of the observable being considered.
 
\item \textbf{Fiducial fraction}. Due to the finite resolution of the detector, events that do not satisfy the particle-level selection can still satisfy
the detector-level selection and be included in the detector-level distribution. The fiducial fraction accounts for
this and is defined as the ratio of simulated events satisfying both the particle- and detector-level selections
to those that satisfy only the detector-level selection, as a function of the detector-level value of the observable being considered.
 
\item \textbf{Purity}. This quantity encapsulates the size of migrations between bins for events as a function of detector-level values of
the observable being considered. It is defined as the fraction of the entries in a detector-level bin that are in the same bin at particle level.
 
\item \textbf{Stability}. This quantity encapsulates the size of migrations between bins for events as a function of particle-level values of
the observable being considered. It is defined as the fraction of the entries in a particle-level bin that are in the same bin at reconstruction level.
\end{itemize}
 
The binning of each distribution is defined so that each bin is expected to contain at least 20 reconstructed events, with a purity of at
least 60$\%$.
 
Figure~\ref{fig:metMono_mig} shows, as an example, the migration matrices for \ptmiss
in the \ptmissjets and \ptrec in the \onegjets region, both for the \onejet phases space,
while Figure~\ref{fig:metMono_grouped_topo} shows the matching efficiency, fiducial fraction, purity
and stability for \ptrec for all auxiliary measurements in
the \onejet phase space.
 
The efficiency is lowest for the \oneeljets region, due to the requirements applied to the real \ptmiss{} and transverse mass.
The highest efficiency is seen in the \ptmissjets region, which has no leptons to reconstruct.
The fiducial fraction is around 0.9 in this region, due to contributions from \wjets events with a particle-level
lepton that is within the fiducial acceptance. These fail to satisfy the particle-level lepton veto,
but satisfy the reconstruction-level lepton veto due to inefficiencies in reconstructing the lepton.
The migration matrices and purity plots are similar between all regions as the migration between \ptmiss bins depends primarily
on the hadronic recoil.
The same qualitative features are present in the \pt of the leading jet in the \vbf phase space (not shown).
 
\begin{figure}
\centering
\subfigure[]{\includegraphics[scale=0.35]{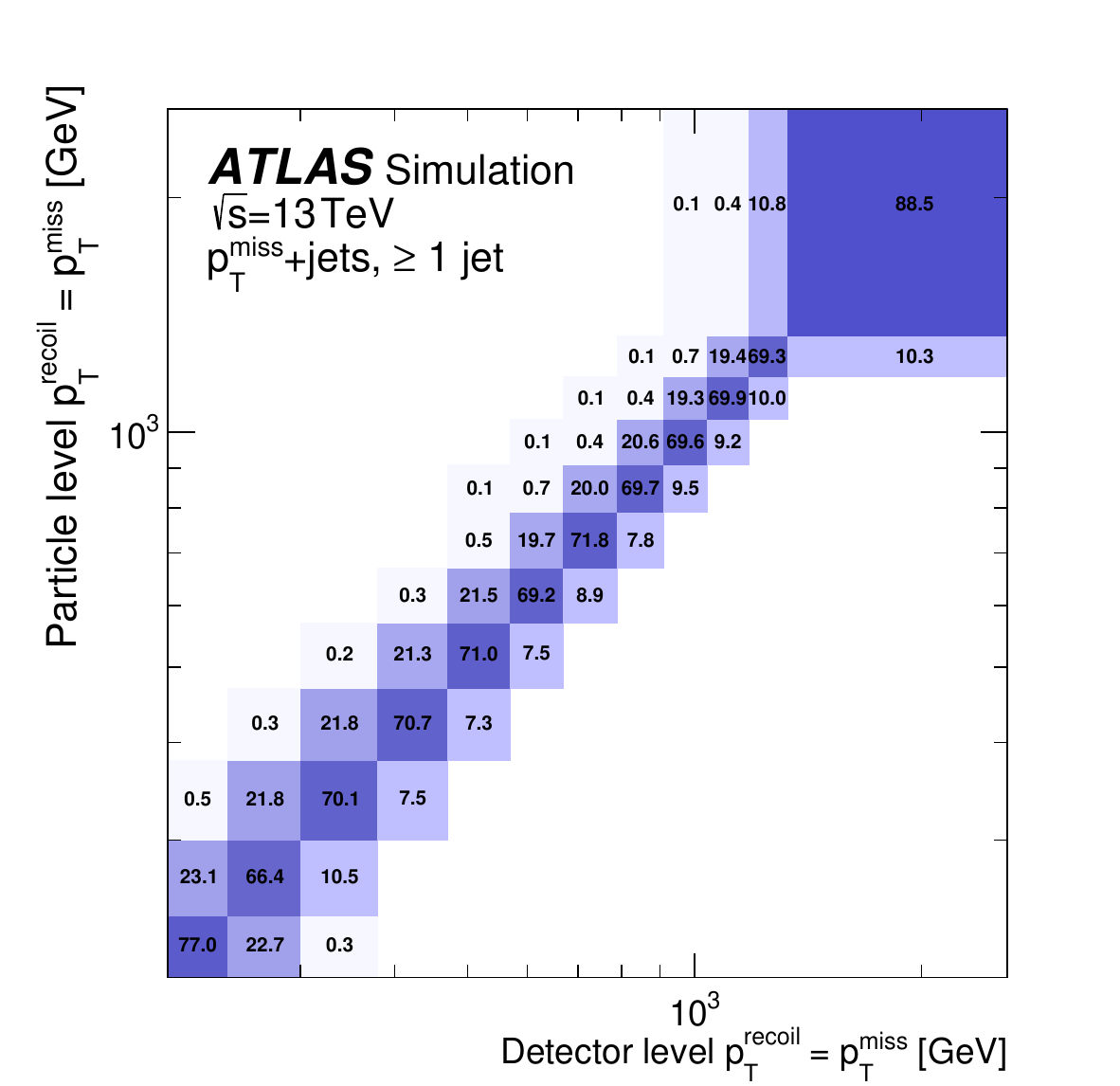}}
\subfigure[]{\includegraphics[scale=0.35]{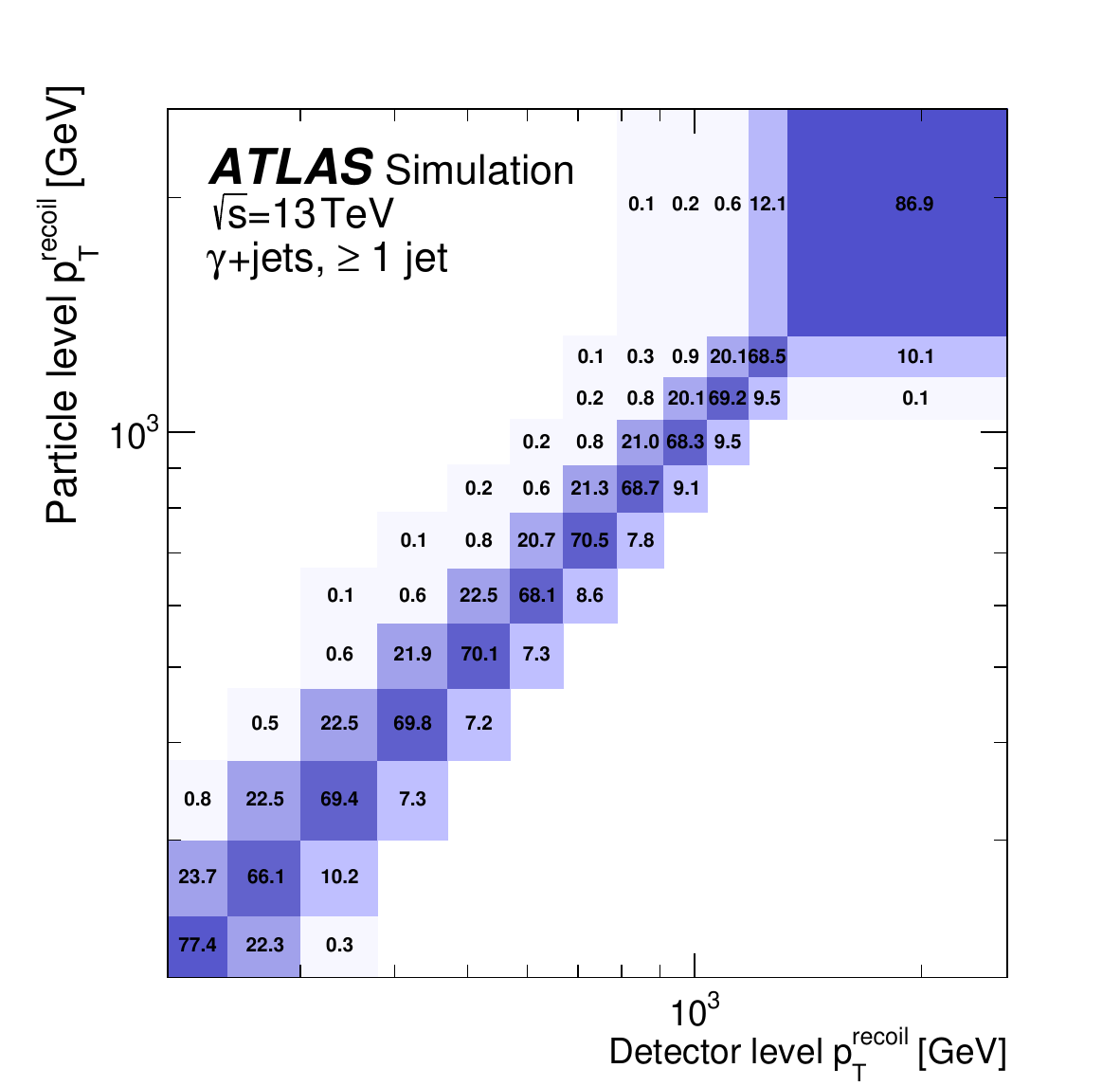}}\\
\caption[]{Migration matrices for (a) the \ptmiss in the \ptmissjets region and (b) \ptrec in \onegjets region, the \onejet phase space,
constructed for all processes that enter the fiducial phase space.
}
\label{fig:metMono_mig}
\end{figure}

\begin{figure}
\centering
\subfigure[]{\includegraphics[scale=0.4]{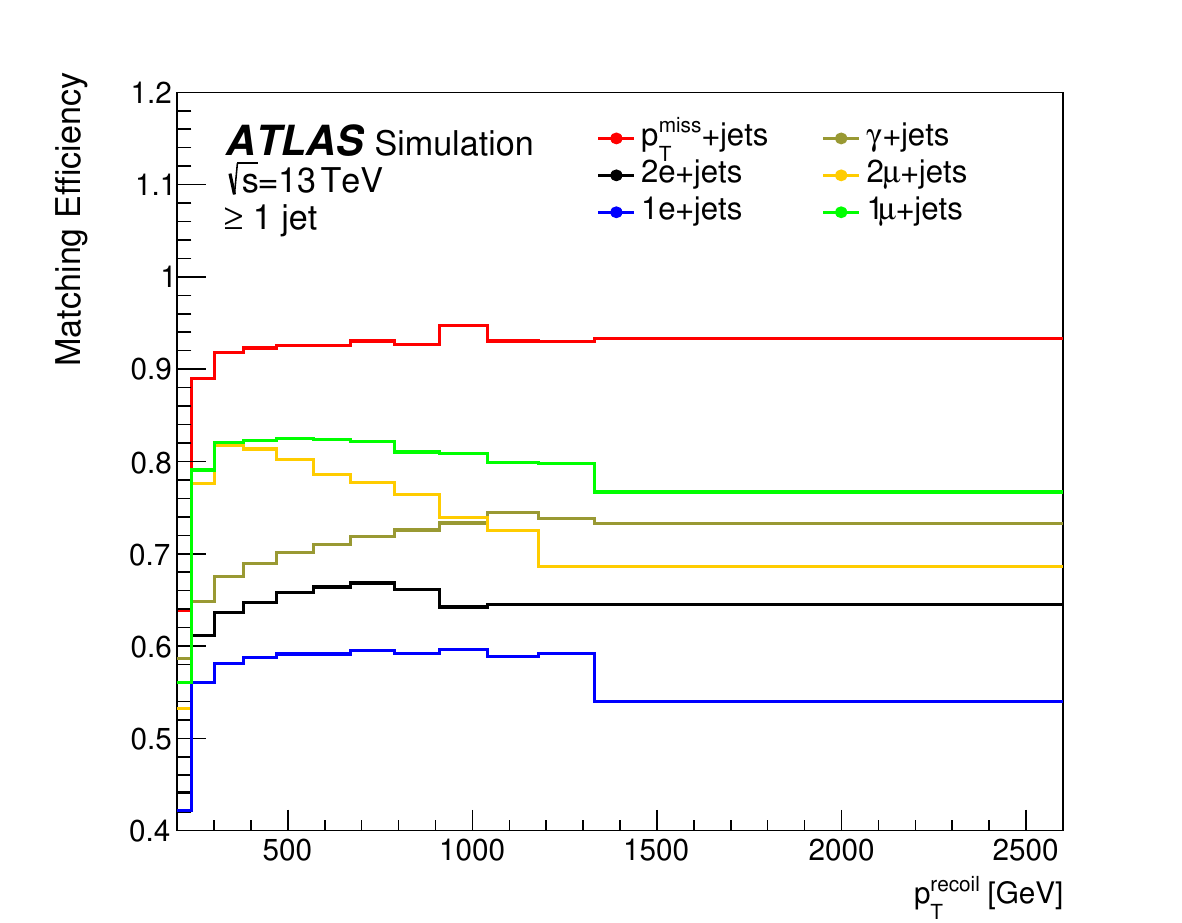}}
\subfigure[]{\includegraphics[scale=0.4]{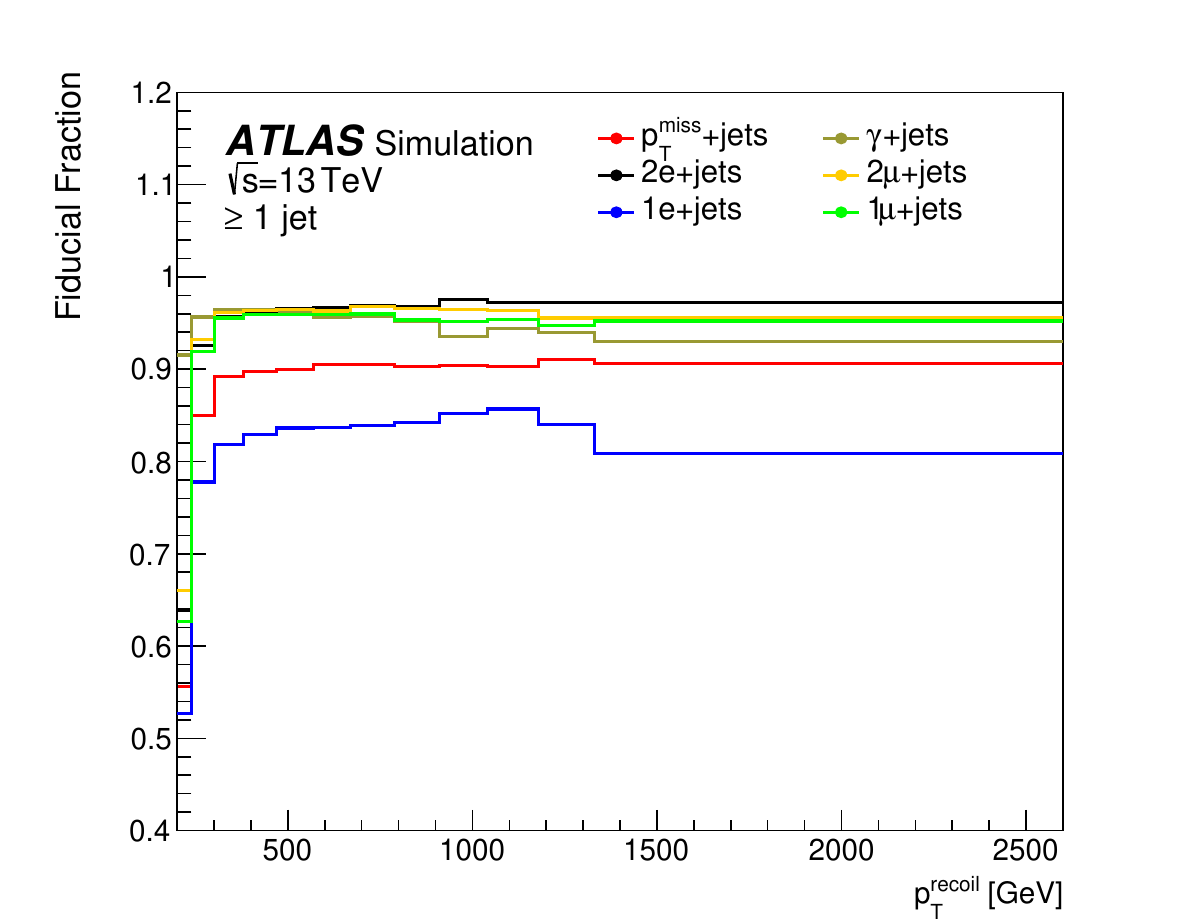}}
\subfigure[]{\includegraphics[scale=0.4]{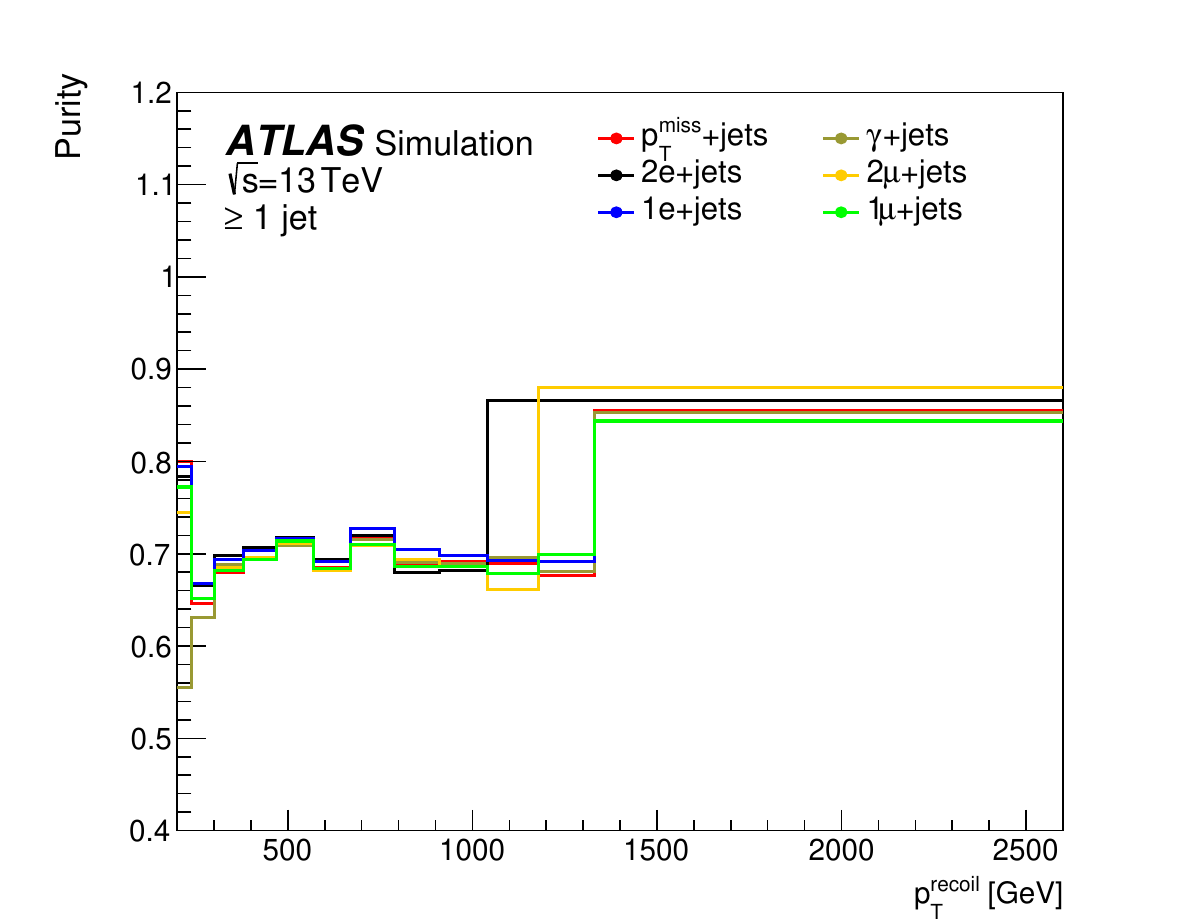}}
\subfigure[]{\includegraphics[scale=0.4]{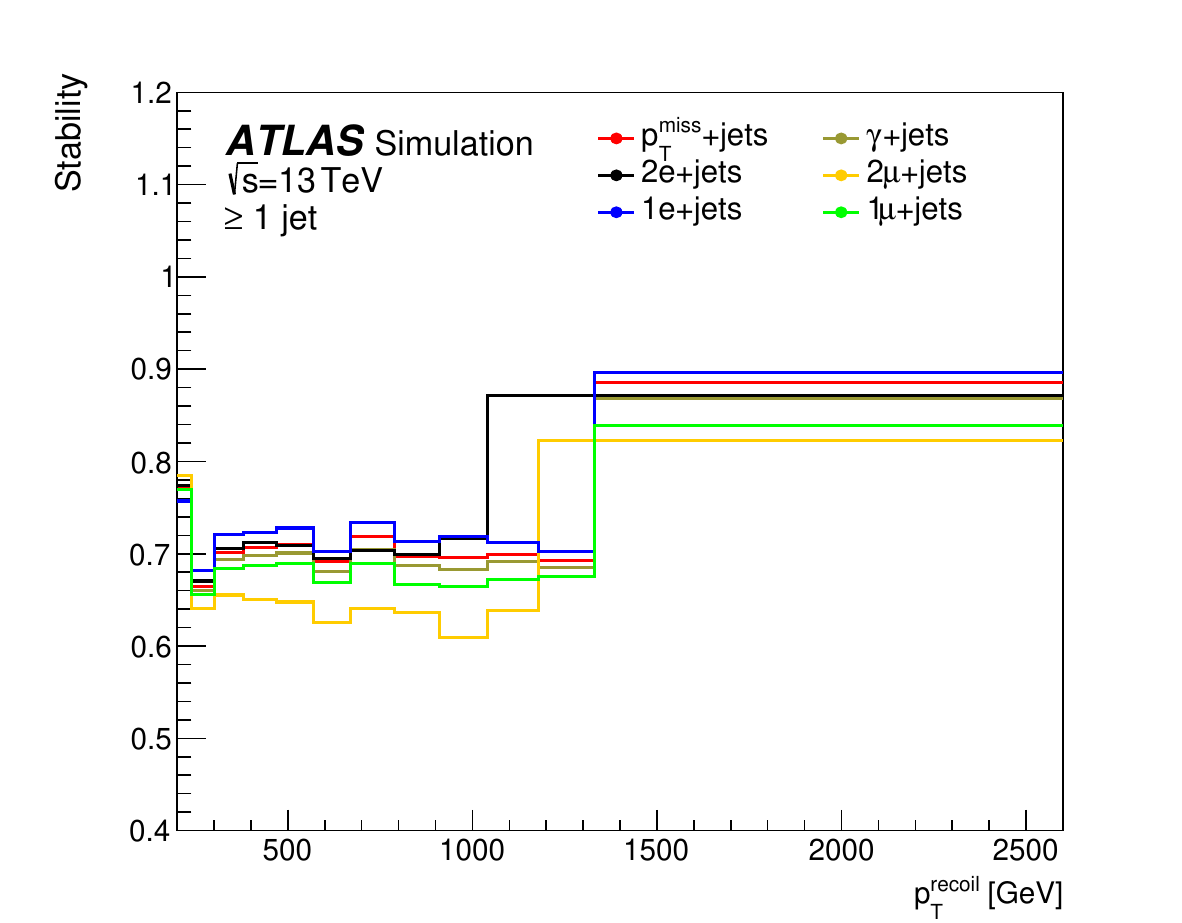}}
\caption[]{(a) Matching efficiency, (b) fiducial fraction, (c) purity and (d) stability for \ptrec, in the \ptmissjets, \onegjets, \oneeljets, \onemujets, \twoeljets, \twomujets
regions of the \onejet phase space, constructed for all processes that enter the fiducial phase space.
The purity and stability are high at high values
of \ptrec because of the large bin widths dictated by the requirement that there be at least 20 events in each bin.
}
\label{fig:metMono_grouped_topo}
\end{figure}
 
Several sources of systematic uncertainty in the particle-level measurements are considered:
\begin{itemize}
\item \textbf{Hidden variables:} While the iterative unfolding handles biases from the assumed prior for the distribution being unfolded, the unfolded result may still
be influenced by the (mis)modelling of other `hidden' event variables, especially if they form part of the selection.
This is studied by reweighting simulated events at particle level so that
the reweighted reconstructed distribution of the hidden variable matches the data.
The variables considered are the leading-jet and leading-lepton kinematics, the number of jets, and the invariant mass of the dilepton system.
Differences in the unfolded results with and without this additional weighting are taken as uncertainties, although they are below the percent level.
 
\item \textbf{Migrations into the fiducial phase space:} The events that satisfy the selection criteria defining the fiducial phase space (Table~\ref{tab:jets-phasespace}) at
reconstruction-level are not identical to those that satisfy them at particle-level. If this difference, or the
underlying distributions for each variable that is used in the selection, is not well modelled then the migrations in and
out of the phase space will not be properly corrected for in the unfolding.
For all observables in all regions, comparisons of data and simulation
are made, and in each case the requirement in question is relaxed in order to study the behaviour of the observable below the selection value.
Simulated events are reweighted such that the reconstructed distribution matches the data. The changes to the
measurement caused by this reweighting are negligible.
 
\item \textbf{Signal injection tests:} Although the unfolding procedure and fiducial definition are designed to minimise dependence on the simulated
distributions, a residual bias may be present due to the absence of BSM effects from the samples used for
unfolding, while such physics may be present in the data.
To test whether this is the case, various BSM processes are injected into the simulated samples, which are then treated as
pseudodata and unfolded with nominal SM simulation.
Three samples of Higgs boson events decaying invisibly were used, with three different Higgs boson masses (75~\GeV, 125~\GeV\ and 750~\GeV),
thus emulating some very extreme Higgs-to-invisible BSM scenarios.
The test is repeated for \ptmiss in the \onejet phase space using a model with
$s$-channel production of a DM particle ($\chi$) via a spin-1 axial-vector mediator $A$,
for $m_\chi = 1~\GeV$ with $m_A = 50~\GeV$ and $700~\GeV$,
and $m_\chi = 355~\GeV$ and $m_A = 700~\GeV$, as well as spin-0 pseudoscalar mediator with $m_\chi = 1~\GeV$ and $m_A = 50~\GeV$.
The maximum bias for any of these scenarios is 10\%, seen at large \ptmiss, and all the models introducing bias
are so extreme that if they were present in reality, the discrepancy
would already be clearly seen in the detector-level data, before unfolding. No additional source of systematic
uncertainty is therefore added from this source.
 
\item \textbf{Sample composition variations:} For the measurement of the \ptmiss cross-sections, the simulated samples used in the unfolding include
all contributing SM processes; the mixture of these processes is constrained by applying the normalisation factors discussed in Section~\ref{sec:smbkg}. Uncertainties are derived by varying the composition within the uncertainties in these normalisation factors. The derived uncertainties are then propagated through the unfolding to the final measurement.
SM processes involving top quarks are among those whose contribution is varied. Events originating from these processes are enriched in the presence of $b$-quarks. This uncertainty therefore also ensures that the resulting measurement can be used safely when comparing to predictions with increased $b$-quark activity.
\end{itemize}
 
For the measurement of \Znunu cross-sections, the contributions from non-\Znunu SM processes are
subtracted before unfolding, with the amount subtracted being constrained using
both the high-\ptmiss measurement region and the control regions. The simulated sample used in the unfolding then includes
only \Znunu processes. The subtraction uncertainties are propagated through the unfolding to the final measurement.
 
\subsection{Detector calibration, resolution and identification uncertainties}
\label{sec:uncertainties}
 
The unfolding procedure relies on knowledge of the detector response, which has uncertainties associated with it.
The impact of these uncertainties is determined by varying the response function in question and re-running the analysis, including
the final unfolding step. The sources of uncertainty considered are given below.
 
\begin{itemize}
 
\item Uncertainties related to jet energy scale (JES) and jet energy resolution
(JER) are derived using dijet samples following the
procedures documented in Ref.~\cite{JETM-2018-05}.
A subclass of JES uncertainties deals with whether the jet is likely to have
been initiated by a quark or a gluon.
The proportion of quark-initiated jets in the measurement regions is
estimated from simulation as a function of transverse momentum and pseudorapidity.
 
\item Uncertainties related to the electron efficiency measurement and
calibration are obtained from tag-and-probe measurements of
$J/\psi$ and $Z\rightarrow e^+e^-$ events, as described in Ref.~\cite{EGAM-2018-01}.
 
\item Muon uncertainties are related to muon calibration and efficiencies are
obtained from $J/\psi$ and $Z\rightarrow \mu^+\mu^-$ events,
as described in Ref.~\cite{MUON-2018-03}.
 
\item $\tau$-lepton calibration uncertainties are accounted for as documented
in Ref.~\cite{ATLAS-CONF-2017-029}.
 
\item Measurements of $Z\rightarrow \ell \ell \gamma$ events are used to study
the performance of photon reconstruction, as documented in Ref.~\cite{EGAM-2018-01}.
 
\item The uncertainty in those (soft) components of \ptmiss not accounted for
already is represented by three systematic uncertainties, detailed
in~Ref.~\cite{PERF-2016-07}.
 
\end{itemize}
 
Figure~\ref{fig:tru_metMonoGROUPED_SYS} shows the breakdown of the statistical and systematic uncertainties
for the \ptmiss observable in the \onejet phase space.
At high \ptmiss, the statistical uncertainty dominates, and over most of the distribution, the JES
is the most significant systematic uncertainty, with the JER next in the region with no leptons or
photons.
Uncertainties associated with lepton and photon identification contribute in the other regions.
The uncertainty due to the unfolding, and the (forward) jet vertex tagging uncertainty, are below 2\%, and much smaller in most cases.

\begin{figure}
\centering
\subfigure[]{\includegraphics[scale=0.39]{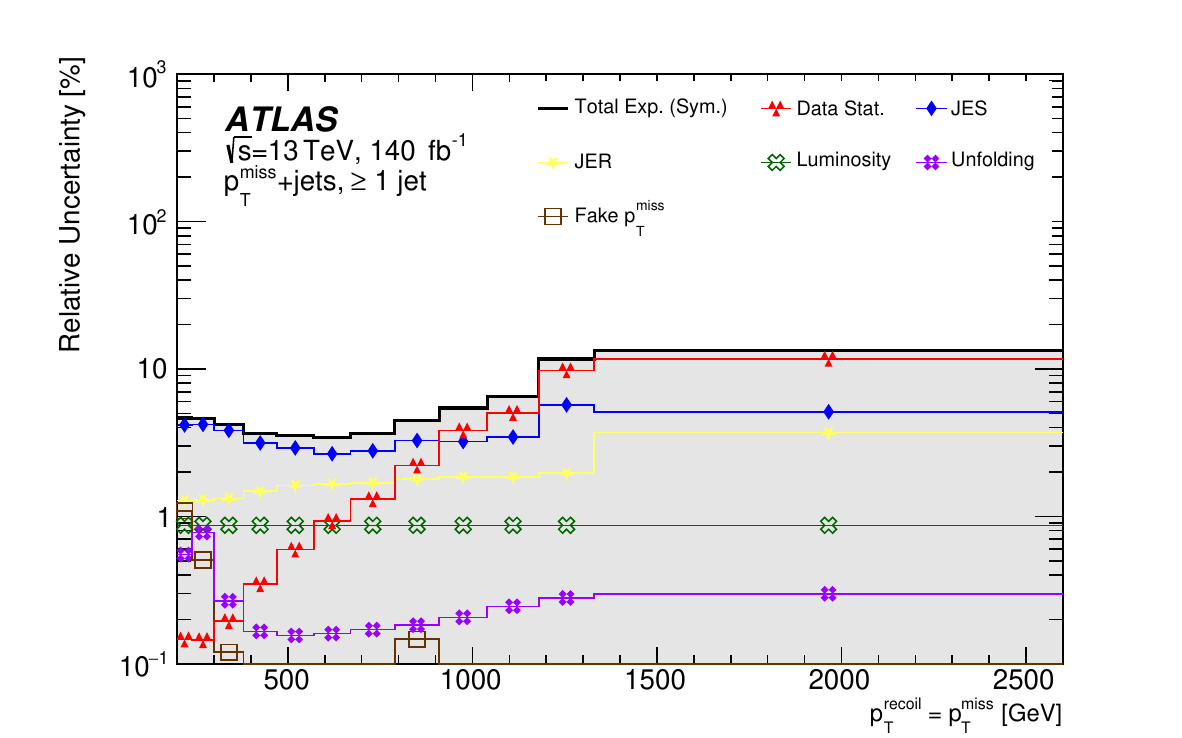} \label{fig:tru_metMonoGROUPED_SYS_a}}
\subfigure[]{\includegraphics[scale=0.39]{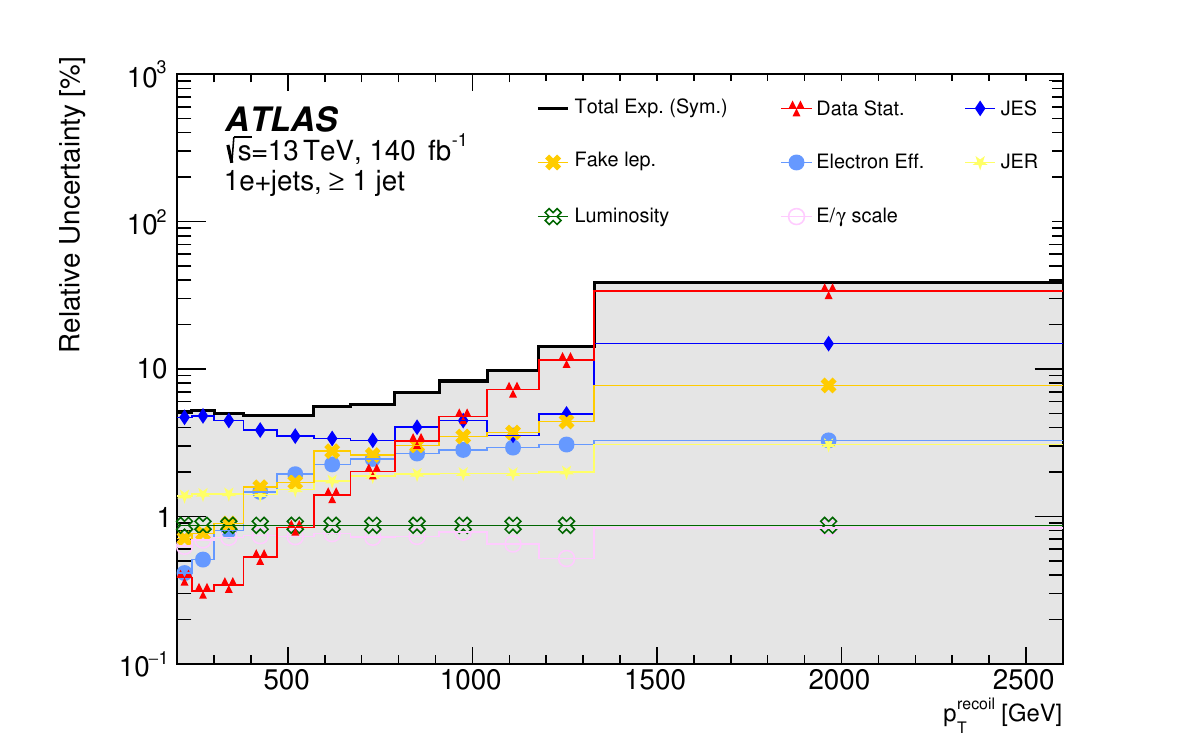} \label{fig:tru_metMonoGROUPED_SYS_b}}
\subfigure[]{\includegraphics[scale=0.39]{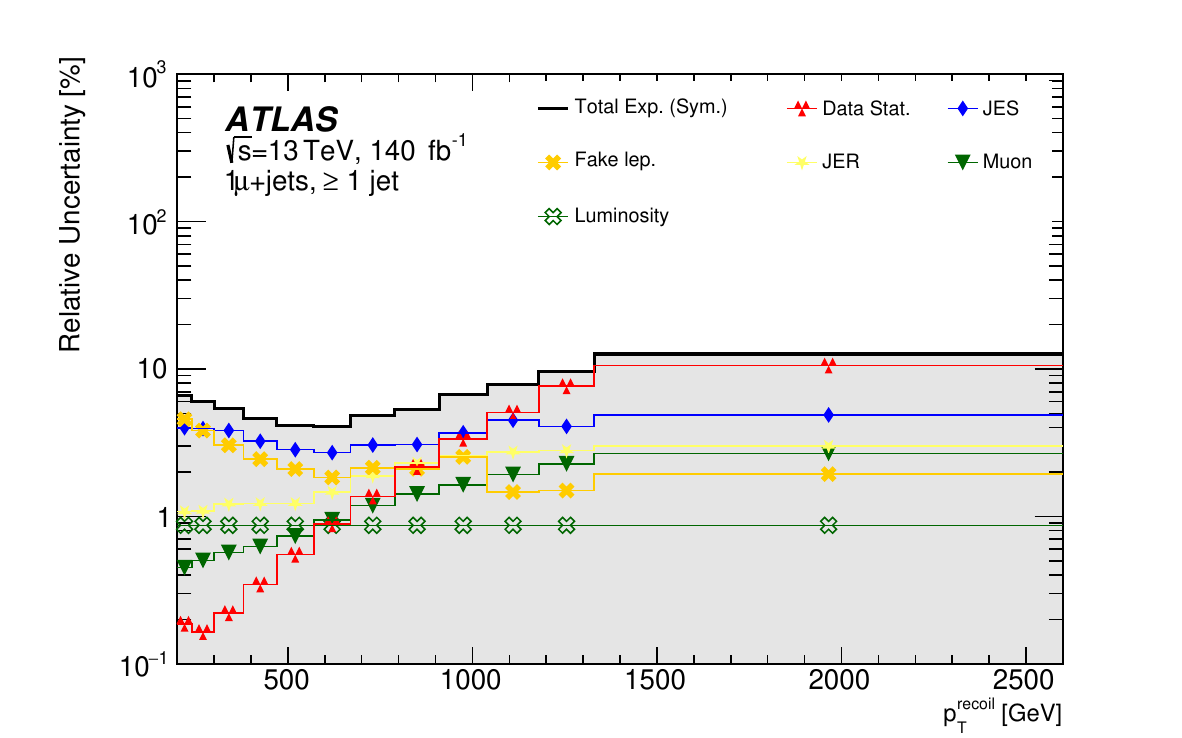} \label{fig:tru_metMonoGROUPED_SYS_c}}
\subfigure[]{\includegraphics[scale=0.39]{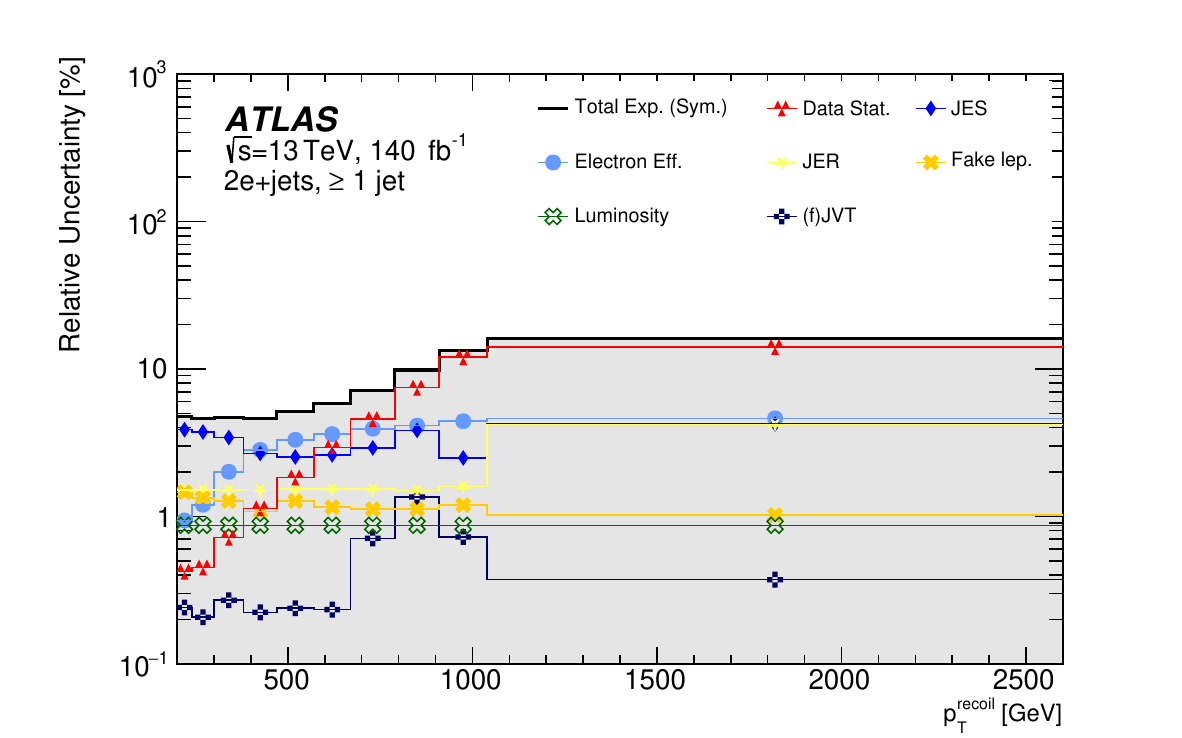} \label{fig:tru_metMonoGROUPED_SYS_d}}
\subfigure[]{\includegraphics[scale=0.39]{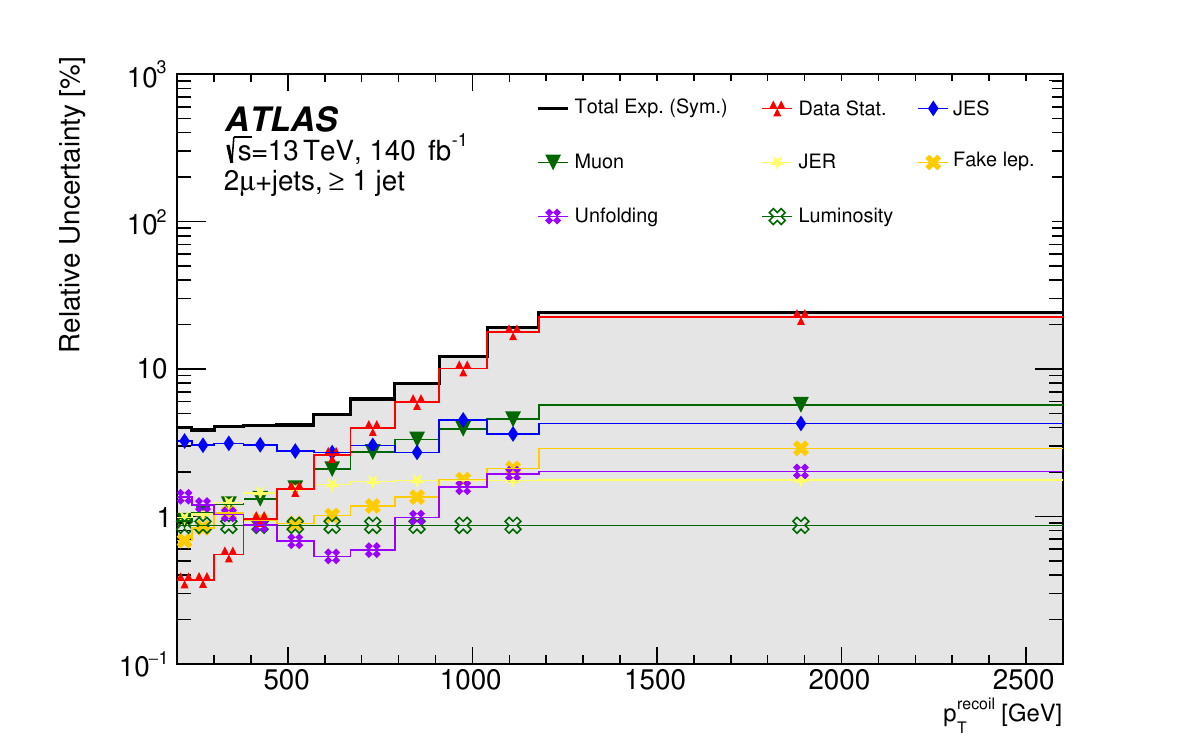} \label{fig:tru_metMonoGROUPED_SYS_e}}
\subfigure[]{\includegraphics[scale=0.39]{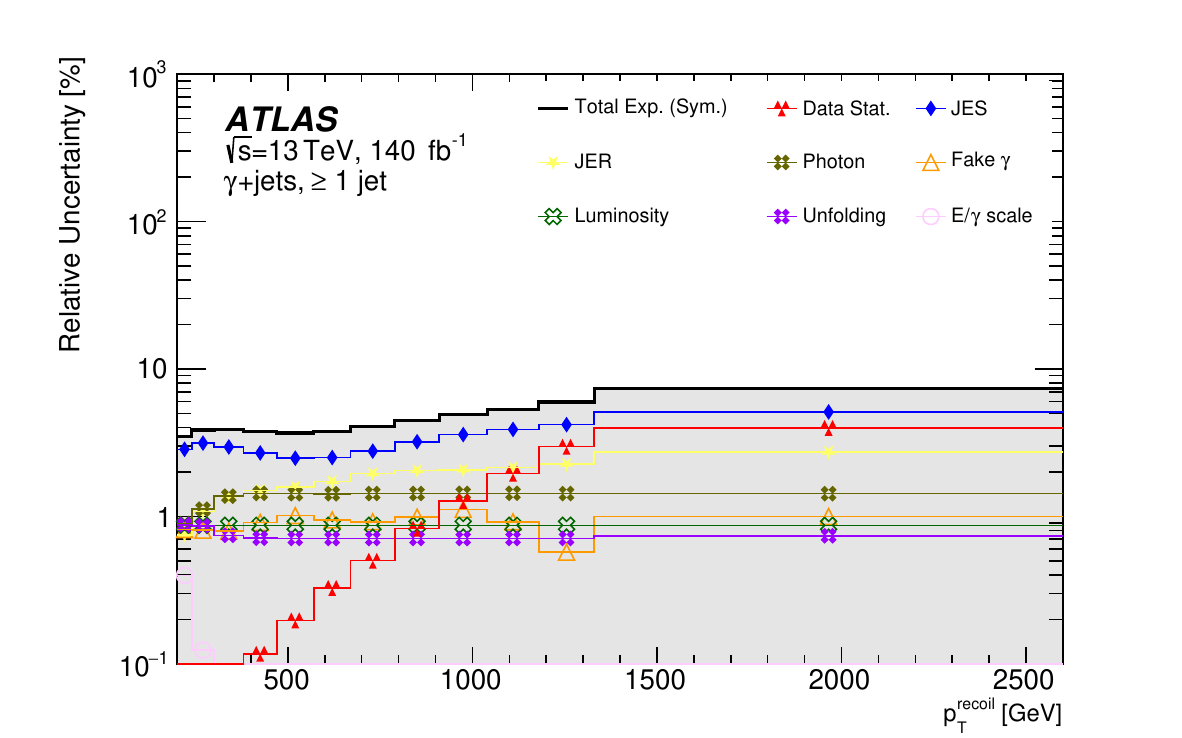} \label{fig:tru_metMonoGROUPED_SYS_f}}
\caption[]{Uncertainty breakdown for \ptrec measurements in the \onejet phase space for
the (a) \ptmissjets, (b) \oneeljets, (c) \onemujets, (d) \twoeljets, (e) \twomujets, and (f) \onegjets regions, showing the statistical uncertainty and
the most significant systematic uncertainties in each case. For illustrative purposes this figure shows the symmetrised uncertainties, calculated as the
average of the asymmetric error in each bin. Total Exp. (Sym.) is the combination
of statistical and systematic uncertainties and indicates the symmetrised total experimental uncertainty.
}
\label{fig:tru_metMonoGROUPED_SYS}
\end{figure}
 
Figure~\ref{fig:tru_metMonoRmissGROUPED_SYS} shows the breakdown of statistical and systematic
uncertainties for the \rmiss ratio of cross-sections, as a function of \ptmiss, in
the \onejet region. At low \ptmiss, the cancellation of JES uncertainties leads to a reduction
in the combined estimate of the experimental uncertainty.
 
\begin{figure}
\centering
\subfigure[]{\includegraphics[scale=0.4]{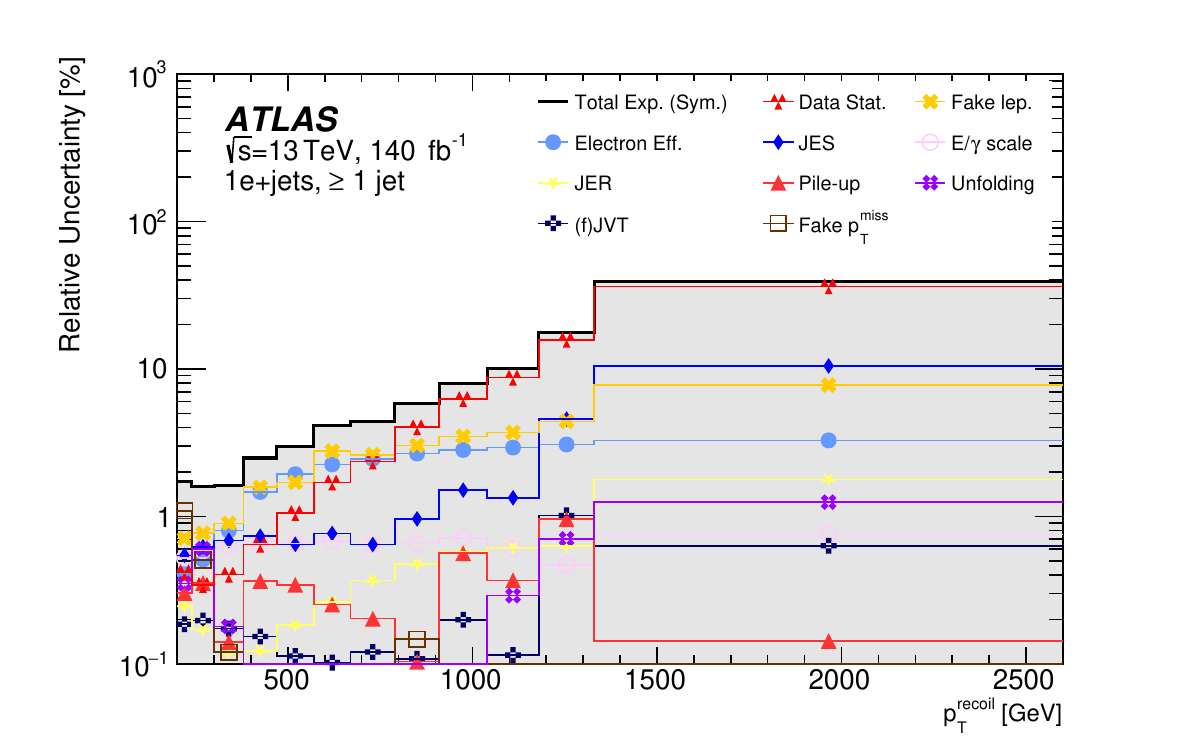}}
\subfigure[]{\includegraphics[scale=0.4]{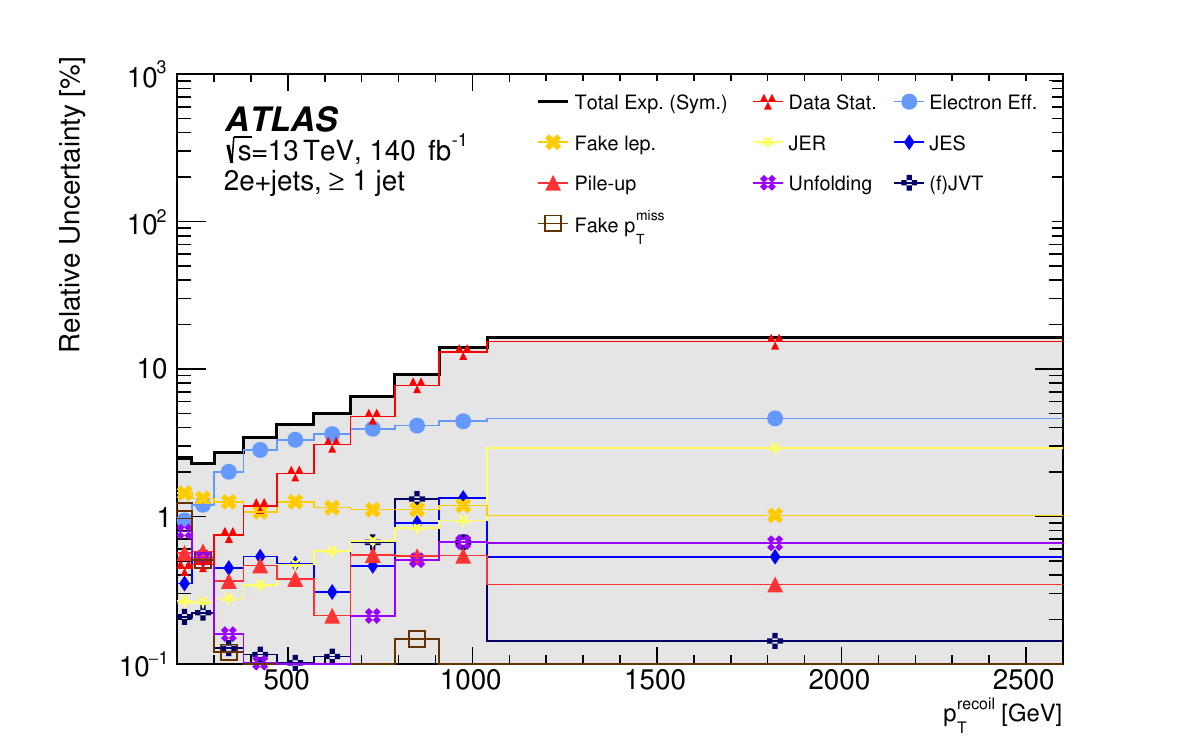}}
\caption[]{Uncertainty breakdown of the \rmiss ratio as a function of \ptrec. As examples, the breakdowns for the
\onejet phase space are shown for the (a) \oneeljets and (b) \twoeljets region. For illustrative purposes this figure shows the
symmetrised uncertainties, calculated as the
average of the asymmetric error in each bin. Total Exp. (Sym.) is the combination
of statistical and systematic uncertainties and indicates the symmetrised total experimental uncertainty.
Compared to the individual measurements in Figure~\ref{fig:tru_metMonoGROUPED_SYS_b} and \ref{fig:tru_metMonoGROUPED_SYS_d},
a cancellation of the JES and JER uncertainties is observed.
}
\label{fig:tru_metMonoRmissGROUPED_SYS}
\end{figure}
 
Finally, uncertainties associated with the estimation of fake backgrounds are accounted for.
For the fake-lepton backgrounds, the dominant source comes from theory uncertainties, such as QCD scale variations, affecting the generator predictions in the regions used to measure the efficiencies. This uncertainty can be up to 100\% of the predicted background yield.
Smaller sources include statistical uncertainties relating to the limited number of events in data in those regions, and uncertainties in the method (evaluated for example by modifying the definition of the regions).
Uncertainties from these sources are typically around 10\% of the predicted yield. After unfolding, the fake leptons uncertainties collectively have an effect of the order of 1\%--4\% depending on the bin.
For the fake-photon background, three sources of uncertainty are considered: the choice of WP used when selecting photons, the correlation between the A, B, C and D regions used for the estimation and the choice of generator for the prompt photon prediction.
Together, these three sources result in an uncertainty of around 30\% in the predicted fake photon yields. This amounts to approximately 1\% in the measured cross-section after unfolding.
The  uncertainties in the multijet and non-collision backgrounds in the signal region represent very small contributions to the final event sample and are subject to statistical variations due to the nature of the evaluation. Therefore, they are conservatively taken to be 100\% of the estimated yield, corresponding to a less than 1\% uncertainty in the final measurement in the most affected bin.


\FloatBarrier
\section{Results and discussion}
\label{sec:results}
 
\subsection{\ptmiss measurements}
 
The measured differential cross-section as a function of the magnitude of the missing transverse momentum, \ptmiss,
is shown in Figure~\ref{fig:ptmiss_final} for the single jet and \vbf phase spaces, in the region with no signal leptons or photons.
In both cases, the cross-section falls by more than five orders of magnitude as \ptmiss increases from 200 to 2500~\GeV.
The cross-section in the \vbf phase space is lower than the single jet phase space due to the jet requirements.
 
Similar behaviour is seen for the transverse momentum of the hadronic system, \ptrec, after the charged lepton requirements are imposed,
as shown in Figure~\ref{fig:ptmiss_inc} for the single jet phase space.
When a single muon is required, the cross-section is similar in magnitude to the zero-lepton/photon cross-section, while requiring two muons
reduces it by about an order of magnitude. The cross-sections after electron requirements are somewhat smaller due to the more
restrictive fiducial requirements imposed on electrons.
 
\begin{figure}
\centering
\subfigure[]{\includegraphics[scale=0.4]{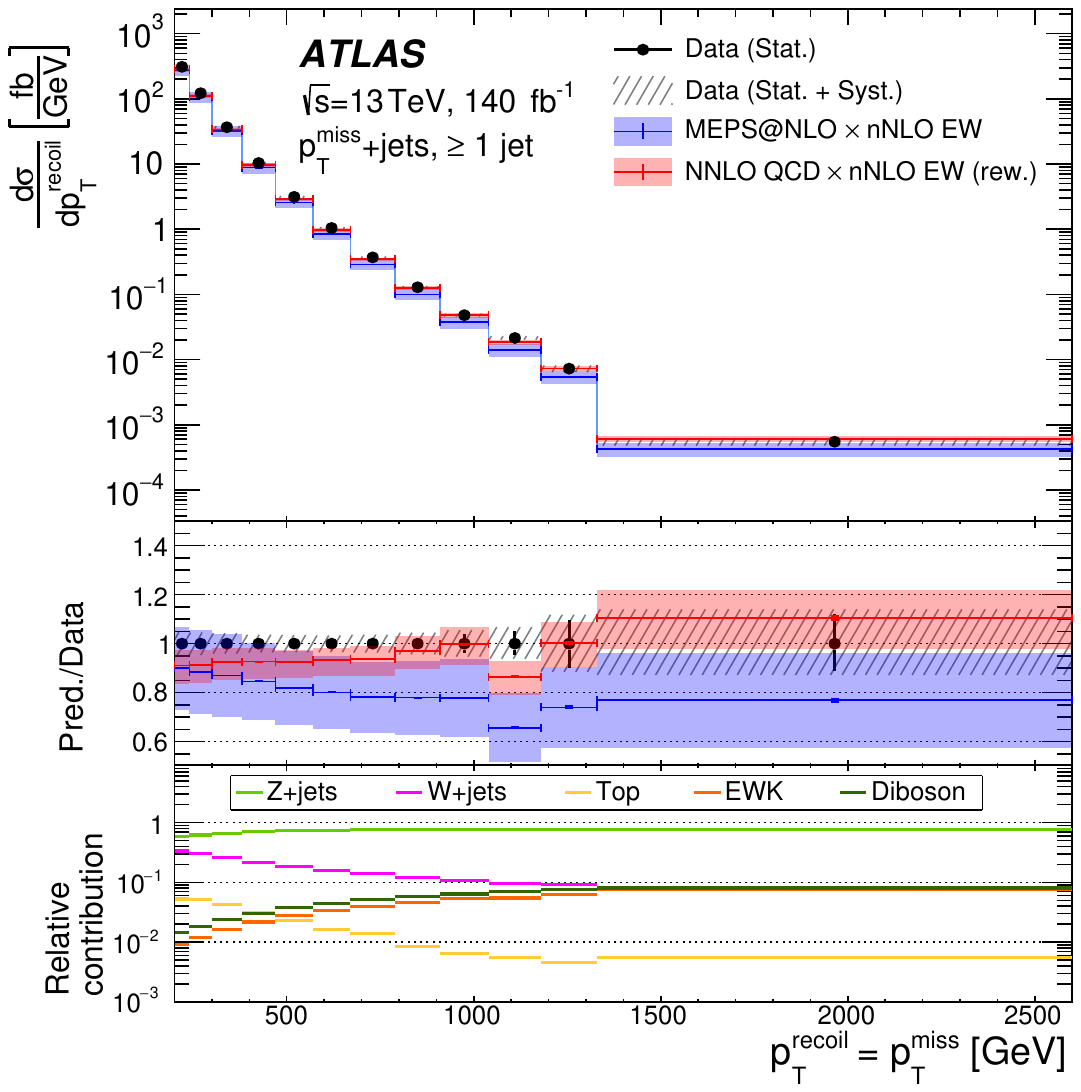}}
\subfigure[]{\includegraphics[scale=0.4]{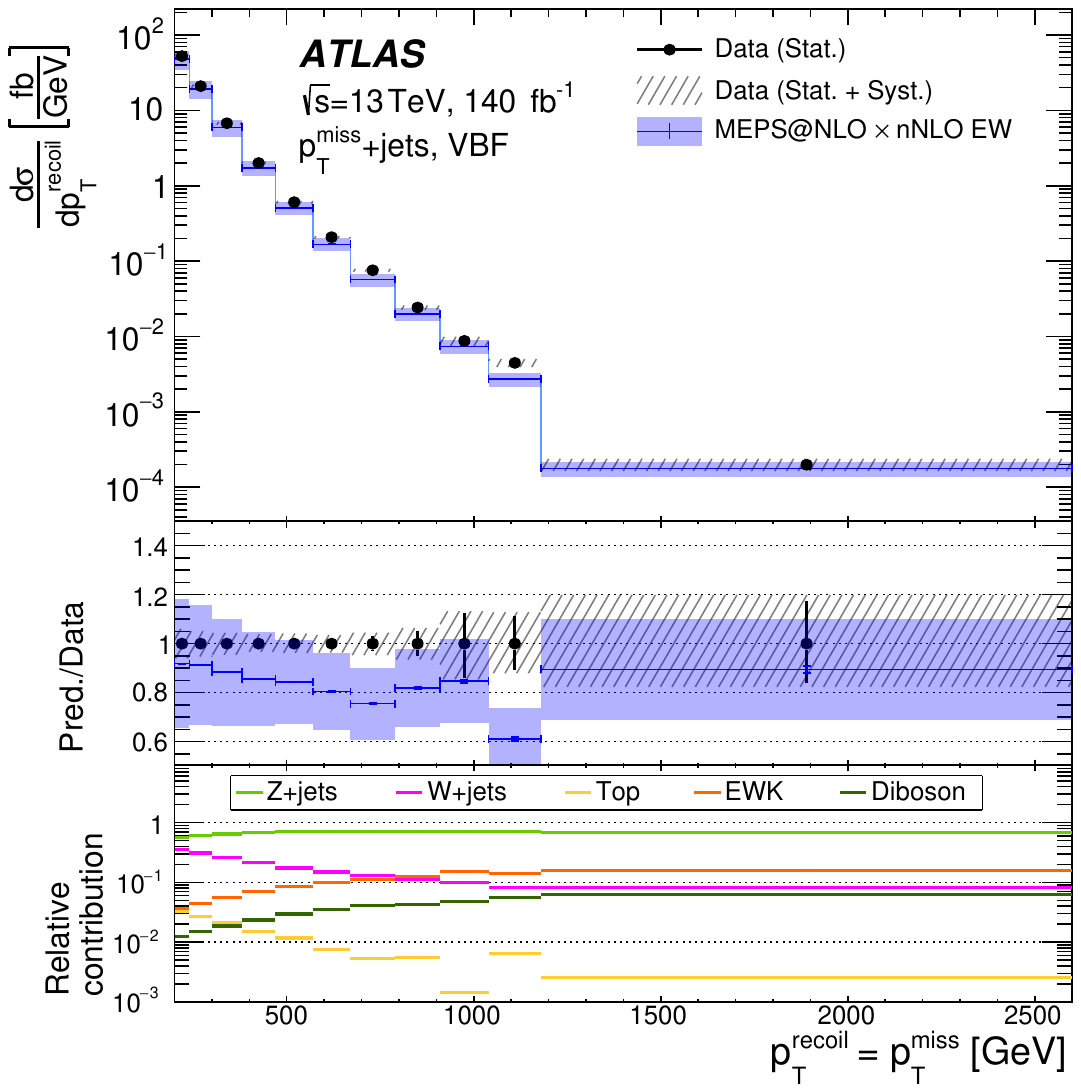}}
\caption[]{The measured \ptmiss differential cross-sections in the \ptmissjets region in (a) \onejet and (b) \vbf phase spaces,
compared with the SM predictions. The middle panels show the ratios of the predictions to the data,
with statistical uncertainties as solid markers and shaded bands to indicate the combined statistical and systematic uncertainties,
while the lower panels show the relative contributions from different SM processes relative to the total MEPS@NLO prediction.}
\label{fig:ptmiss_final}
\end{figure}
 
The SM predictions described in Section~\ref{sec:sim} are also shown in Figure~\ref{fig:ptmiss_final} and Figure~\ref{fig:ptmiss_inc}.
Apart from a difference between the normalisations, they describe the data well in all regions;
a quantitative study is presented in Section~\ref{sec:smstudy}.
Also shown are the subcomponents of the \MEPSatNLO prediction.
With no leptons or photons present, the dominant contribution is \Znunu, with a top-quark contribution of a few per cent at
low \ptmiss that falls to the per-mille level at higher values.
The contribution from \wjets is around 25\% at low \ptmiss, but falls rapidly with \ptmiss to
form about 10\% of the cross-section at higher values.
The contribution from electroweak production mechanisms is around 1\% at low \ptmiss, but rises rapidly with \ptmiss to
form about 10\% of the cross-section at higher values.
 
\begin{figure}
\centering
\subfigure[]{\includegraphics[scale=0.4]{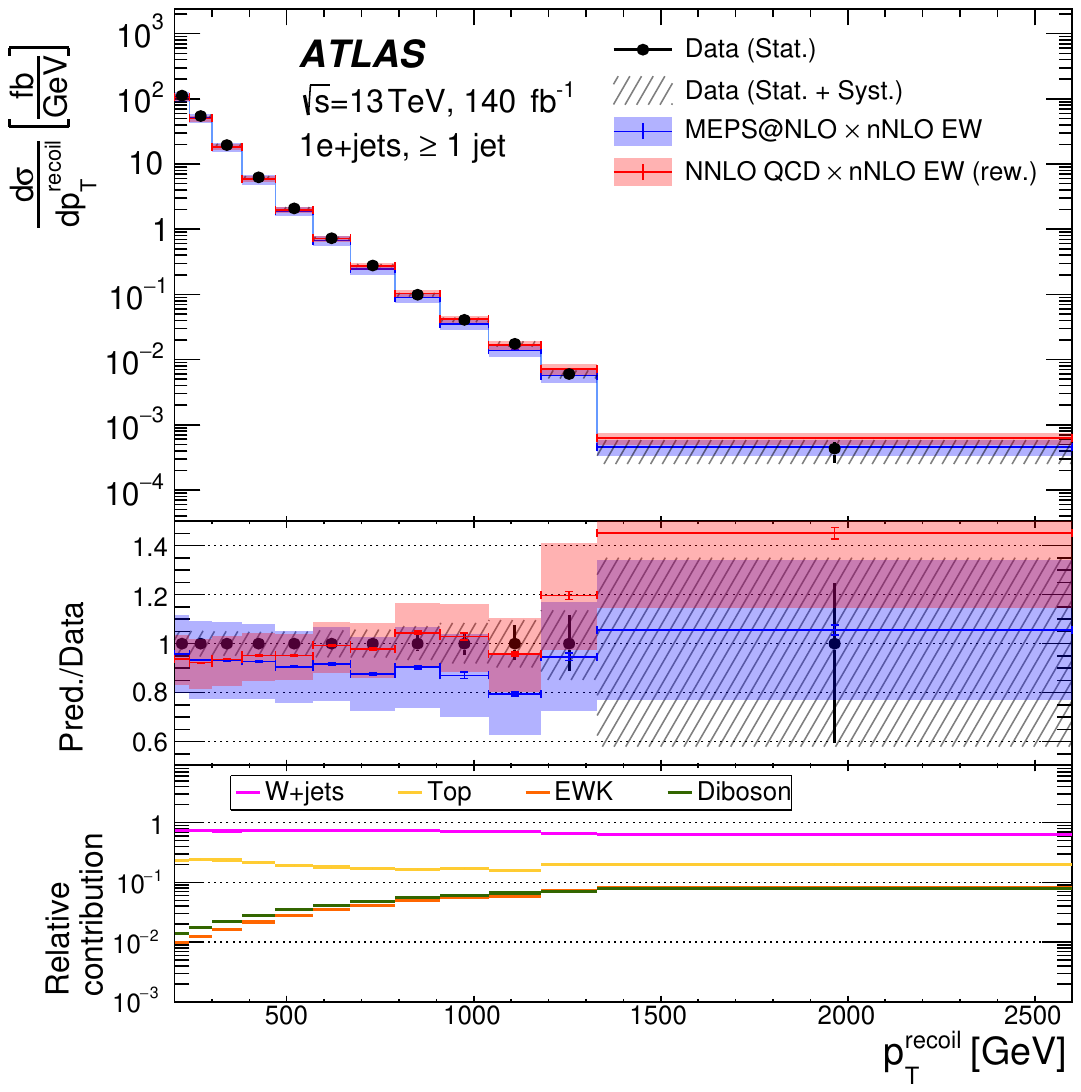}}
\subfigure[]{\includegraphics[scale=0.4]{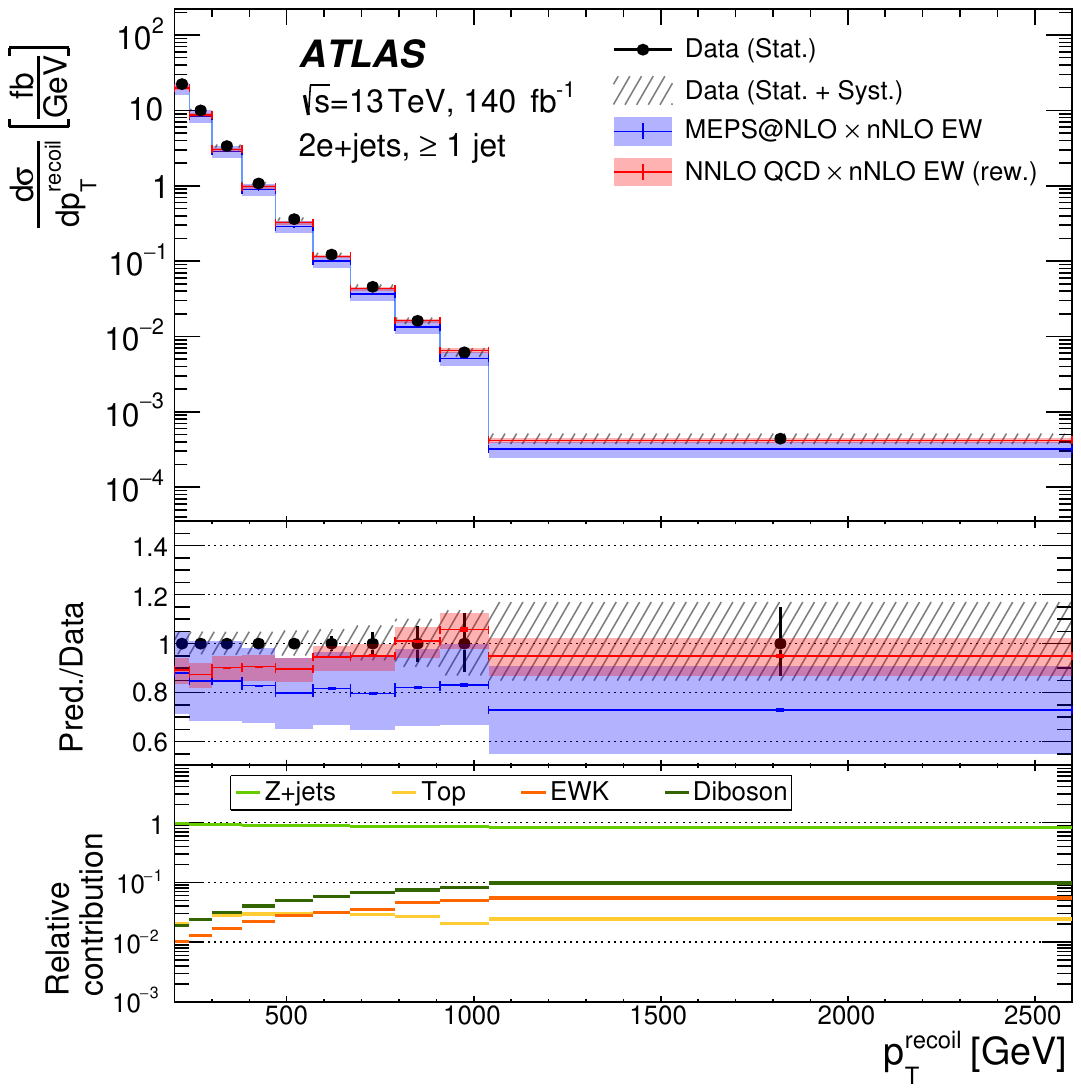}}
\subfigure[]{\includegraphics[scale=0.4]{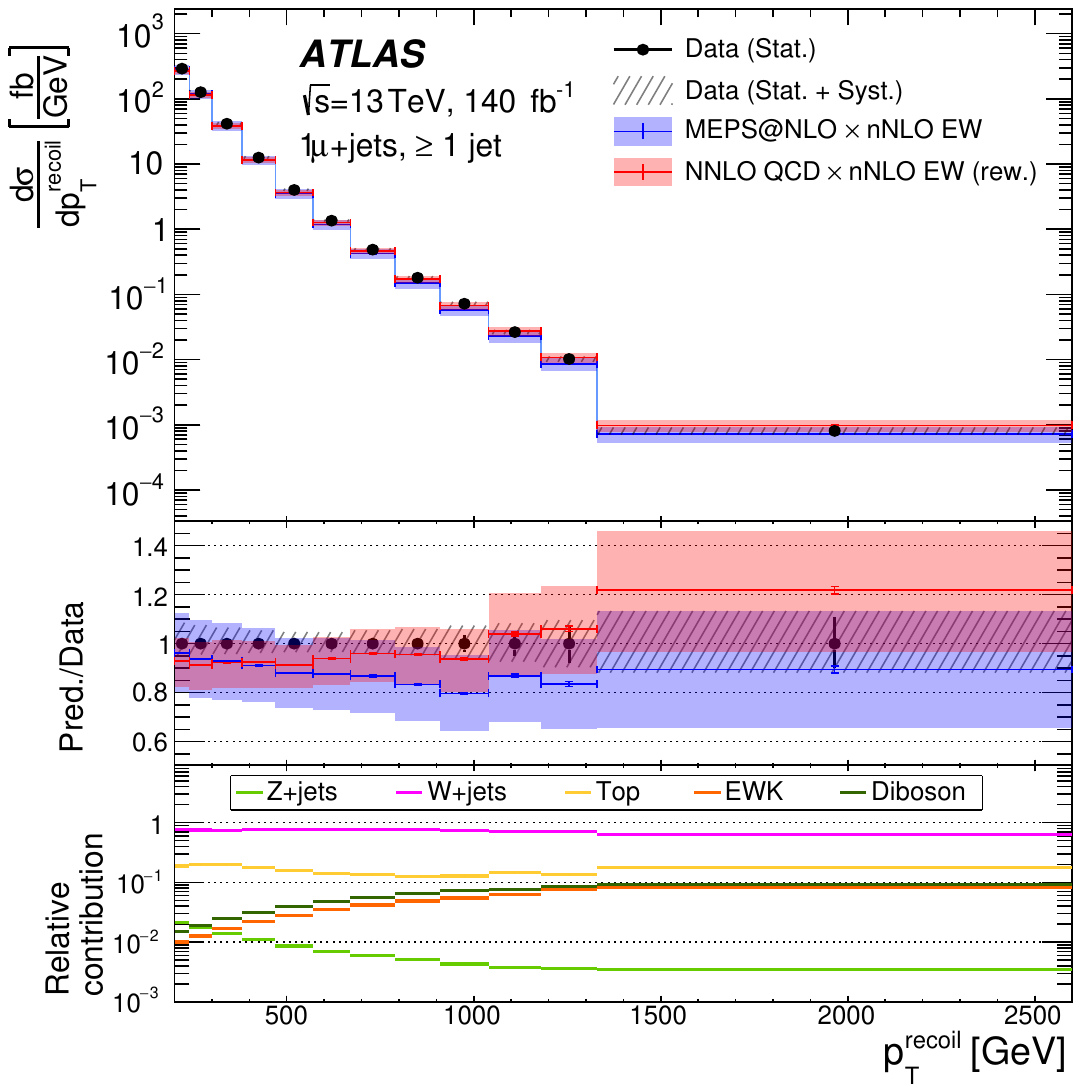}}
\subfigure[]{\includegraphics[scale=0.4]{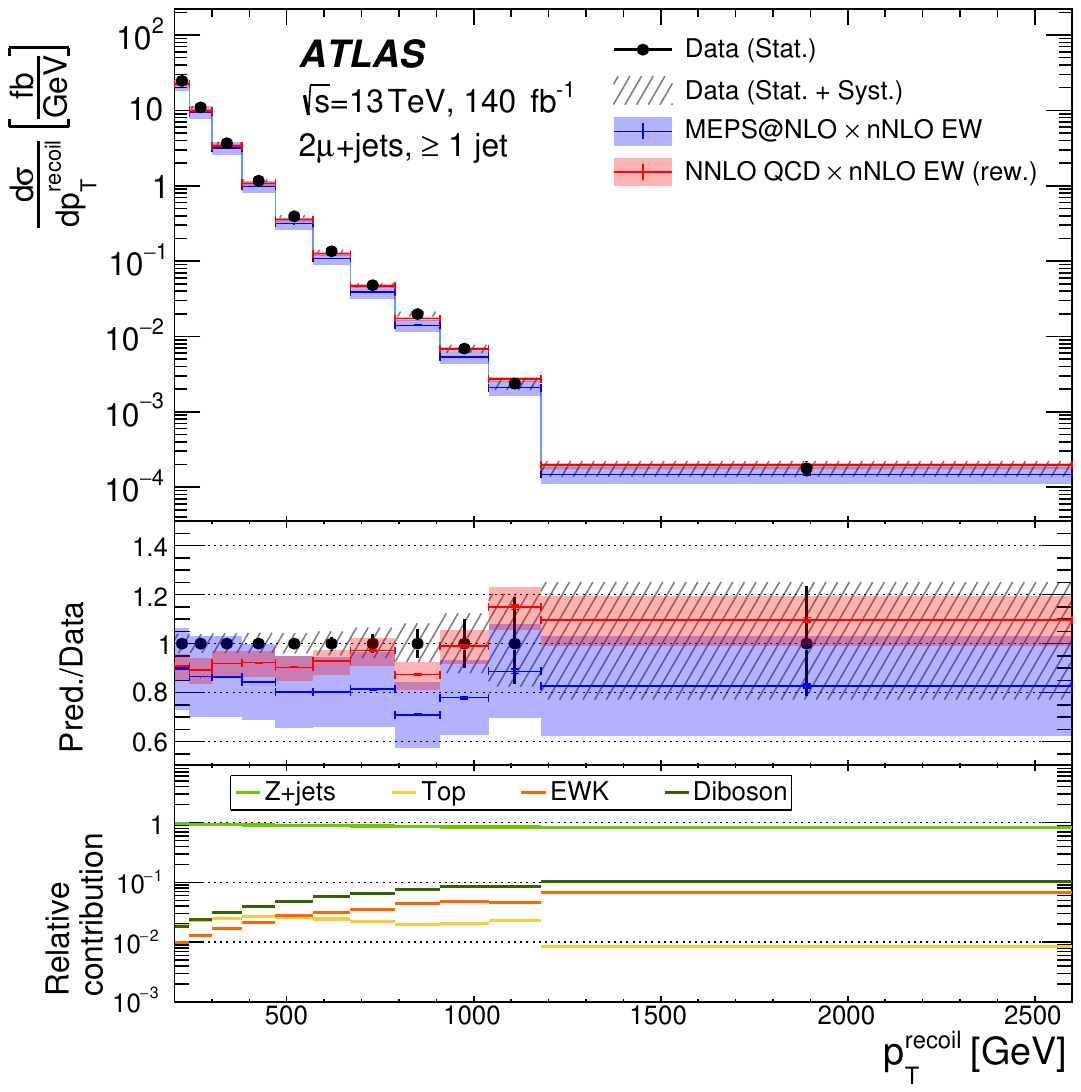}}
\caption[]{The measured \ptrec differential cross-sections in the inclusive jet phase space compared with the SM predictions: (a) \oneeljets (b) \twoeljets (c) \onemujets and (d) \twomujets.
The middle panels show the ratios of the predictions to the data, with statistical uncertainties as solid markers and shaded bands to indicate the combined statistical and systematic uncertainties,
while the lower panels show the relative contributions from different SM processes relative to the total MEPS@NLO prediction.}
\label{fig:ptmiss_inc}
\end{figure}

For the one-electron and one-muon phase spaces, the top-quark contribution is around 15\% at low \ptrec, falling to a few percent at high values.
It is never more than a few percent of the two-charged-lepton cross-sections. For all the charged-lepton cross-sections, the electroweak
contribution is around 1\% at low \ptrec, rising with \ptrec to be just below 10\% of the cross-section.

The measured differential cross-section as a function of the \mjj and \dphijj is shown for the \vbf selection in Figure~\ref{fig:vbf_0l_2m}, along with the SM
predictions for different sub-processes, for the  \ptmissjets and the \twomujets regions.
The cross-section falls rapidly with dijet mass, and the electroweak contribution rises from around
1\% to 50\% as \mjj rises from 600~\GeV\ to 6~\TeV.
The overall \dphijj distribution peaks mildly at $\pm \pi/2$. The predicted
fractional contribution from top has a peak at zero, and both the diboson and top contributions rise towards $\pm\pi$.
The  electroweak contribution has maxima at $\pm 0.8\pi$.
 
\begin{figure}
\centering
\subfigure[]{\includegraphics[scale=0.40]{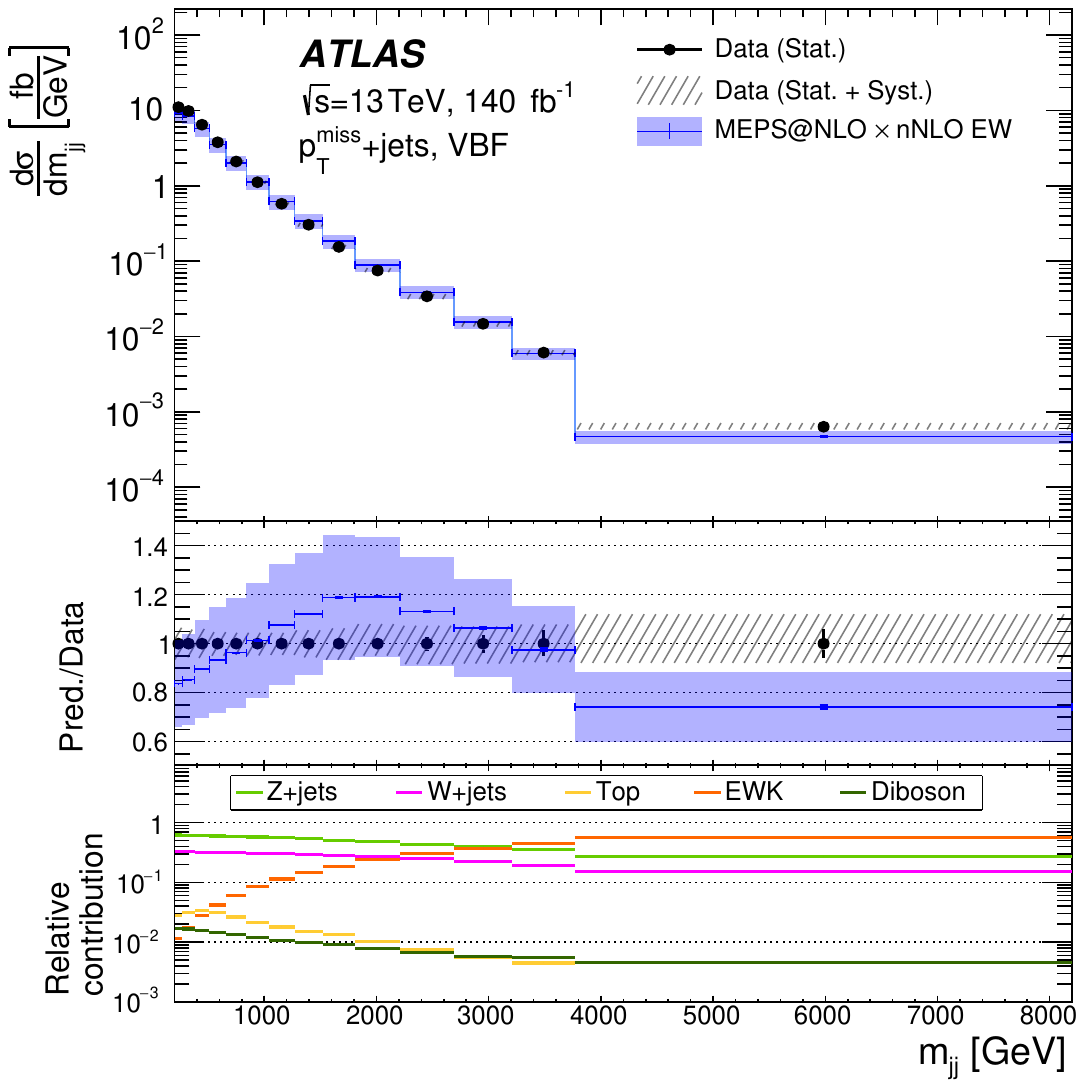}}
\subfigure[]{\includegraphics[scale=0.40]{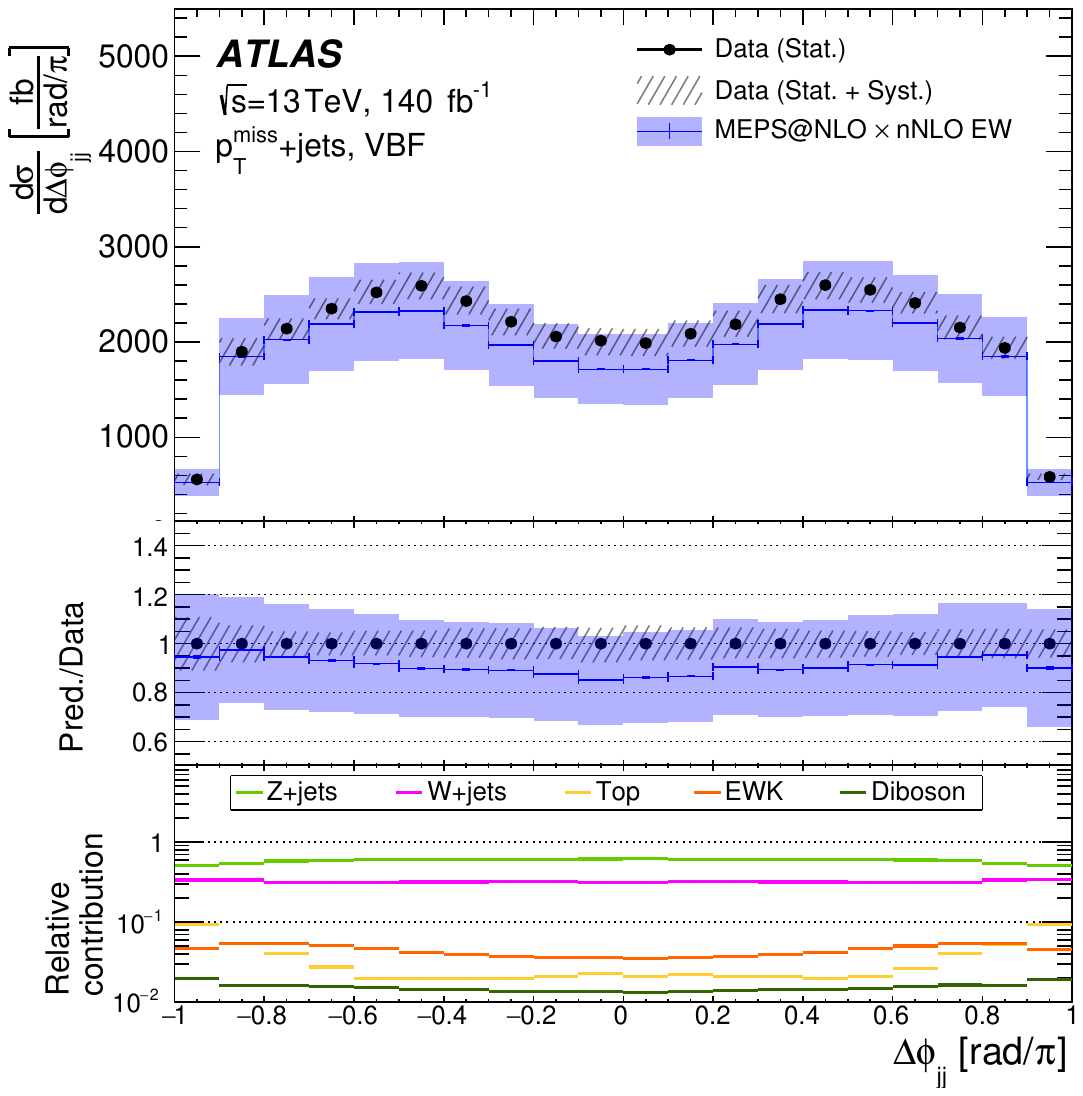}}
\subfigure[]{\includegraphics[scale=0.40]{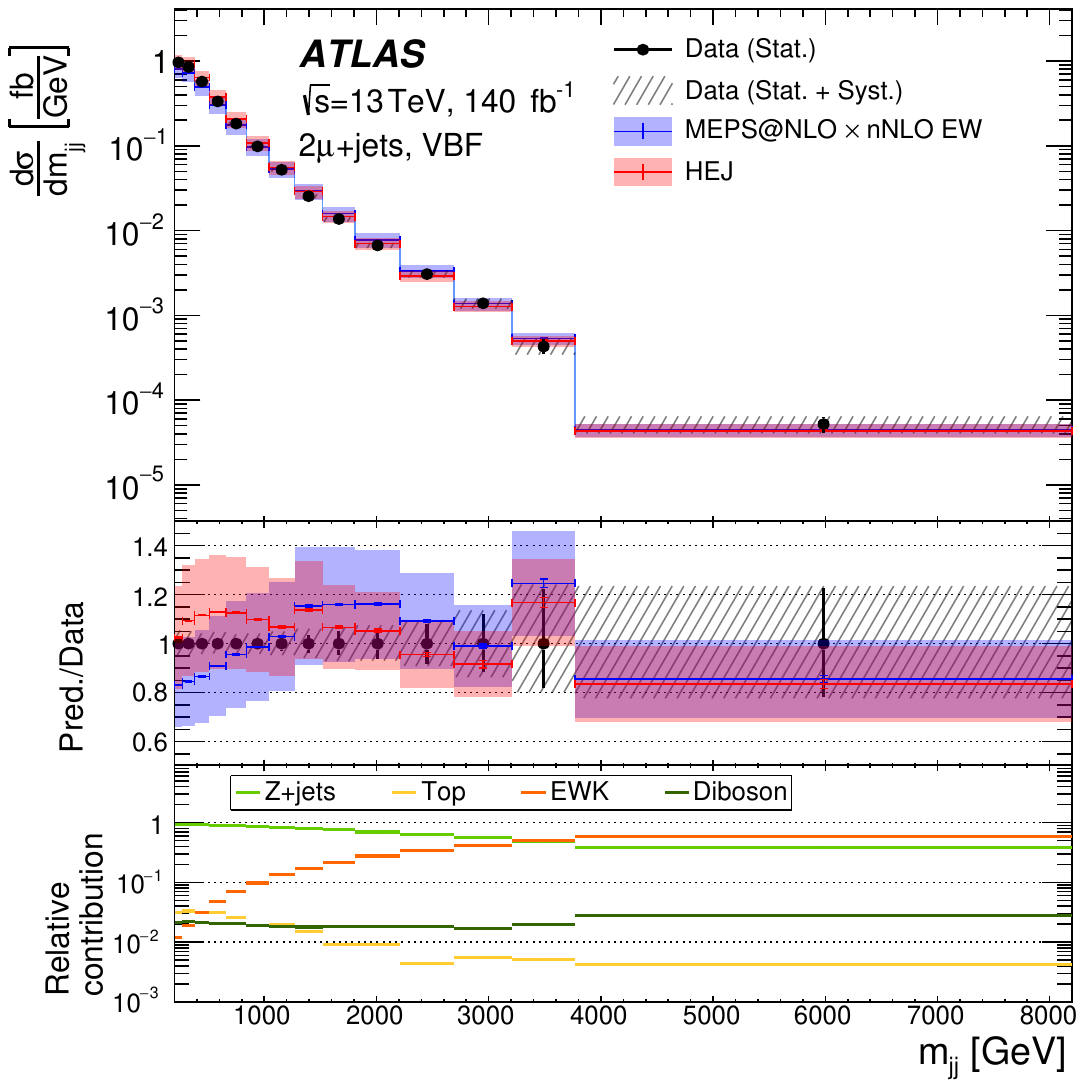}}
\subfigure[]{\includegraphics[scale=0.40]{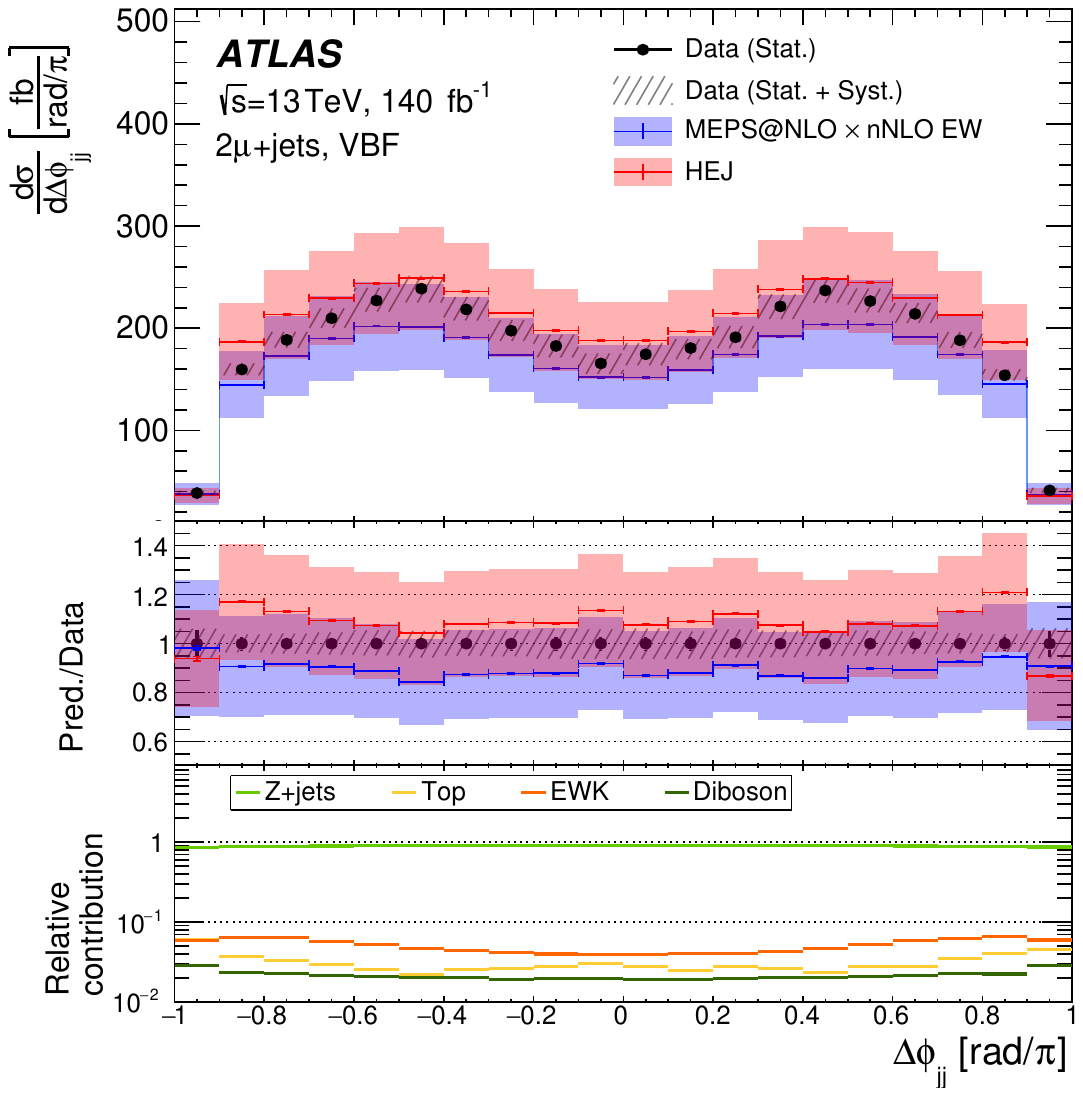}}
\caption{The measured \mjj and \dphijj distributions in the \vbf phase space compared with the SM predictions, for (a, b)
the   \ptmissjets and (c, d) the \twomujets regions, for illustration. 
The middle panels show the ratios of the predictions to the data, with statistical uncertainties as solid markers and shaded bands to indicate the combined statistical and systematic uncertainties, while the lower panels show the relative contributions from different SM processes relative to the total MEPS@NLO prediction.}
\label{fig:vbf_0l_2m}
\end{figure}
 
The description of the data by the SM predictions is generally good, except for the \mjj distribution, where the SM lies below the data at low values,
but falls less steeply, to lie above the data around 2~\TeV. This is discussed further in Section~\ref{sec:smstudy}. A resummed calculation using \HEJ
is also shown, which describes the \mjj distribution somewhat better.
 
Figure~\ref{fig:ptmiss_inc_rmiss} shows some examples of the \rmiss ratios of the results presented so far. 
The ratios tend to be flat or slowly falling across the measured spectra.
The \rmiss ratios benefit from a cancellation of discrepancies in modelling and some systematic uncertainties, and thus the agreement between
data and theory is improved compared to that for the cross-sections, most notably for the \mjj observable.

\begin{figure}
\centering
\subfigure[]{\includegraphics[scale=0.4]{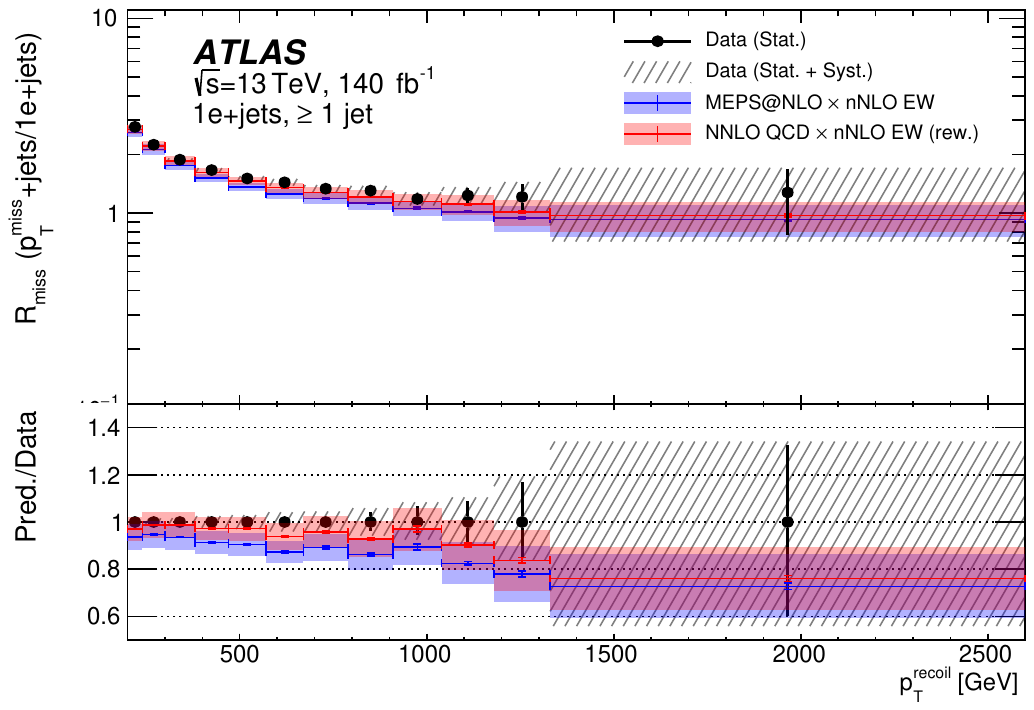}}
\subfigure[]{\includegraphics[scale=0.4]{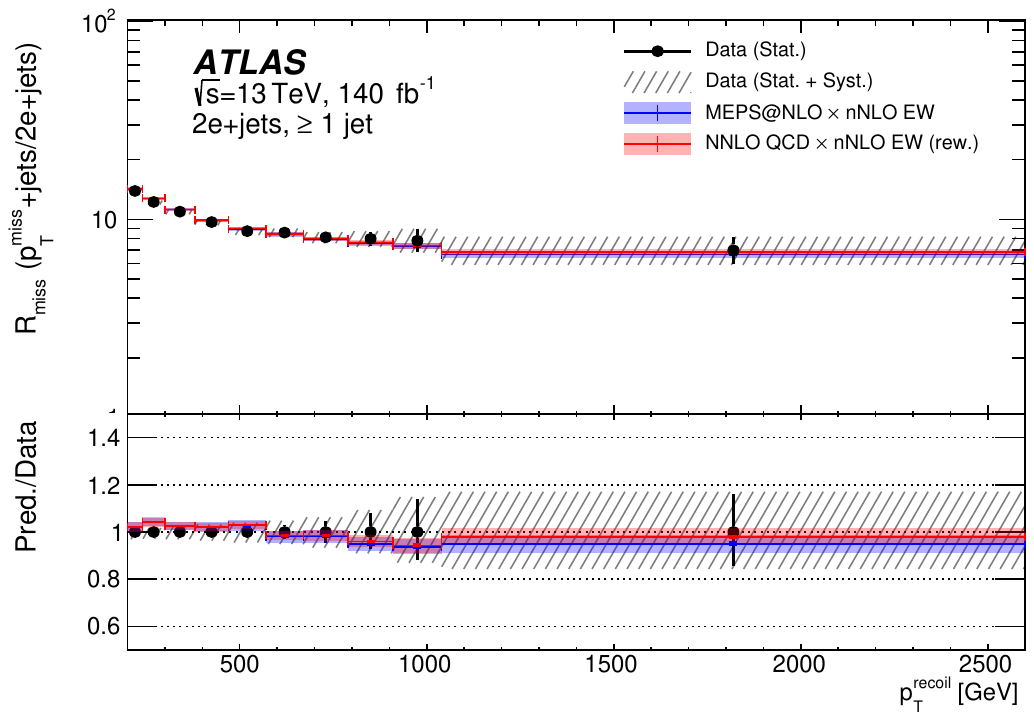}}
\subfigure[]{\includegraphics[scale=0.40]{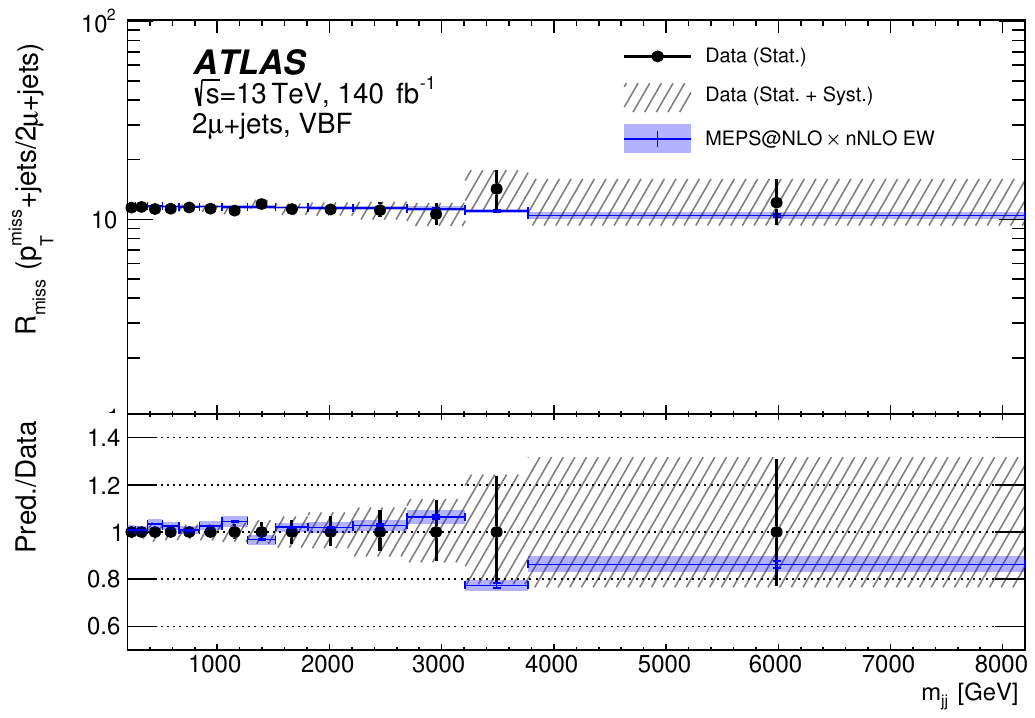}}
\subfigure[]{\includegraphics[scale=0.40]{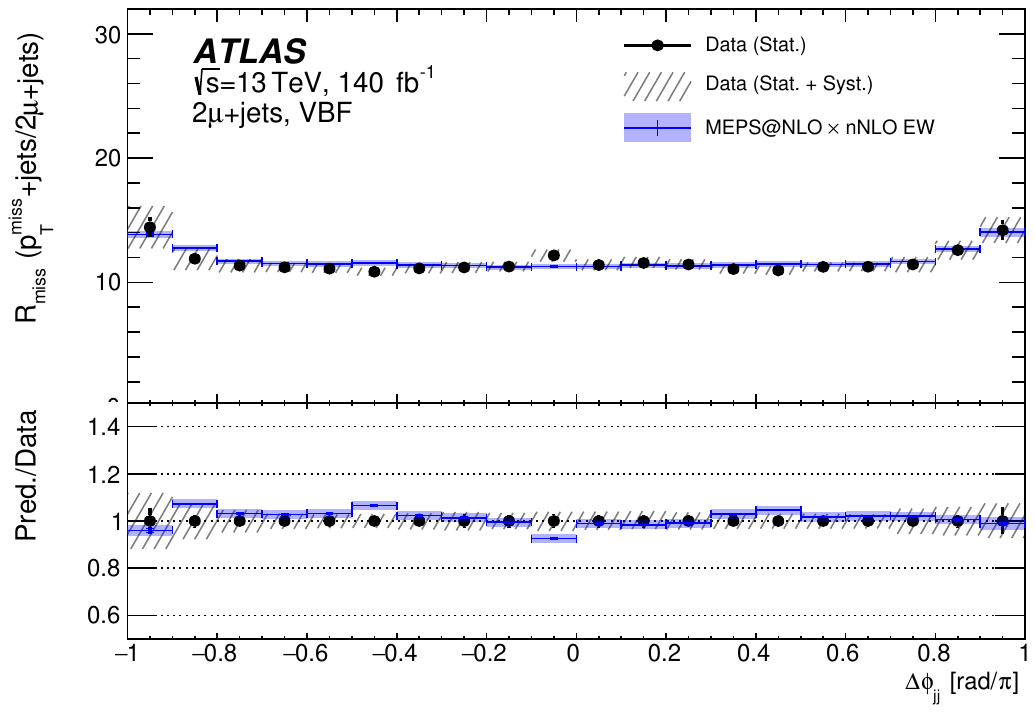}}
\caption[]{Comparison with SM predictions of the \rmiss ratios of the measured differential cross-sections for
(a) \oneeljets and (b) \twoeljets as a function of \ptrec in the inclusive jet phase space and
for \twomujets as a function of (c) \mjj and (d) \dphijj in the VBF phase space.
The bottom panels show the ratios of the predictions to the data with their statistical uncertainties as solid markers and shaded bands to indicate the combined statistical and systematic uncertainties.}
\label{fig:ptmiss_inc_rmiss}
\end{figure}
 
\subsection{$Z \rightarrow \nu\bar{\nu}$ measurement}
 
The \zjets process dominates the zero-lepton phase space.
As discussed in Section~\ref{sec:correction}, non-\zjets SM processes can be subtracted from the data before detector corrections to
extract a measurement of \zjets.
This method gives consistent results with the inclusive measurements when the particle-level predictions for the subtracted processes are added back in after unfolding.
The differential cross-section for \zjets is shown in Figure~\ref{fig:zpt} as a function of \ptZ in the
\onejet and \vbf phase spaces, and as a function of \mjj and \dphijj in the \vbf phase space. The level of description by the SM is similar to
that for the inclusive measurement.
 
\begin{figure}
\centering
\subfigure[]{\includegraphics[scale=0.4]{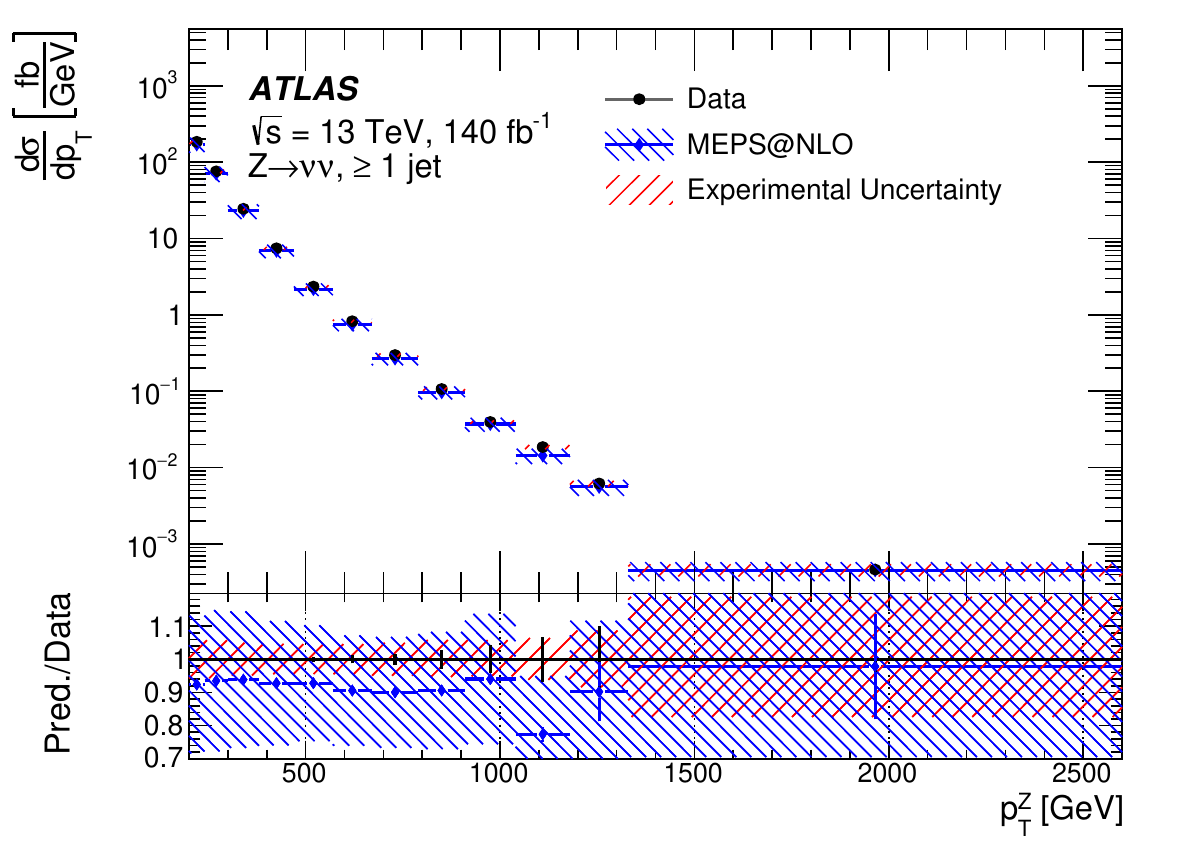}}
\subfigure[]{\includegraphics[scale=0.4]{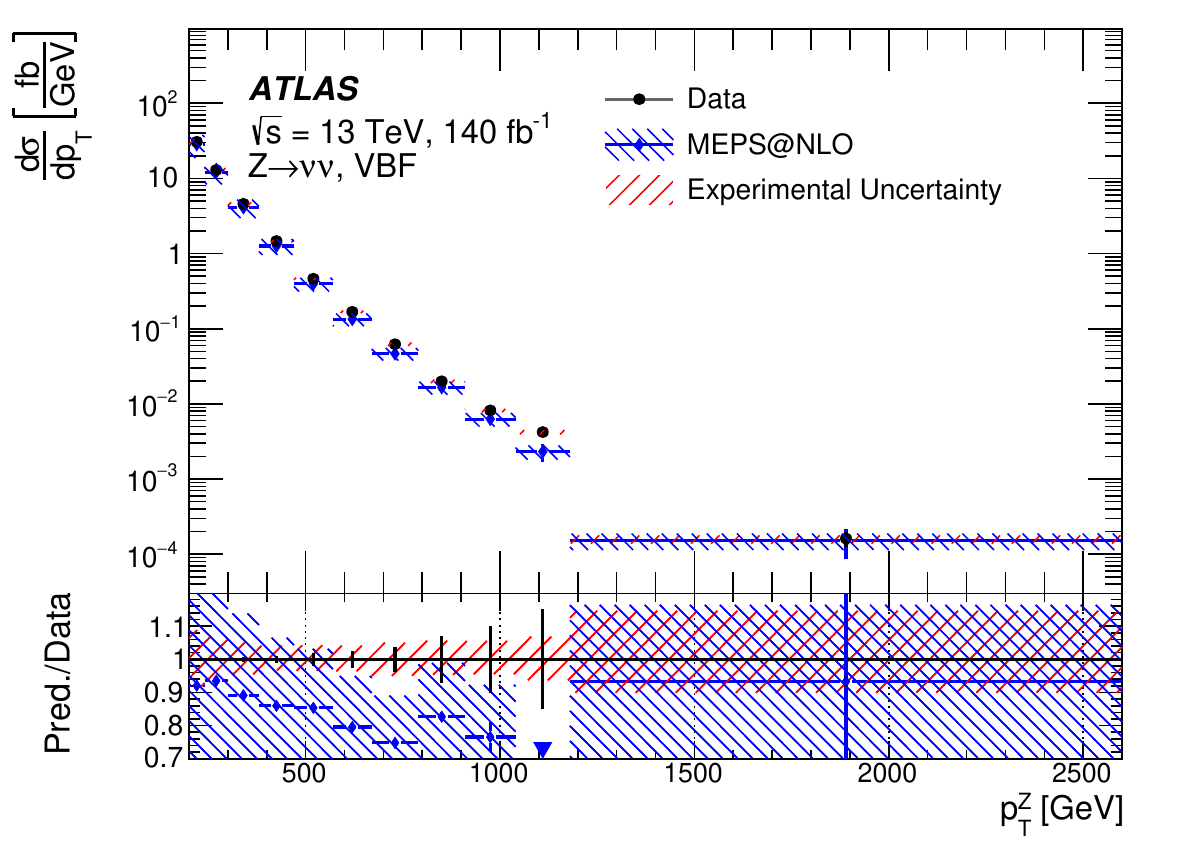}}
\subfigure[]{\includegraphics[scale=0.4]{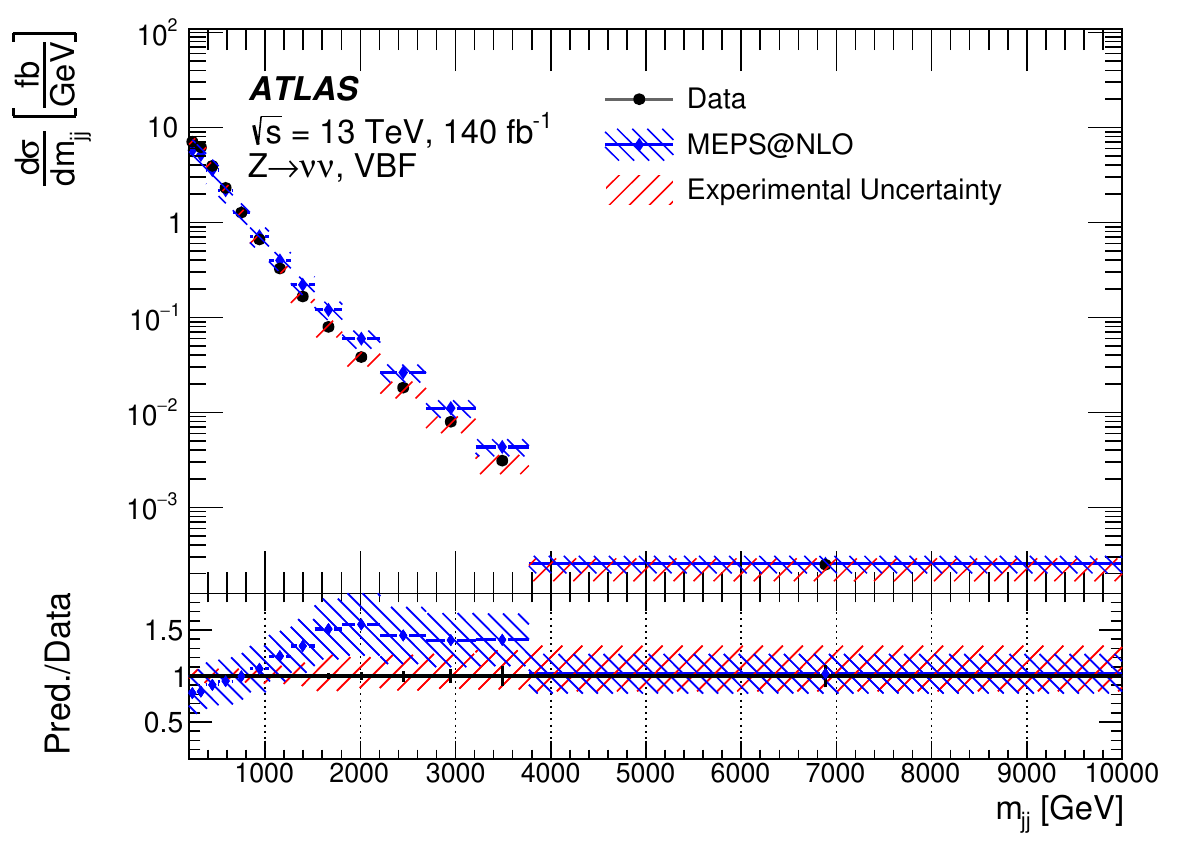}}
\subfigure[]{\includegraphics[scale=0.4]{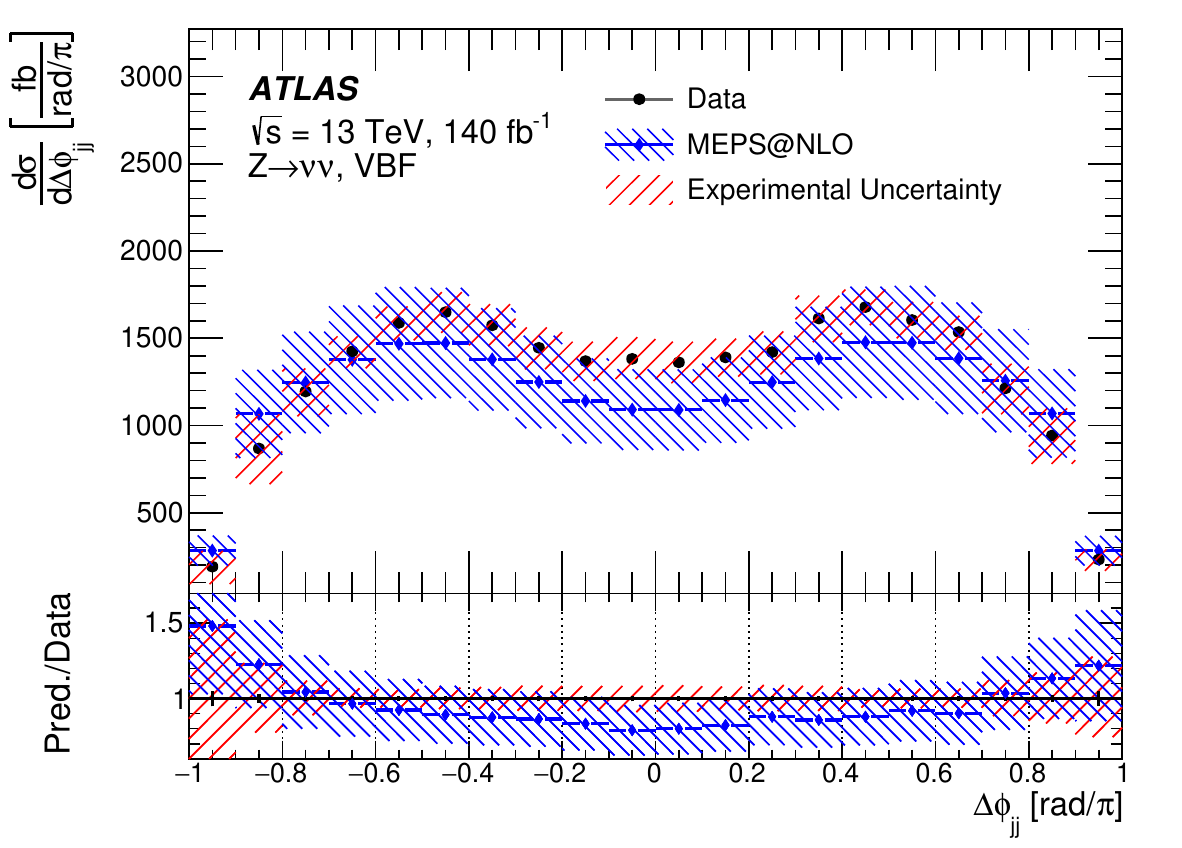}}
\caption{The measured \Znunu cross-section, differential in \ptZ in the (a) single jet and (b) \vbf phase spaces, and
differential in (c) \mjj and (d) \dphijj in the \vbf phase space, compared with the SM predictions.
The lower panels show the ratios of the predictions to the data, along with the data statistical uncertainties (black bars)
and systematic uncertainties (red hatched band).}
\label{fig:zpt}
\end{figure}
 
The production of an isolated photon in association with jets, \onegjets, in a similar kinematic region, shares several theoretical and experiment
uncertainties with the \zjets process. The measurements of this final state are shown in Figure~\ref{fig:gpt}.
 
\begin{figure}
\centering
\subfigure[]{\includegraphics[scale=0.4]{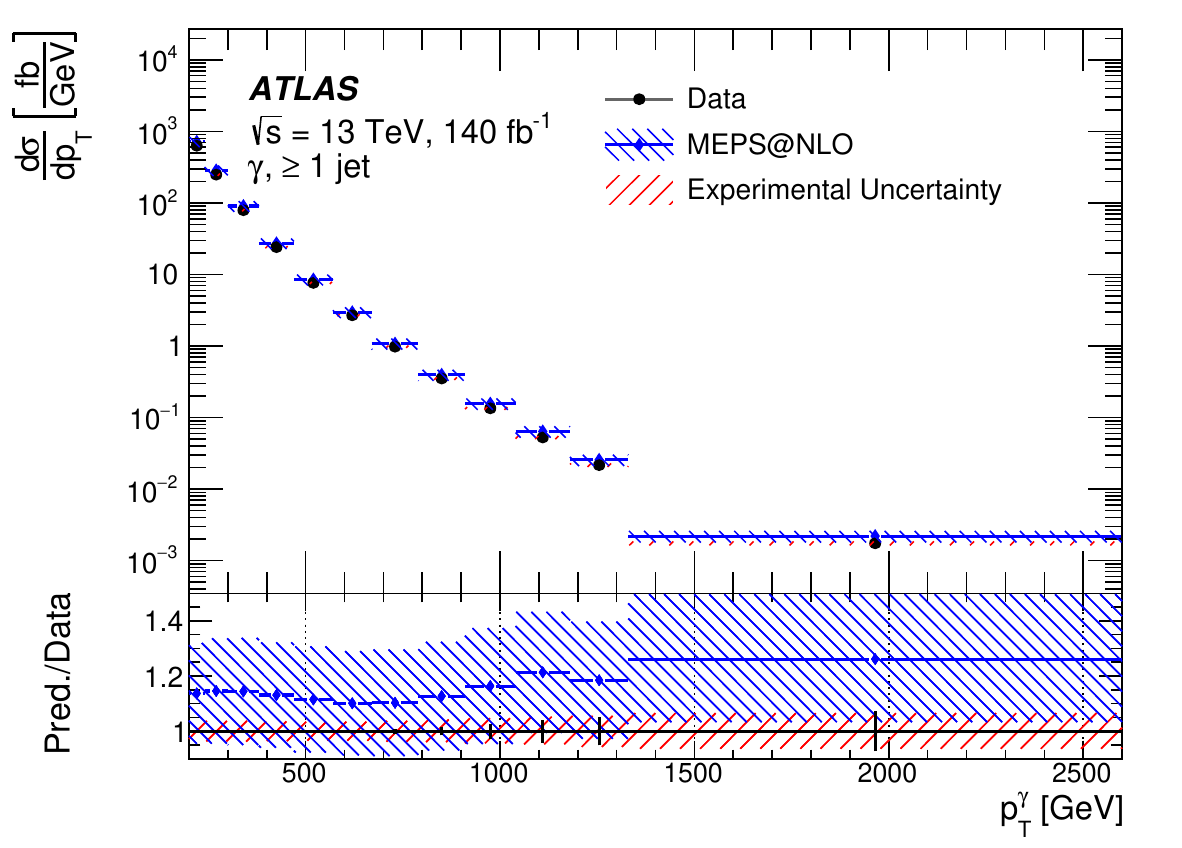}}
\subfigure[]{\includegraphics[scale=0.4]{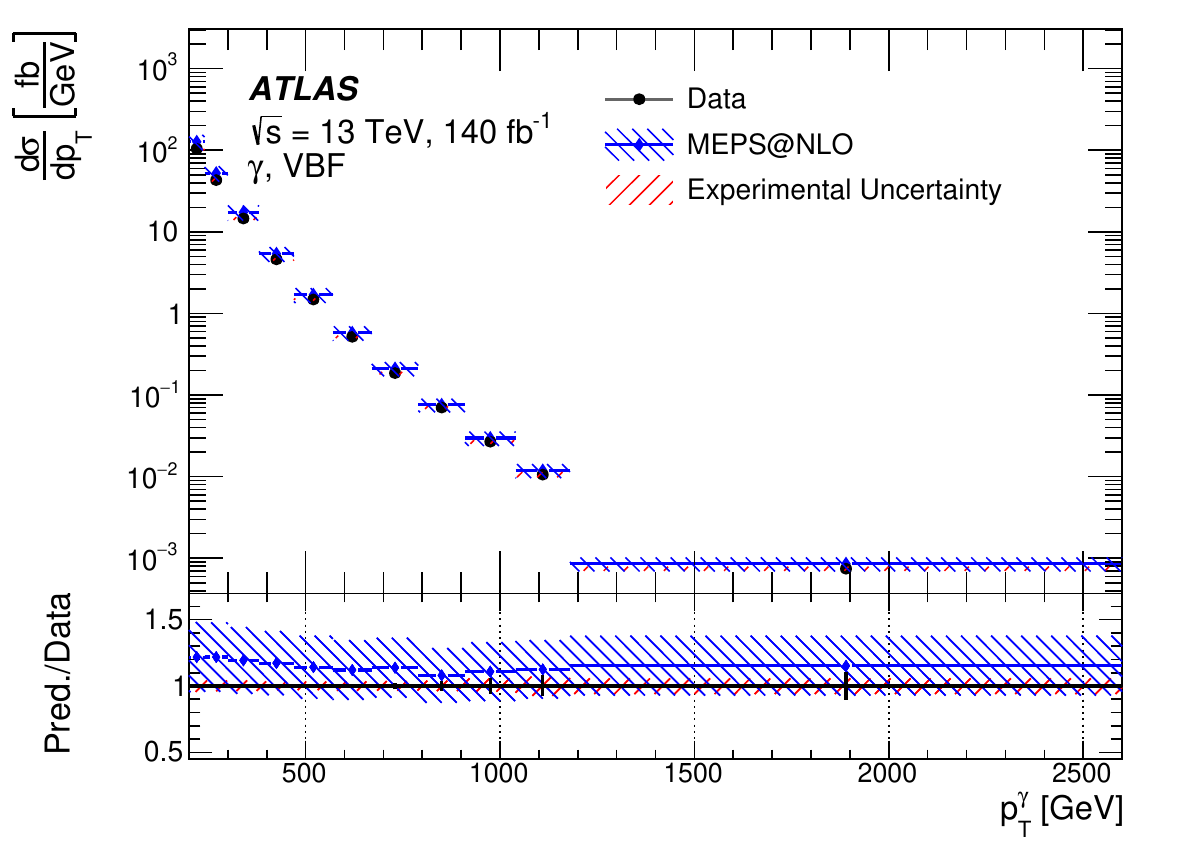}}
\subfigure[]{\includegraphics[scale=0.4]{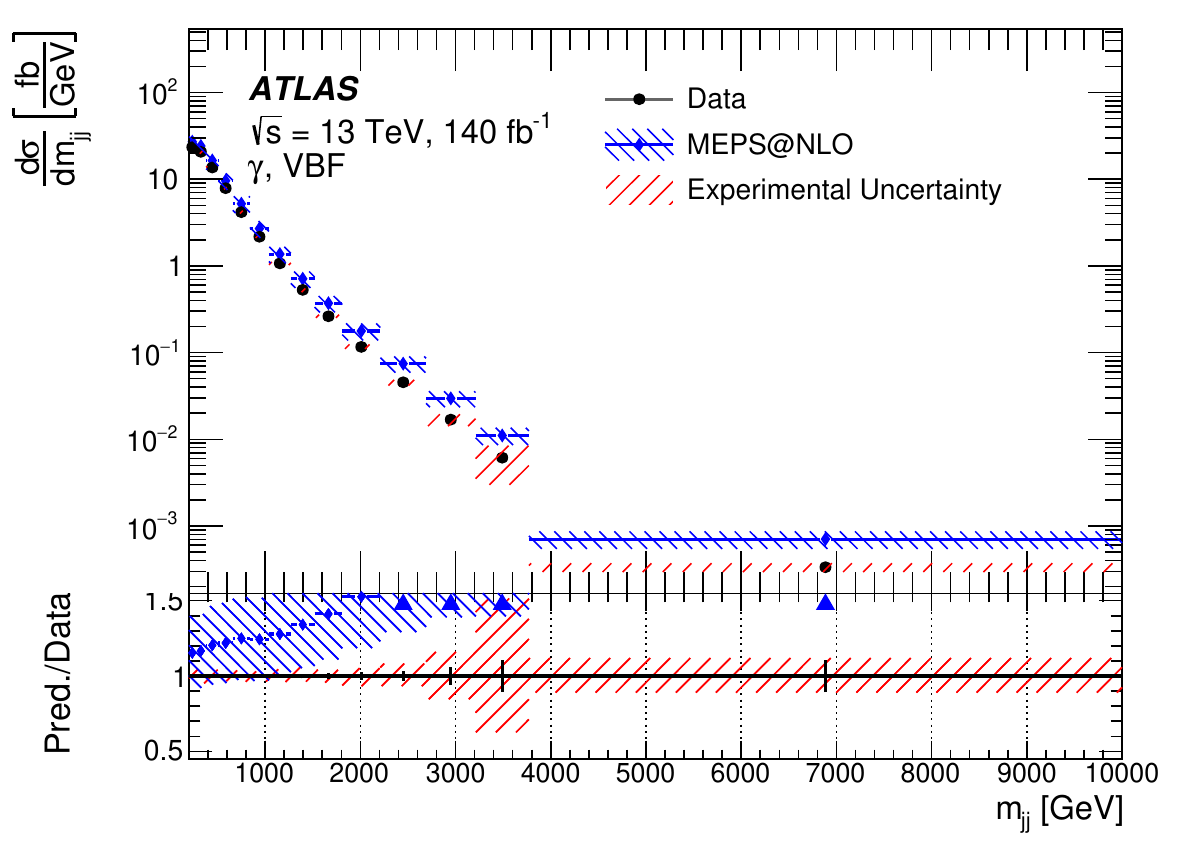}}
\subfigure[]{\includegraphics[scale=0.4]{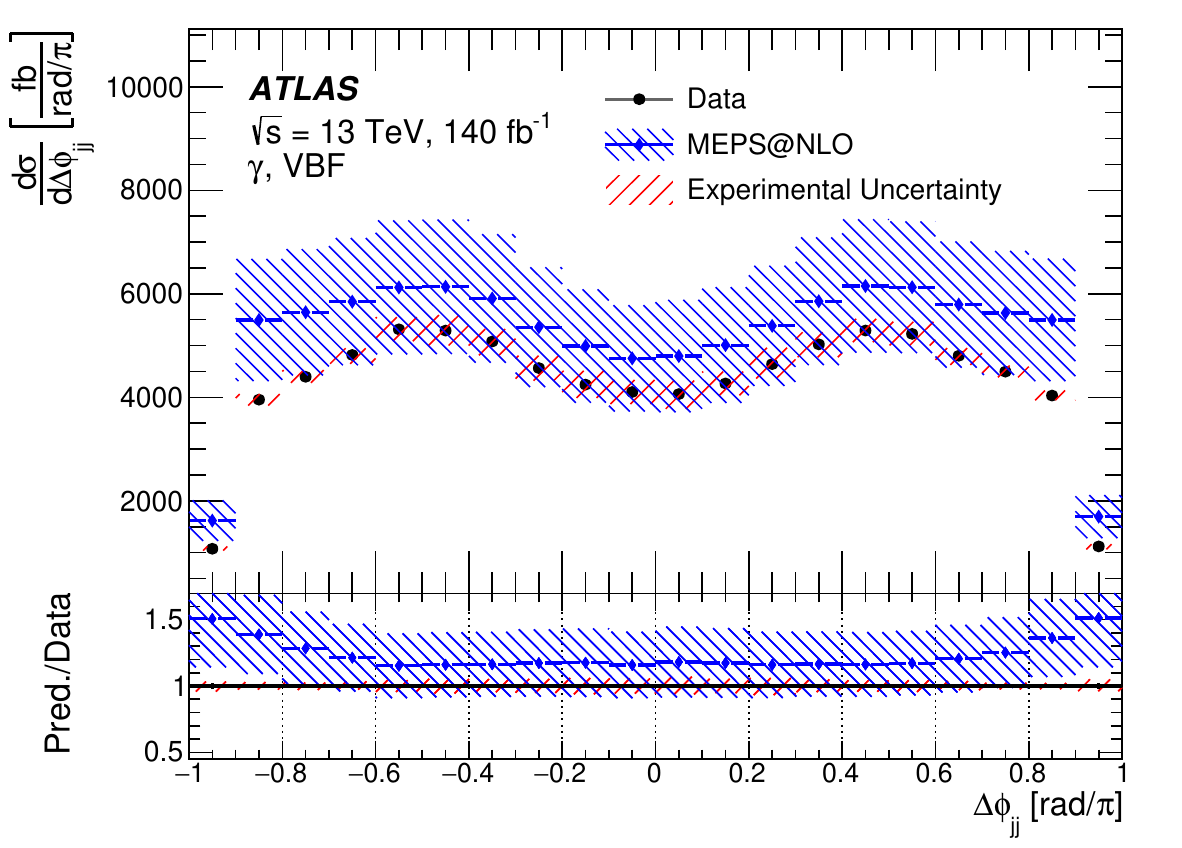}}
\caption{The measured \onegjets cross-section, differential in \ptg in the (a) single jet and (b) \vbf phase spaces, and
differential in (c) \mjj and (d) \dphijj in the \vbf phase space, compared with the SM predictions.
The lower panels show the ratios of the predictions to the data, along with the data statistical uncertainties (black bars)
and systematic uncertainties (red hatched band).}
\label{fig:gpt}
\end{figure}
 
The \onegjets cross-section is generally a factor of about five above the \zjets cross-section, with similar features.
The prediction is 10\%--20\% above the data, although generally lies within
the uncertainties in the calculation for the cross-section measurements. The discrepancy remains in the \rmiss measurements, since this normalisation issue only applies in the denominator and therefore does not cancel out.
Similar shifts were observed in independent studies~\cite{Chen:2019zmr}, and were found to be induced by photon isolation criteria.
 
\subsection{Quantitative comparison to SM predictions}
\label{sec:smstudy}
 
To quantify the level of agreement or disagreement between the measurement and the SM predictions, fits are performed by minimising the negative logarithm of the likelihood.
Experimental and theoretical uncertainties are either added to a covariance matrix (if their impact is below one percent everywhere in the fitted distributions) or otherwise assigned to a nuisance parameter (NP) that is allowed to float according to the estimated uncertainty.
The likelihood is evaluated for the resulting level of agreement, under the condition that the SM is the correct
underlying model and taking into account the residuals, the covariance matrix and pulls on the nuisance parameters.
Fits to the differential cross-sections are performed using the inclusive measurements, in all phase spaces, using the MEPS@NLO prediction. Since the \onegjets measurements show a strong normalisation offset not present in the other regions, only the  \ptmissjets, \oneeljets, \twoeljets, \onemujets and \twomujets regions (or their respective \rmiss ratios) are used in the fits.
Since the phase spaces are not orthogonal, statistical correlations exist between some measurements. These are evaluated using the Bootstrap method~\cite{ATL-PHYS-PUB-2021-011}, and are accounted for in the correlation matrix in the fitting procedure.
 
The level of agreement between the fitted results and the SM for \ptrec is reasonable, with a $\chi^2/\text{d.o.f.} \approx 101/57$.
For the differential cross-section as a function of \dphijj, the post-fit distributions also show reasonable agreement
between the measurement and the SM\@.
For \mjj however, the agreement is not good, due to the poor modelling of this distribution seen in Figure~\ref{fig:vbf_0l_2m}.
Because of this, the combined fit using all distributions for all observables, regions and phase spaces simultaneously also shows poor agreement,
with a $\chi^2/\text{d.o.f.}  \approx 390/70$.
 
Fits are also performed to ratios of the measurements, \rmiss, defined as the fiducial cross-section differential in each kinematic variable
for \ptmissjets events, divided by the same cross-section for events in each of the \oneeljets, \onemujets, \twoeljets and \twomujets regions. In this case,
some of the uncertainties largely or completely cancel out, and thus do not have an associated nuisance parameter. In addition, the modelling discrepancy in
\mjj is seen in all regions and so cancels out to a large extent in \rmiss. The ratio plots are therefore expected to give improved fit results
compared to the cross-section fits. Indeed a $\chi^2/\text{d.o.f.} \approx 323/220$ is obtained for the combined \rmiss fit to all distributions in all regions,
with the individual fit to \rmiss as a function of \mjj having $\chi^2/\text{d.o.f.} \approx 62/56$, indicating that indeed the discrepancy between
data and the SM predictions cancels out between the different regions.
 
For re-interpretation, the \rmiss measurements are always used. Specifically, the fits use the \rmiss distributions as a function of \ptmiss: either using just the inclusive jet phase space ($\chi^2/\text{d.o.f.} \approx 48/45$) or both the inclusive jet and \vbf phase spaces ($\chi^2/\text{d.o.f.} \approx 110/84$). Both of these options display good agreement between the SM predictions and the measurements, and therefore can be safely used for re-interpretation and establishing constraints on new physics models.
 
\FloatBarrier


 
\section{Implications for physics beyond the Standard Model}
\label{sec:interpretation}
 
Part of the motivation of these particle-level measurements is that they can easily be confronted with new SM predictions, and predictions
from extensions to the SM\@. Since the measurements of \ptmiss and \rmiss are consistent with the SM, they can thus be used to set limits on BSM
physics, particularly models that contain a DM candidate, the presence of which could affect the \ptmiss
distribution. This procedure was already demonstrated in the previous measurement~\cite{EXOT-2016-03} using a subset of the current data,
and is extended and updated here.

A common benchmark simplified model involves extending the SM with an additional U(1) gauge symmetry, in which a DM candidate, $\chi$,
is a Dirac fermion that has charges only under this gauge group~\cite{Abdallah:2015ter}.
If SM quarks are also charged under this gauge group, DM particles can be produced at
the LHC via the gauge boson, $Z^\prime$, associated with the new U(1) symmetry, which is massive assuming the U(1) symmetry is
spontaneously broken.
This  model was searched for previously by ATLAS~\cite{ATLAS:2021kxv} using the same data sample as the current analysis,
and by CMS~\cite{CMS:2021far}.
 
The case where $\chi$ has axial-vector couplings $g_{\chi}=1.0$ to the $Z^\prime$, and the coupling between the quarks and the $Z^\prime$, $g_{q}=0.25$,
is studied in this analysis in the ($m_{\chi}$, $m_{Z^\prime}$) plane, to compare the sensitivity of the current measurement
to that of the search results.
The fitting procedure described in the previous section is used to evaluate the likelihoods associated with a fit to \rmiss
when excluding or including a BSM contribution from this model. During fitting, the predicted \rmiss is recalculated to include the signal.
The limits obtained from the likelihood ratio, shown in Figure~\ref{fig:interpret:AVDM}, are very similar to those previously published in the ATLAS search
analysis~\cite{ATLAS:2021kxv}, with mediator masses up to about 2.1~\TeV\ excluded in the region $m_{Z^\prime}>2m_{\chi}$. The residual differences
are attributed to the slightly different kinematic selections used, and the requirement for a minimum number of events in the high \ptmiss
bin for unfolding.
This demonstrates that particle-level measurements can be used for searches and setting constraints with only a minor penalty in sensitivity, and with the advantage that it can be done in future without the need to repeat complex and time-consuming detector simulation.
 
\begin{figure}
\centering
\includegraphics[height=200pt]{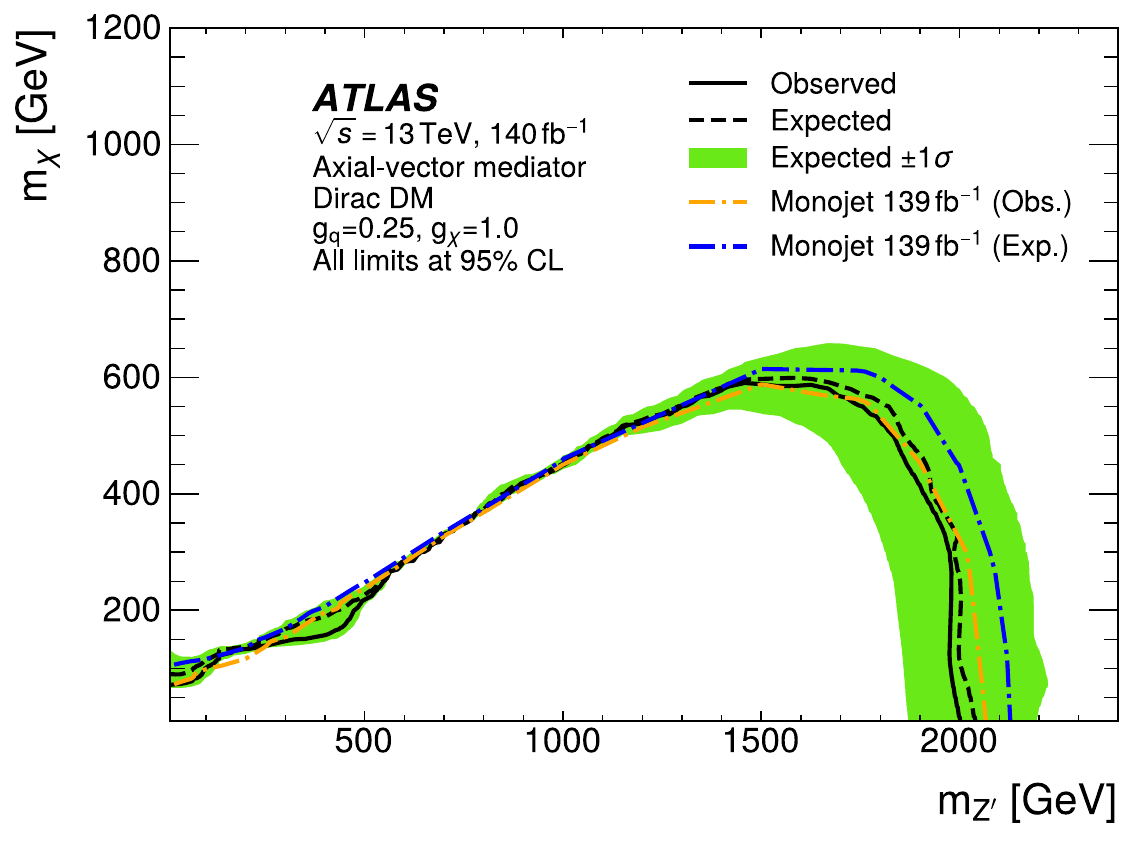}
\caption{
Exclusion limits at 95\% in the plane of Dark Matter mass and mediator mass for a simplified DM model with
an axial-vector coupling to the SM\@. The limits from this analysis, evaluated using the particle-level \rmiss measurements, are
compared with the limits from the ATLAS monojet search~\cite{ATLAS:2021kxv}. The region below the solid line is excluded.
}
\label{fig:interpret:AVDM}
\end{figure}
 
A more complicated model for DM involves the introduction of an additional Higgs doublet, along with a pseudoscalar, $a$,
which couples to DM~\cite{Bauer:2017ota,LHCDarkMatterWorkingGroup:2018ufk}.
Through mixing with the pseudoscalar component of the Higgs doublet, the pseudoscalar $a$ acts as a mediator between the SM
and the dark sector.
The model, referred to as \THDMa, has a rich phenomenology with a wide variety of possible final states
produced~\cite{Pani:2017qyd,CidVidal:2018eel,ATL-PHYS-PUB-2018-027,Haisch:2018djm,Butterworth:2020vnb,Robens:2021lov},
many of which involve \ptmiss produced in association with jets or other SM objects.
In this analysis, a signal scan is  conducted in the \MaTanB plane where $\MA=\SI{600}{\GeV}$, as shown in Figure~\ref{fig:interpret:2HDMa}.
The scan makes use of the full set of \rmiss measurements, in both the \onejet and \vbf phase spaces, taking into
account statistical and systematic correlations.
 
\begin{figure}
\centering
\subfigure[]{\includegraphics[height=200pt]{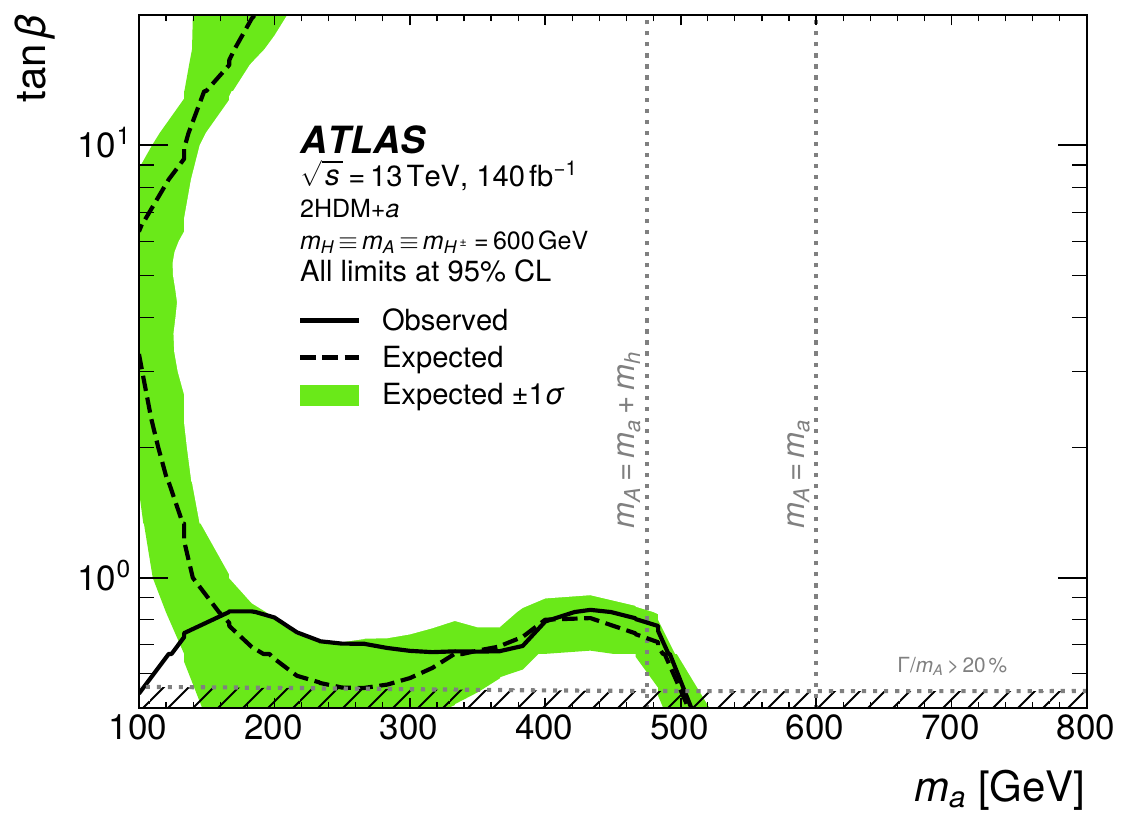}}
\caption{
Exclusion limits at 95\% in the \MaTanB plane for the \THDMa model.
A region of \tanB below 0.7 is excluded for \Ma values between approximately 150~\GeV\ and 500~\GeV.
The regions shaded with diagonal lines indicate the region where the width of any of the Higgs bosons exceeds \SI{20}{\%} of its mass;
in this region the narrow-width approximation is violated and predictions become less precise. Moreover, the two vertical dashed lines represent scenarios where the mass of the neutral pseudoscalar $A$ is either equal to the mass of the pseudoscalar $a$ or a sum of the masses of $a$ and Higgs boson $h$. The region below the solid line is excluded.
}
\label{fig:interpret:2HDMa}
\end{figure}
 
The scan reveals two major regions of sensitivity:
\begin{itemize}
\item
For $\tanB<0.7$, masses of the pseudoscalar~$a$ up to \SI{520}{\GeV} are excluded because of loop-induced production of $a$
and its subsequent decay into DM particles, $pp\to a(\to\xx)+$jets.
The sensitivity is larger at $\Ma>\SI{350}{\GeV}\approx2m_\mathrm{top}$ because here the $a$ can be produced resonantly from top quarks.
For $\tanB\gg10$, there is a second island of sensitivity because of $b$-quark induced production of $a$ and its subsequent decay into DM particles.
\item
At small \Ma, the expected exclusion limits are generally stronger because of processes almost independent of \tanB, e.g., $pp\to H\to aZ$ and $pp\to H^\pm\to aW^\pm$.
However, the sensitivity to these processes is not large enough to close the sensitivity gap between small and large values of \tanB.
\end{itemize}
Qualitatively the sensitivity is similar to the existing exclusion from \ptmiss-based searches in different final states~\cite{ATLAS:2023rvb}.
Differences in the exclusion limits originate from differences in the SM calculations used, and from the use of the
\vbf phase space in addition to the \onejet region.
 
Overall these studies show that the inclusive, particle-level measurement provides good sensitivity to BSM physics, and is amenable
to reinterpretation in terms of different models.
 
\FloatBarrier


\section{Conclusion}
\label{sec:conclusion}
 
Inclusive measurements of \ptmiss are made using 140~fb$^{-1}$ of $pp$ collision data at $\sqrt{s} = 13$~\TeV\ collected with the ATLAS detector during Run 2 of the LHC\@. The measurements
are made in fiducial regions closely reflecting the detector acceptance, and are corrected for detector effects
within these regions, yielding differential cross-sections defined in terms of final-state particles.
Differential cross-sections are measured as a function of \ptmiss in a \onejet and a \vbf phase space.
The latter is defined by the presence of two jets, and the cross-section is also measured as a function of the azimuthal
angular distance between the jets, \dphijj, and the dijet invariant mass, \mjj.
The cross-section for \Znunu production is determined, differential in the \ptmiss and, in the \vbf
phase space, in \dphijj and \mjj.
 
Measurements of lepton+jets, dilepton+jets and $\gamma$+jets final states are also made in the same kinematic regions.
Many uncertainties, both theoretical and experimental, are correlated between these measurements and the \ptmiss measurements,
and therefore cancel out in the ratio of cross-sections, \rmiss.
 
Quantitative comparisons with state-of-the-art SM predictions show a reasonable description of all measured cross-sections as a function of
most observables, except \mjj. The discrepancy in the shape of the distribution of this observable is present in the lepton+jets, dilepton+jets and
$\gamma$+jets measurements as well
as in the \ptmiss region. It therefore cancels in \rmiss, which is well described by the predictions.
The resummed calculation in the HEJ prediction,
available for the leptonic measurements, provides a better description of \mjj.
 
The measurements are designed to be readily reinterpreted, and the effectiveness of this is illustrated by comparisons with two DM models.
Specifically, the measured \rmiss distribution is used to reproduce limits on a simplified DM model, obtaining results consistent
with a previously published search using the same data set.
Limits are also set on a model involving an additional Higgs doublet and a pseudoscalar coupling to a
DM particle, where again they are similar to those obtained in searches, with extended sensitivity in some regions due to the use
of the \vbf phase space in addition to the \onejet region.
The derived constraints are found to be only marginally weaker than for dedicated searches, while eliminating the need for
complicated detector simulations. The published results can consequently be directly used for future interpretations.
Information about uncertainties and correlations is provided on HEPData,
along with
a Rivet analysis, to facilitate the use of this LHC Run 2 measurement in future studies with other new physics models and
improved SM predictions, as they become available.


\section*{Acknowledgements}


We thank CERN for the very successful operation of the LHC and its injectors, as well as the support staff at
CERN and at our institutions worldwide without whom ATLAS could not be operated efficiently.
 
The crucial computing support from all WLCG partners is acknowledged gratefully, in particular from CERN, the ATLAS Tier-1 facilities at TRIUMF/SFU (Canada), NDGF (Denmark, Norway, Sweden), CC-IN2P3 (France), KIT/GridKA (Germany), INFN-CNAF (Italy), NL-T1 (Netherlands), PIC (Spain), RAL (UK) and BNL (USA), the Tier-2 facilities worldwide and large non-WLCG resource providers. Major contributors of computing resources are listed in Ref.~\cite{ATL-SOFT-PUB-2023-001}.
 
We gratefully acknowledge the support of ANPCyT, Argentina; YerPhI, Armenia; ARC, Australia; BMWFW and FWF, Austria; ANAS, Azerbaijan; CNPq and FAPESP, Brazil; NSERC, NRC and CFI, Canada; CERN; ANID, Chile; CAS, MOST and NSFC, China; Minciencias, Colombia; MEYS CR, Czech Republic; DNRF and DNSRC, Denmark; IN2P3-CNRS and CEA-DRF/IRFU, France; SRNSFG, Georgia; BMBF, HGF and MPG, Germany; GSRI, Greece; RGC and Hong Kong SAR, China; ISF and Benoziyo Center, Israel; INFN, Italy; MEXT and JSPS, Japan; CNRST, Morocco; NWO, Netherlands; RCN, Norway; FCT, Portugal; MNE/IFA, Romania; MESTD, Serbia; MSSR, Slovakia; ARRS and MIZ\v{S}, Slovenia; DSI/NRF, South Africa; MICINN, Spain; SRC and Wallenberg Foundation, Sweden; SERI, SNSF and Cantons of Bern and Geneva, Switzerland; MOST, Taipei; TENMAK, T\"urkiye; STFC, United Kingdom; DOE and NSF, United States of America.
 
Individual groups and members have received support from BCKDF, CANARIE, CRC and DRAC, Canada; CERN-CZ, PRIMUS 21/SCI/017 and UNCE SCI/013, Czech Republic; COST, ERC, ERDF, Horizon 2020, ICSC-NextGenerationEU and Marie Sk{\l}odowska-Curie Actions, European Union; Investissements d'Avenir Labex, Investissements d'Avenir Idex and ANR, France; DFG and AvH Foundation, Germany; Herakleitos, Thales and Aristeia programmes co-financed by EU-ESF and the Greek NSRF, Greece; BSF-NSF and MINERVA, Israel; Norwegian Financial Mechanism 2014-2021, Norway; NCN and NAWA, Poland; La Caixa Banking Foundation, CERCA Programme Generalitat de Catalunya and PROMETEO and GenT Programmes Generalitat Valenciana, Spain; G\"{o}ran Gustafssons Stiftelse, Sweden; The Royal Society and Leverhulme Trust, United Kingdom.
 
In addition, individual members wish to acknowledge support from CERN: European Organization for Nuclear Research (CERN PJAS); Chile: Agencia Nacional de Investigaci\'on y Desarrollo (FONDECYT 1190886, FONDECYT 1210400, FONDECYT 1230812, FONDECYT 1230987); China: National Natural Science Foundation of China (NSFC - 12175119, NSFC 12275265, NSFC-12075060); Czech Republic: PRIMUS Research Programme (PRIMUS/21/SCI/017); EU: H2020 European Research Council (ERC - 101002463); European Union: European Research Council (ERC - 948254, ERC 101089007), Horizon 2020 Framework Programme (MUCCA - CHIST-ERA-19-XAI-00), European Union, Future Artificial Intelligence Research (FAIR-NextGenerationEU PE00000013), Italian Center for High Performance Computing, Big Data and Quantum Computing (ICSC, NextGenerationEU); France: Agence Nationale de la Recherche (ANR-20-CE31-0013, ANR-21-CE31-0013, ANR-21-CE31-0022), Investissements d'Avenir Labex (ANR-11-LABX-0012); Germany: Baden-Württemberg Stiftung (BW Stiftung-Postdoc Eliteprogramme), Deutsche Forschungsgemeinschaft (DFG - 469666862, DFG - CR 312/5-2); Italy: Istituto Nazionale di Fisica Nucleare (ICSC, NextGenerationEU); Japan: Japan Society for the Promotion of Science (JSPS KAKENHI JP21H05085, JSPS KAKENHI JP22H01227, JSPS KAKENHI JP22H04944, JSPS KAKENHI JP22KK0227); Netherlands: Netherlands Organisation for Scientific Research (NWO Veni 2020 - VI.Veni.202.179); Norway: Research Council of Norway (RCN-314472); Poland: Polish National Agency for Academic Exchange (PPN/PPO/2020/1/00002/U/00001), Polish National Science Centre (NCN 2021/42/E/ST2/00350, NCN OPUS nr 2022/47/B/ST2/03059, NCN UMO-2019/34/E/ST2/00393, UMO-2020/37/B/ST2/01043, UMO-2021/40/C/ST2/00187, UMO-2022/47/O/ST2/00148); Slovenia: Slovenian Research Agency (ARIS grant J1-3010); Spain: BBVA Foundation (LEO22-1-603), Generalitat Valenciana (Artemisa, FEDER, IDIFEDER/2018/048), La Caixa Banking Foundation (LCF/BQ/PI20/11760025), Ministry of Science and Innovation (MCIN \& NextGenEU PCI2022-135018-2, MICIN \& FEDER PID2021-125273NB, RYC2019-028510-I, RYC2020-030254-I, RYC2021-031273-I, RYC2022-038164-I), PROMETEO and GenT Programmes Generalitat Valenciana (CIDEGENT/2019/023, CIDEGENT/2019/027); Sweden: Swedish Research Council (VR 2018-00482, VR 2022-03845, VR 2022-04683, VR grant 2021-03651), Knut and Alice Wallenberg Foundation (KAW 2017.0100, KAW 2018.0157, KAW 2018.0458, KAW 2019.0447, KAW 2022.0358); Switzerland: Swiss National Science Foundation (SNSF - PCEFP2\_194658); United Kingdom: Leverhulme Trust (Leverhulme Trust RPG-2020-004); United States of America: U.S. Department of Energy (ECA DE-AC02-76SF00515), Neubauer Family Foundation.



\printbibliography

@article{LUCID2,
    author = "Avoni, G. and others",
    title = "{The new LUCID-2 detector for luminosity measurement and monitoring in ATLAS}",
    doi = "10.1088/1748-0221/13/07/P07017",
    journal = "JINST",
    volume = "13",
    number = "07",
    pages = "P07017",
    year = "2018"
}

@article{Cacciari:2011ma,
    author = "Cacciari, Matteo and Salam, Gavin P. and Soyez, Gregory",
    title = "{FastJet User Manual}",
    eprint = "1111.6097",
    archivePrefix = "arXiv",
    primaryClass = "hep-ph",
    reportNumber = "CERN-PH-TH-2011-297",
    doi = "10.1140/epjc/s10052-012-1896-2",
    journal = "Eur. Phys. J. C",
    volume = "72",
    pages = "1896",
    year = "2012"
}

@article{CMS:2020ulv,
    author  = "{CMS Collaboration}",
    collaboration = "CMS",
    title = "{Search for dark matter produced in association with a leptonically decaying Z boson in proton-proton collisions at $\sqrt{s} =$ 13 TeV}",
    eprint = "2008.04735",
    archivePrefix = "arXiv",
    primaryClass = "hep-ex",
    reportNumber = "CMS-EXO-19-003, CERN-EP-2020-136",
    doi = "10.1140/epjc/s10052-020-08739-5",
    journal = "Eur. Phys. J. C",
    volume = "81",
    number = "1",
    pages = "13",
    year = "2021",
    related = "CMS:2020ulv-err",
    relatedstring = "Erratum:",
}

@article{ATLAS:2023rvb,
    author = "{ATLAS Collaboration}",
    collaboration = "ATLAS",
    title = "{Combination and summary of ATLAS dark matter searches interpreted in a 2HDM with a pseudo-scalar mediator using 139 fb$^{-1}$ of $\sqrt{s} = 13$ TeV $pp$ collision data}",
    eprint = "2306.00641",
    archivePrefix = "arXiv",
    primaryClass = "hep-ex",
    reportNumber = "CERN-EP-2023-088",
    doi = "10.48550/arXiv.2306.00641",
    month = "6",
    year = "2023"
}

@article{Pani:2017qyd,
    author = "Pani, Priscilla and Polesello, Giacomo",
    title = "{Dark matter production in association with a single top-quark at the LHC in a two-Higgs-doublet model with a pseudoscalar mediator}",
    eprint = "1712.03874",
    archivePrefix = "arXiv",
    primaryClass = "hep-ph",
    doi = "10.1016/j.dark.2018.04.006",
    journal = "Phys. Dark Univ.",
    volume = "21",
    pages = "8--15",
    year = "2018"
}

@article{CidVidal:2018eel,
    author = "Cid Vidal, Xabier and others",
    editor = "Dainese, Andrea and Mangano, Michelangelo and Meyer, Andreas B. and Nisati, Aleandro and Salam, Gavin and Vesterinen, Mika Anton",
    title = "{Beyond the Standard Model physics at the HL-LHC and HE-LHC}",
    eprint = "1812.07831",
    archivePrefix = "arXiv",
    primaryClass = "hep-ph",
    reportNumber = "CERN-LPCC-2018-05",
    doi = "10.23731/CYRM-2019-007.585",
    journal = "CERN Yellow Rep. Monogr.",
    volume = "7",
    pages = "585--865",
    year = "2019"
}

@article{Haisch:2018djm,
    author = "Haisch, Ulrich and Polesello, Giacomo",
    title = "{Searching for heavy Higgs bosons in the $t \bar t Z$ and $t b W$ final states}",
    eprint = "1807.07734",
    archivePrefix = "arXiv",
    primaryClass = "hep-ph",
    reportNumber = "CERN-TH-2018-155",
    doi = "10.1007/JHEP09(2018)151",
    journal = "JHEP",
    volume = "09",
    pages = "151",
    year = "2018"
}

@article{Butterworth:2020vnb,
    author = "Butterworth, J. M. and Habedank, M. and Pani, P. and Vaitkus, A.",
    title = "{A study of collider signatures for two Higgs doublet models with a Pseudoscalar mediator to Dark Matter}",
    eprint = "2009.02220",
    archivePrefix = "arXiv",
    primaryClass = "hep-ph",
    reportNumber = "MCnet-20, DESY-21-040",
    doi = "10.21468/SciPostPhysCore.4.1.003",
    journal = "SciPost Phys. Core",
    volume = "4",
    pages = "003",
    year = "2021"
}

@article{Robens:2021lov,
    author = "Robens, Tania",
    title = "{The THDMa Revisited}",
    eprint = "2106.02962",
    archivePrefix = "arXiv",
    primaryClass = "hep-ph",
    reportNumber = "RBI-ThPhys-2021-25",
    doi = "10.3390/sym13122341",
    journal = "Symmetry",
    volume = "13",
    number = "12",
    pages = "2341",
    year = "2021"
}

@article{Bauer:2017ota,
    author = "Bauer, Martin and Haisch, Ulrich and Kahlhoefer, Felix",
    title = "{Simplified dark matter models with two Higgs doublets: I. Pseudoscalar mediators}",
    eprint = "1701.07427",
    archivePrefix = "arXiv",
    primaryClass = "hep-ph",
    reportNumber = "CERN-TH-2017-011, DESY-17-010",
    doi = "10.1007/JHEP05(2017)138",
    journal = "JHEP",
    volume = "05",
    pages = "138",
    year = "2017"
}

@article{LHCDarkMatterWorkingGroup:2018ufk,
    author = "Abe, Tomohiro and others",
    collaboration = "LHC Dark Matter Working Group",
    title = "{LHC Dark Matter Working Group: Next-generation spin-0 dark matter models}",
    eprint = "1810.09420",
    archivePrefix = "arXiv",
    primaryClass = "hep-ex",
    reportNumber = "CERN-LPCC-2018-02",
    doi = "10.1016/j.dark.2019.100351",
    journal = "Phys. Dark Univ.",
    volume = "27",
    pages = "100351",
    year = "2020"
}

@article{CMS:2021far,
    author         = "{CMS Collaboration}",
    collaboration = "CMS",
    title = "{Search for new particles in events with energetic jets and large missing transverse momentum in proton-proton collisions at $ \sqrt{s} $ = 13 TeV}",
    eprint = "2107.13021",
    archivePrefix = "arXiv",
    primaryClass = "hep-ex",
    reportNumber = "CMS-EXO-20-004, CERN-EP-2021-136",
    doi = "10.1007/JHEP11(2021)153",
    journal = "JHEP",
    volume = "11",
    pages = "153",
    year = "2021"
}

@article{ATLAS:2021kxv,
    author         = "{ATLAS Collaboration}",
    collaboration = "ATLAS",
    title = "{Search for new phenomena in events with an energetic jet and missing transverse momentum in $pp$ collisions at $\sqrt {s}$ =13  TeV with the ATLAS detector}",
    eprint = "2102.10874",
    archivePrefix = "arXiv",
    primaryClass = "hep-ex",
    reportNumber = "CERN-EP-2020-238",
    doi = "10.1103/PhysRevD.103.112006",
    journal = "Phys. Rev. D",
    volume = "103",
    number = "11",
    pages = "112006",
    year = "2021"
}

@article{Abdallah:2015ter,
    author = "Abdallah, Jalal and others",
    title = "{Simplified Models for Dark Matter Searches at the LHC}",
    eprint = "1506.03116",
    archivePrefix = "arXiv",
    primaryClass = "hep-ph",
    reportNumber = "FERMILAB-PUB-15-283-CD, CERN-PH-TH-2015-139",
    doi = "10.1016/j.dark.2015.08.001",
    journal = "Phys. Dark Univ.",
    volume = "9-10",
    pages = "8--23",
    year = "2015"
}

@article{Maguire:2017ypu,
    author = "Maguire, Eamonn and Heinrich, Lukas and Watt, Graeme",
    editor = "Mount, Richard and Tull, Craig",
    title = "{HEPData: a repository for high energy physics data}",
    eprint = "1704.05473",
    archivePrefix = "arXiv",
    primaryClass = "hep-ex",
    reportNumber = "IPPP-17-31",
    doi = "10.1088/1742-6596/898/10/102006",
    journal = "J. Phys. Conf. Ser.",
    volume = "898",
    number = "10",
    pages = "102006",
    year = "2017"
}

@article{Bertone:2004pz,
    author = "Bertone, Gianfranco and Hooper, Dan and Silk, Joseph",
    title = "{Particle dark matter: Evidence, candidates and constraints}",
    eprint = "hep-ph/0404175",
    archivePrefix = "arXiv",
    reportNumber = "FERMILAB-PUB-04-047-A",
    doi = "10.1016/j.physrep.2004.08.031",
    journal = "Phys. Rept.",
    volume = "405",
    pages = "279--390",
    year = "2005"
}

@article{Bertone:2017bme,
    author = "{V. Bertone, S. Carrazza, N. P. Hartland, and J. Rojo}",
    collaboration = "NNPDF",
    title = "{Illuminating the photon content of the proton within a global PDF analysis}",
    eprint = "1712.07053",
    archivePrefix = "arXiv",
    primaryClass = "hep-ph",
    reportNumber = "Nikhef/2017-064, CERN-TH-2017-235, NIKHEF-2017-064",
    doi = "10.21468/SciPostPhys.5.1.008",
    journal = "SciPost Phys.",
    volume = "5",
    number = "1",
    pages = "008",
    year = "2018"
}

@article{Bernlochner:2018opw,
      author         = "Bernlochner, Florian U. and Englert, Christoph and Hays,
                        Chris and Lohwasser, Kristin and Mildner, Hannes and
                        Pilkington, Andrew and Price, Darren D. and Spannowsky,
                        Michael",
      title          = "{Angles on CP-violation in Higgs boson interactions}",
      journal        = "Phys. Lett. B",
      volume         = "790",
      year           = "2019",
      pages          = "372-379",
      doi            = "10.1016/j.physletb.2019.01.043",
      eprint         = "1808.06577",
      archivePrefix  = "arXiv",
      primaryClass   = "hep-ph",
      reportNumber   = "IPPP/18/68",
      SLACcitation   = "%%CITATION = ARXIV:1808.06577;%%"
}

@article{Allison:2006ve,
      author         = "Allison, John and others",
      title          = "{Geant4 developments and applications}",
      journal        = "IEEE Trans. Nucl. Sci.",
      volume         = "53",
      year           = "2006",
      pages          = "270",
      doi            = "10.1109/TNS.2006.869826",
      reportNumber   = "SLAC-PUB-11870",
      SLACcitation   = "%%CITATION = IETNA,53,270;%%"
}

@article{Bierlich:2019rhm,
    author = "Bierlich, Christian and others",
    title = "{Robust Independent Validation of Experiment and Theory: Rivet version 3}",
    eprint = "1912.05451",
    archivePrefix = "arXiv",
    primaryClass = "hep-ph",
    reportNumber = "MCnet-19-26",
    doi = "10.21468/SciPostPhys.8.2.026",
    journal = "SciPost Phys.",
    volume = "8",
    pages = "026",
    year = "2020"
}

@article{Lindert:2017olm,
      author         = "Lindert, J. M. and others",
      title          = "{Precise predictions for $V+$ jets dark matter
                        backgrounds}",
      journal        = "Eur. Phys. J. C",
      volume         = "77",
      year           = "2017",
      number         = "12",
      pages          = "829",
      doi            = "10.1140/epjc/s10052-017-5389-1",
      eprint         = "1705.04664",
      archivePrefix  = "arXiv",
      primaryClass   = "hep-ph",
      reportNumber   = "CERN-TH-2017-102, CERN-LPCC-2017-02, IPPP-17-38,
                        FERMILAB-PUB-17-152-T, ZU--TH-12-17",
      SLACcitation   = "%%CITATION = ARXIV:1705.04664;%%"
}

@article{Gehrmann-DeRidder:2015wbt,
    author = "Gehrmann-De Ridder, A. and Gehrmann, T. and Glover, E. W. N. and Huss, A. and Morgan, T. A.",
    title = "{Precise QCD predictions for the production of a Z boson in association with a hadronic jet}",
    eprint = "1507.02850",
    archivePrefix = "arXiv",
    primaryClass = "hep-ph",
    reportNumber = "IPPP-15-44, ZU-TH-23-15",
    doi = "10.1103/PhysRevLett.117.022001",
    journal = "Phys. Rev. Lett.",
    volume = "117",
    number = "2",
    pages = "022001",
    year = "2016"
}

@article{Gehrmann-DeRidder:2016cdi,
    author = "Gehrmann-De Ridder, Aude and Gehrmann, T. and Glover, E. W. N. and Huss, A. and Morgan, T. A.",
    title = "{The NNLO QCD corrections to Z boson production at large transverse momentum}",
    eprint = "1605.04295",
    archivePrefix = "arXiv",
    primaryClass = "hep-ph",
    reportNumber = "IPPP-16-39, NSF-KITP-16-067, ZU-TH-18-16",
    doi = "10.1007/JHEP07(2016)133",
    journal = "JHEP",
    volume = "07",
    pages = "133",
    year = "2016"
}

@article{Boughezal:2016isb,
    author = "Boughezal, Radja and Liu, Xiaohui and Petriello, Frank",
    title = "{Phenomenology of the Z boson plus jet process at NNLO}",
    eprint = "1602.08140",
    archivePrefix = "arXiv",
    primaryClass = "hep-ph",
    doi = "10.1103/PhysRevD.94.074015",
    journal = "Phys. Rev. D",
    volume = "94",
    number = "7",
    pages = "074015",
    year = "2016"
}

@article{Boughezal:2016dtm,
    author = "Boughezal, Radja and Liu, Xiaohui and Petriello, Frank",
    title = "{W-boson plus jet differential distributions at NNLO in QCD}",
    eprint = "1602.06965",
    archivePrefix = "arXiv",
    primaryClass = "hep-ph",
    doi = "10.1103/PhysRevD.94.113009",
    journal = "Phys. Rev. D",
    volume = "94",
    number = "11",
    pages = "113009",
    year = "2016"
}

@article{Denner:2011vu,
    author = {Denner, Ansgar and Dittmaier, Stefan and Kasprzik, Tobias and M\"uck, Alexander},
    title = "{Electroweak corrections to dilepton + jet production at hadron colliders}",
    eprint = "1103.0914",
    archivePrefix = "arXiv",
    primaryClass = "hep-ph",
    doi = "10.1007/JHEP06(2011)069",
    journal = "JHEP",
    volume = "06",
    pages = "069",
    year = "2011"
}

@article{Denner:2012ts,
    author = {Denner, Ansgar and Dittmaier, Stefan and Kasprzik, Tobias and M\"uck, Alexander},
    title = "{Electroweak corrections to monojet production at the Tevatron and the LHC}",
    eprint = "1211.5078",
    archivePrefix = "arXiv",
    primaryClass = "hep-ph",
    doi = "10.1140/epjc/s10052-013-2297-x",
    journal = "Eur. Phys. J. C",
    volume = "73",
    number = "2",
    pages = "2297",
    year = "2013"
}

@article{Denner:2009gj,
    author = {Denner, Ansgar and Dittmaier, Stefan and Kasprzik, Tobias and M\"uck, Alexander},
    title = "{Electroweak corrections to W + jet hadroproduction including leptonic W-boson decays}",
    eprint = "0906.1656",
    archivePrefix = "arXiv",
    primaryClass = "hep-ph",
    reportNumber = "MPP-2009-72, PSI-PR-09-08",
    doi = "10.1088/1126-6708/2009/08/075",
    journal = "JHEP",
    volume = "08",
    pages = "075",
    year = "2009"
}

@article{Kuhn:2004em,
    author = {K\"uhn, Johann H. and Kulesza, A. and Pozzorini, S. and Schulze, M.},
    title = "{Logarithmic electroweak corrections to hadronic Z+1 jet production at large transverse momentum}",
    eprint = "hep-ph/0408308",
    archivePrefix = "arXiv",
    reportNumber = "TTP04-18, SFB-CPP-04-32",
    doi = "10.1016/j.physletb.2005.01.059",
    journal = "Phys. Lett. B",
    volume = "609",
    pages = "277--285",
    year = "2005"
}

@article{Kuhn:2007qc,
    author = {K\"uhn, Johann H. and Kulesza, A. and Pozzorini, S. and Schulze, M.},
    title = "{Electroweak corrections to large transverse momentum production of W bosons at the LHC}",
    eprint = "hep-ph/0703283",
    archivePrefix = "arXiv",
    reportNumber = "TTP07-07, SFB-CP-07-12, MPP-2007-35, DESY-07-041",
    doi = "10.1016/j.physletb.2007.06.028",
    journal = "Phys. Lett. B",
    volume = "651",
    pages = "160--165",
    year = "2007"
}

@article{Kuhn:2005az,
    author = {K\"uhn, Johann H. and Kulesza, A. and Pozzorini, S. and Schulze, M.},
    title = "{One-loop weak corrections to hadronic production of Z bosons at large transverse momenta}",
    eprint = "hep-ph/0507178",
    archivePrefix = "arXiv",
    reportNumber = "TTP05-10, SFB-CPP-05-30",
    doi = "10.1016/j.nuclphysb.2005.08.019",
    journal = "Nucl. Phys. B",
    volume = "727",
    pages = "368--394",
    year = "2005"
}

@article{Kuhn:2007cv,
    author = {K\"uhn, Johann H. and Kulesza, A. and Pozzorini, S. and Schulze, M.},
    title = "{Electroweak corrections to hadronic production of W bosons at large transverse momenta}",
    eprint = "0708.0476",
    archivePrefix = "arXiv",
    primaryClass = "hep-ph",
    reportNumber = "TTP07-19, SFB-CPP-07-43, MPP-2007-102, DESY-07-112",
    doi = "10.1016/j.nuclphysb.2007.12.029",
    journal = "Nucl. Phys. B",
    volume = "797",
    pages = "27--77",
    year = "2008"
}

@article{Lindert:2022ejn,
    author = {Lindert, Jonas M. and Pozzorini, Stefano and Sch\"onherr, Marek},
    title = "{Precise predictions for V + 2 jet backgrounds in searches for invisible Higgs decays}",
    eprint = "2204.07652",
    archivePrefix = "arXiv",
    primaryClass = "hep-ph",
    reportNumber = "IPPP/22/22, ZU-TH 15/22",
    doi = "10.1007/JHEP01(2023)070",
    journal = "JHEP",
    volume = "01",
    pages = "070",
    year = "2023",
    addendum = "{We are very grateful to Jonas~Lindert for insights and providing dedicated calculations for this analysis.}"
}

@article{Andersen:2009nu,
    author = "Andersen, Jeppe R. and Smillie, Jennifer M.",
    title = "{Constructing All-Order Corrections to Multi-Jet Rates}",
    eprint = "0908.2786",
    archivePrefix = "arXiv",
    primaryClass = "hep-ph",
    reportNumber = "CERN-PH-TH-2009-153, DCPT-09-122, IPPP-09-61",
    doi = "10.1007/JHEP01(2010)039",
    journal = "JHEP",
    volume = "01",
    pages = "039",
    year = "2010"
}

@article{Andersen:2011hs,
    author = "Andersen, Jeppe R. and Smillie, Jennifer M.",
    title = "{Multiple Jets at the LHC with High Energy Jets}",
    eprint = "1101.5394",
    archivePrefix = "arXiv",
    primaryClass = "hep-ph",
    reportNumber = "CP3-ORIGINS-2011-02, EDINBURGH-2011-03",
    doi = "10.1007/JHEP06(2011)010",
    journal = "JHEP",
    volume = "06",
    pages = "010",
    year = "2011"
}

@article{Andersen:2023kuj,
    author = "Andersen, Jeppe R. and Duclou\'e, Bertrand and Elrick, Conor and Hassan, Hitham and Maier, Andreas and Nail, Graeme and Paltrinieri, J\'er\'emy and Papaefstathiou, Andreas and Smillie, Jennifer M.",
    title = "{HEJ 2.2: W boson pairs and Higgs boson plus jet production at high energies}",
    eprint = "2303.15778",
    archivePrefix = "arXiv",
    primaryClass = "hep-ph",
    reportNumber = "IPPP/23/18, DCPT/23/36, DESY-23-038",
    doi = "10.21468/SciPostPhysCodeb.21",
    month = "3",
    year = "2023",
    addendum = "{We are very grateful to Jeppe~Andersen, Conor~Elrick, Andreas~Maier and Jennifer~Smillie for providing dedicated predictions for this analysis.}"
}

@article{Trimble:1987ee,
    author = "Trimble, Virginia",
    title = "{Existence and Nature of Dark Matter in the Universe}",
    doi = "10.1146/annurev.aa.25.090187.002233",
    journal = "Ann. Rev. Astron. Astrophys.",
    volume = "25",
    pages = "425--472",
    year = "1987"
}

@article{Feng:2010gw,
    author = "Feng, Jonathan L.",
    title = "{Dark Matter Candidates from Particle Physics and Methods of Detection}",
    eprint = "1003.0904",
    archivePrefix = "arXiv",
    primaryClass = "astro-ph.CO",
    reportNumber = "UCI-TR-2009-13",
    doi = "10.1146/annurev-astro-082708-101659",
    journal = "Ann. Rev. Astron. Astrophys.",
    volume = "48",
    pages = "495--545",
    year = "2010"
}

@article{Chen:2019zmr,
  author = {Chen, Xuan and Gehrmann, Thomas and Glover, Nigel and H\"ofer, Marius and Huss, Alexander},
  title = "{Isolated photon and photon+jet production at NNLO QCD accuracy}",
  eprint = "1904.01044",
  archivePrefix = "arXiv",
  primaryClass = "hep-ph",
  reportNumber = "IPPP-19-23, ZU-TH-13-19, IPPP/19/23, ZU-TH 13/19, CERN-TH-2019-034",
  doi = "10.1007/JHEP04(2020)166",
  journal = "JHEP",
  volume = "04",
  pages = "166",
  year = "2020"
}

@Booklet{ATL-SOFT-PUB-2023-001,
%    author         = "{ATLAS Collaboration}",
%    title          = "{Evaluating statistical uncertainties and correlations using the bootstrap method}",
%    howpublished   = "{ATL-PHYS-PUB-2021-011}",
%    url            = "https://cds.cern.ch/record/2759945",
%    year           = "2021",
%}

@article{ATLAS:2023ynf,
  author = "{ATLAS Collaboration}",
  collaboration = "ATLAS",
  title = "{Measurement of the Z boson invisible width at $\sqrt{s}=13$ TeV with the ATLAS detector}",
  eprint = "2312.02789",
  archivePrefix = "arXiv",
  primaryClass = "hep-ex",
  reportNumber = "CERN-EP-2023-232",
  month = "12",
  doi = "10.48550/arXiv.2312.02789",
  year = "2023"
}

@Article{Cacciari:2008gp,
     author    = "Cacciari, Matteo and Salam, Gavin P. and Soyez, Gregory",
     title     = "{The anti-\(k_{t}\) jet clustering algorithm}",
     journal   = "JHEP",
     volume    = "04",
     year      = "2008",
     pages     = "063",
     eprint    = "0802.1189",
     archivePrefix = "arXiv",
     primaryClass  =  "hep-ph",
     doi       = "10.1088/1126-6708/2008/04/063",
     SLACcitation  = "%%CITATION = 0802.1189;%%"
}

@Article{Butterworth:2015oua,
      author         = "Butterworth, Jon and others",
      title          = "{PDF4LHC recommendations for LHC Run II}",
      journal        = "J. Phys. G",
      volume         = "43",
      year           = "2016",
      pages          = "023001",
      doi            = "10.1088/0954-3899/43/2/023001",
      eprint         = "1510.03865",
      archivePrefix  = "arXiv",
      primaryClass   = "hep-ph",
      reportNumber   = "OUTP-15-17P, SMU-HEP-15-12, TIF-UNIMI-2015-14,
                        LCTS-2015-27, CERN-PH-TH-2015-249",
      SLACcitation   = "%%CITATION = ARXIV:1510.03865;%%"
}

@article{Ball:2012cx,
      author         = "{NNPDF Collaboration} and Ball, Richard D. and others",
      title          = "{Parton distributions with LHC data}",
      journal        = "Nucl. Phys. B",
      volume         = "867",
      year           = "2013",
      pages          = "244",
      doi            = "10.1016/j.nuclphysb.2012.10.003",
      eprint         = "1207.1303",
      archivePrefix  = "arXiv",
      primaryClass   = "hep-ph",
      reportNumber   = "EDINBURGH-2012-08, IFUM-FT-997, FR-PHENO-2012-014,
                        RWTH-TTK-12-25, CERN-PH-TH-2012-037, SFB-CPP-12-47\,
                        --CERN-PH-TH-2012-037",
      SLACcitation   = "%%CITATION = ARXIV:1207.1303;%%"
}

@article{Ball:2014uwa,
      author         = "{NNPDF Collaboration} and Ball, Richard D. and others",
      title          = "{Parton distributions for the LHC run II}",
      journal        = "JHEP",
      volume         = "04",
      year           = "2015",
      pages          = "040",
      doi            = "10.1007/JHEP04(2015)040",
      eprint         = "1410.8849",
      archivePrefix  = "arXiv",
      primaryClass   = "hep-ph",
      reportNumber   = "EDINBURGH-2014-15, IFUM-1034-FT, CERN-PH-TH-2013-253,
                        OUTP-14-11P, CAVENDISH-HEP-14-11",
      SLACcitation   = "%%CITATION = ARXIV:1410.8849;%%"
}

@article{Czakon:2012pz,
      author         = "Czakon, Michal and Mitov, Alexander",
      title          = "{NNLO corrections to top pair production at hadron
                        colliders: the quark-gluon reaction}",
      journal        = "JHEP",
      volume         = "01",
      pages          = "080",
      doi            = "10.1007/JHEP01(2013)080",
      year           = "2013",
      eprint         = "1210.6832",
      archivePrefix  = "arXiv",
      primaryClass   = "hep-ph",
      SLACcitation   = "%%CITATION = ARXIV:1210.6832;%%",
}

@Article{Sjostrand:2007gs,
	author    = "Sj{\"o}strand, T. and Mrenna, S. and Skands, P.",
	title     = "{A brief introduction to PYTHIA 8.1}",
	journal   = "Comput. Phys. Commun.",
	volume    = "178",
	year      = "2008",
	pages     = "852-867",
	eprint    = "0710.3820",
	archivePrefix = "arXiv",
	primaryClass  =  "hep-ph",
	doi       = "10.1016/j.cpc.2008.01.036",
	SLACcitation  = "%%CITATION = 0710.3820;%%"
}

@article{Sjostrand:2014zea,
      author         = "Sj{\"o}strand, Torbj{\"o}rn and Ask, Stefan and Christiansen,
                        Jesper R. and Corke, Richard and Desai, Nishita and Ilten,
                        Philip and Mrenna, Stephen and Prestel, Stefan and
                        Rasmussen, Christine O. and Skands, Peter Z.",
      title          = "{An introduction to PYTHIA 8.2}",
      journal        = "Comput. Phys. Commun.",
      volume         = "191",
      year           = "2015",
      pages          = "159",
      doi            = "10.1016/j.cpc.2015.01.024",
      eprint         = "1410.3012",
      archivePrefix  = "arXiv",
      primaryClass   = "hep-ph",
      reportNumber   = "LU-TP-14-36, MCNET-14-22, CERN-PH-TH-2014-190,
                        FERMILAB-PUB-14-316-CD, DESY-14-178, SLAC-PUB-16122,
                        --FERMILAB-PUB-14-316-CD",
      SLACcitation   = "%%CITATION = ARXIV:1410.3012;%%"
}

@article{Sjostrand:1984ic,
      author         = "Sj{\"o}strand, Torbjorn",
      title          = "{Jet fragmentation of multiparton
                  configurations in a string framework}",
      journal        = "Nucl. Phys. B",
      volume         = "248",
      year           = "1984",
      pages          = "469-502",
      doi            = "10.1016/0550-3213(84)90607-2",
      reportNumber   = "DESY-T-84-01",
      SLACcitation   = "%%CITATION = NUPHA,B248,469;%%"                                                                                                       
}

@article{Andersson:1983ia,
      author         = "Andersson, Bo and Gustafson, G. and Ingelman, G. and
                        Sj{\"o}strand, T.",
      title          = "{Parton fragmentation and string dynamics}",
      journal        = "Phys. Rept.",
      volume         = "97",
      year           = "1983",
      pages          = "31-145",
      doi            = "10.1016/0370-1573(83)90080-7",
      reportNumber   = "LU-TP-83-10",
      SLACcitation   = "%%CITATION = PRPLC,97,31;%%"
}

@Article{Frixione:2008yi,
     author    = "Frixione, Stefano and Laenen, Eric and Motylinski, Patrick and White, Chris and Webber, Bryan R.",
     title     = "{Single-top hadroproduction in association with a \(W\) boson}",
     journal   = "JHEP",
     volume    = "07",
     year      = "2008",
     pages     = "029",
     eprint    = "0805.3067",
     archivePrefix = "arXiv",
     primaryClass  =  "hep-ph",
     doi       = "10.1088/1126-6708/2008/07/029"
}

@article{Frixione:1998jh,
      author         = "Frixione, Stefano",
      title          = "{Isolated photons in perturbative QCD}",
      journal        = "Phys. Lett. B",
      volume         = "429",
      year           = "1998",
      pages          = "369-374",
      doi            = "10.1016/S0370-2693(98)00454-7",
      eprint         = "hep-ph/9801442",
      archivePrefix  = "arXiv",
      XprimaryClass   = "hep-ph",
      reportNumber   = "ETH-TH-97-40",
      SLACcitation   = "%%CITATION = HEP-PH/9801442;%%"
}

@Article{Nason:2004rx,
      author         = "Nason, Paolo",
      title          = "{A new method for combining NLO QCD with shower Monte Carlo algorithms}",
      journal        = "JHEP",
      volume         = "11",
      pages          = "040",
      doi            = "10.1088/1126-6708/2004/11/040",
      year           = "2004",
      eprint         = "hep-ph/0409146",
      archivePrefix  = "arXiv",
}

@Article{Frixione:2007vw,
      author         = "Frixione, Stefano and Nason, Paolo and Oleari, Carlo",
      title          = "{Matching NLO QCD computations with parton shower
                        simulations: the POWHEG method}",
      journal        = "JHEP",
      volume         = "11",
      pages          = "070",
      doi            = "10.1088/1126-6708/2007/11/070",
      year           = "2007",
      eprint         = "0709.2092",
      archivePrefix  = "arXiv",
      primaryClass   = "hep-ph",
}

@Article{Alioli:2010xd,
      author         = "Alioli, Simone and Nason, Paolo and Oleari, Carlo and Re,
                        Emanuele",
      title          = "{A general framework for implementing NLO calculations in
                        shower Monte Carlo programs: the POWHEG BOX}",
      journal        = "JHEP",
      volume         = "06",
      pages          = "043",
      doi            = "10.1007/JHEP06(2010)043",
      year           = "2010",
      eprint         = "1002.2581",
      archivePrefix  = "arXiv",
      primaryClass   = "hep-ph",
}

@article{Alioli:2009je,
      author         = "Alioli, Simone and Nason, Paolo and Oleari, Carlo and Re,
                        Emanuele",
      title          = "{NLO single-top production matched with shower in POWHEG:
                        \(s\)- and \(t\)-channel contributions}",
      journal        = "JHEP",
      volume         = "09",
      year           = "2009",
      pages          = "111",
      doi            = "10.1088/1126-6708/2009/09/111",
      eprint         = "0907.4076",
      archivePrefix  = "arXiv",
      primaryClass   = "hep-ph",
      SLACcitation   = "%%CITATION = ARXIV:0907.4076;%%",
      related        = "Alioli:2009je-err",
      relatedstring  = "Erratum:",
}

@article{Frederix:2012dh,
      author         = "Frederix, Rikkert and Re, Emanuele and Torrielli, Paolo",
      title          = "{Single-top \(t\)-channel hadroproduction in the four-flavour
                        scheme with POWHEG and aMC@NLO}",
      journal        = "JHEP",
      volume         = "09",
      year           = "2012",
      pages          = "130",
      doi            = "10.1007/JHEP09(2012)130",
      eprint         = "1207.5391",
      archivePrefix  = "arXiv",
      primaryClass   = "hep-ph",
      reportNumber   = "ZU-TH-14-12, IPPP-12-54, DCPT-12-108,
                        CERN-PH-TH-2012-206, MCNET-12-10, LPN12-084",
      SLACcitation   = "%%CITATION = ARXIV:1207.5391;%%"
}

@article{Aliev:2010zk,
      author         = "Aliev, M. and Lacker, H. and Langenfeld, U. and Moch, S.
                        and Uwer, P. and Wiedermann, M.",
      title          = "{HATHOR -- HAdronic Top and Heavy quarks crOss section
                        calculatoR}",
      journal        = "Comput. Phys. Commun.",
      volume         = "182",
      year           = "2011",
      pages          = "1034-1046",
      doi            = "10.1016/j.cpc.2010.12.040",
      eprint         = "1007.1327",
      archivePrefix  = "arXiv",
      primaryClass   = "hep-ph",
      reportNumber   = "DESY-10-091, HU-EP-10-33, SFB-CPP-10-60",
      SLACcitation   = "%%CITATION = ARXIV:1007.1327;%%"
}

@article{Re:2010bp,
      author         = "Re, Emanuele",
      title          = "{Single-top \(Wt\)-channel production matched with parton
                        showers using the POWHEG method}",
      journal        = "Eur. Phys. J. C",
      volume         = "71",
      year           = "2011",
      pages          = "1547",
      doi            = "10.1140/epjc/s10052-011-1547-z",
      eprint         = "1009.2450",
      archivePrefix  = "arXiv",
      primaryClass   = "hep-ph",
      reportNumber   = "IPPP-10-74, DCPT-10-148",
      SLACcitation   = "%%CITATION = ARXIV:1009.2450;%%"
}

@article{Kant:2014oha,
      author         = "Kant, P. and Kind, O. M. and Kintscher, T. and Lohse, T.
                        and Martini, T. and Mölbitz, S. and Rieck, P. and Uwer, P.",
      title          = "{HatHor for single top-quark production: Updated
                        predictions and uncertainty estimates for single top-quark
                        production in hadronic collisions}",
      journal        = "Comput. Phys. Commun.",
      volume         = "191",
      year           = "2015",
      pages          = "74-89",
      doi            = "10.1016/j.cpc.2015.02.001",
      eprint         = "1406.4403",
      archivePrefix  = "arXiv",
      primaryClass   = "hep-ph",
      reportNumber   = "HU-EP-14-22",
      SLACcitation   = "%%CITATION = ARXIV:1406.4403;%%"
}

@article{Beneke:2011mq,
      author         = "Beneke, M. and Falgari, P. and Klein, S. and Schwinn, C.",
      title          = "{Hadronic top-quark pair production with NNLL threshold
                        resummation}",
      journal        = "Nucl. Phys. B",
      volume         = "855",
      year           = "2012",
      pages          = "695-741",
      doi            = "10.1016/j.nuclphysb.2011.10.021",
      eprint         = "1109.1536",
      archivePrefix  = "arXiv",
      primaryClass   = "hep-ph",
      reportNumber   = "TTK-11-38, ITP-UU-11-26, SPIN-11-19, FR-PHENO-2011-015,
                        SFB-CPP-11-49",
      SLACcitation   = "%%CITATION = ARXIV:1109.1536;%%"
}

@article{Cacciari:2011hy,
      author         = "Cacciari, Matteo and Czakon, Michal and Mangano,
                        Michelangelo and Mitov, Alexander and Nason, Paolo",
      title          = "{Top-pair production at hadron colliders with
                        next-to-next-to-leading logarithmic soft-gluon
                        resummation}",
      journal        = "Phys. Lett. B",
      volume         = "710",
      pages          = "612-622",
      doi            = "10.1016/j.physletb.2012.03.013",
      year           = "2012",
      eprint         = "1111.5869",
      primaryClass   = "hep-ph",
      archivePrefix  = "arXiv",
      reportNumber   = "CERN-PH-TH-2011-277, TTK-11-54",
      SLACcitation   = "%%CITATION = ARXIV:1111.5869;%%",
}

@article{Czakon:2012zr,
      author         = "Czakon, Michal and Mitov, Alexander",
      title          = "{NNLO corrections to top-pair production at hadron
                        colliders: the all-fermionic scattering channels}",
      journal        = "JHEP",
      volume         = "12",
      pages          = "054",
      doi            = "10.1007/JHEP12(2012)054",
      year           = "2012",
      eprint         = "1207.0236",
      archivePrefix  = "arXiv",
      primaryClass   = "hep-ph",
      SLACcitation   = "%%CITATION = ARXIV:1207.0236;%%",
}

@article{Czakon:2013goa,
      author         = "Czakon, Michal and Fiedler, Paul and Mitov, Alexander",
      title          = "{Total Top-Quark Pair-Production Cross Section at
                        Hadron Colliders Through \(O(\alpha_S^4)\)}",
      journal        = "Phys. Rev. Lett.",
      volume         = "110",
      pages          = "252004",
      doi            = "10.1103/PhysRevLett.110.252004",
      year           = "2013",
      eprint         = "1303.6254",
      archivePrefix  = "arXiv",
      primaryClass   = "hep-ph",
      reportNumber   = "CERN-PH-TH-2013-056, TTK-13-08",
      SLACcitation   = "%%CITATION = ARXIV:1303.6254;%%",
}

@article{Baernreuther:2012ws,
      author         = "B{\"a}rnreuther, Peter and Czakon, Michal and Mitov,
                        Alexander",
      title          = "{Percent-Level-Precision Physics at the
                  Tevatron: Next-to-Next-to-Leading Order QCD
                  Corrections to \(q \bar{q} \to t \bar{t} + X\)}",
      journal        = "Phys. Rev. Lett.",
      volume         = "109",
      pages          = "132001",
      doi            = "10.1103/PhysRevLett.109.132001",
      year           = "2012",
      eprint         = "1204.5201",
      archivePrefix  = "arXiv",
      primaryClass   = "hep-ph",
      SLACcitation   = "%%CITATION = ARXIV:1204.5201;%%",
}

@article{Frixione:2007nw,
      author         = "Frixione, Stefano and Ridolfi, Giovanni and Nason, Paolo",
      title          = "{A positive-weight next-to-leading-order Monte Carlo for
                        heavy flavour hadroproduction}",
      journal        = "JHEP",
      volume         = "09",
      pages          = "126",
      doi            = "10.1088/1126-6708/2007/09/126",
      year           = "2007",
      eprint         = "0707.3088",
      archivePrefix  = "arXiv",
      primaryClass   = "hep-ph",
      SLACcitation   = "%%CITATION = ARXIV:0707.3088;%%",
}

@Article{Gleisberg:2008ta,
      author         = "Gleisberg, T. and H{\"o}che, S. and Krauss, F. and Sch{\"o}nherr, M. and Schumann, S. and Siegert, F. and Winter J.",
      title          = "{Event generation with SHERPA 1.1}",
      journal        = "JHEP",
      volume         = "02",
      pages          = "007",
      doi            = "10.1088/1126-6708/2009/02/007",
      year           = "2009",
      eprint         = "0811.4622",
      archivePrefix  = "arXiv",
      primaryClass   = "hep-ph",
}

@Article{Hoeche:2009rj,
      author         = "H{\"o}che, Stefan and Krauss, Frank and Schumann, Steffen and Siegert, Frank",
      title          = "{QCD matrix elements and truncated showers}",
      journal        = "JHEP",
      volume         = "05",
      pages          = "053",
      doi            = "10.1088/1126-6708/2009/05/053",
      year           = "2009",
      eprint         = "0903.1219",
      archivePrefix  = "arXiv",
      primaryClass   = "hep-ph",
}

@Article{Gleisberg:2008fv,
      author         = "Gleisberg, Tanju and H{\"o}che, Stefan",
      title          = "{Comix, a new matrix element generator}",
      journal        = "JHEP",
      volume         = "12",
      pages          = "039",
      doi            = "10.1088/1126-6708/2008/12/039",
      year           = "2008",
      eprint         = "0808.3674",
      archivePrefix  = "arXiv",
      primaryClass   = "hep-ph",
}

@Article{Schumann:2007mg,
      author         = "Schumann, Steffen and Krauss, Frank",
      title          = "{A parton shower algorithm based on Catani--Seymour dipole factorisation}",
      journal        = "JHEP",
      volume         = "03",
      pages          = "038",
      doi            = "10.1088/1126-6708/2008/03/038",
      year           = "2008",
      eprint         = "0709.1027",
      archivePrefix  = "arXiv",
      primaryClass   = "hep-ph",
}

@article{Hoeche:2012yf,
      author         = "H{\"o}che, Stefan and Krauss, Frank and Sch{\"o}nherr, Marek and
                        Siegert, Frank",
      title          = "{QCD matrix elements + parton showers. The NLO case}",
      journal        = "JHEP",
      volume         = "04",
      year           = "2013",
      pages          = "027",
      doi            = "10.1007/JHEP04(2013)027",
      eprint         = "1207.5030",
      archivePrefix  = "arXiv",
      primaryClass   = "hep-ph",
      reportNumber   = "SLAC-PUB-15191, IPPP-12-52, DCPT-12-104, LPN12-081,
                        FR-PHENO-2012-017, MCNET-12-09, --FR-PHENO-2012-017",
      SLACcitation   = "%%CITATION = ARXIV:1207.5030;%%"
}

@article{Bothmann:2019yzt,
      author = "Bothmann, Enrico and others",
      title = "{Event generation with Sherpa 2.2}",
      journal = "SciPost Phys.",
      volume = "7",
      year = "2019",
      number = "3",
      pages = "034",
      doi = "10.21468/SciPostPhys.7.3.034",
      reportNumber = "FERMILAB-PUB-19-218-T, SLAC-PUB-17433, IPPP/19/42, MCNET-19-11",
      eprint = "1905.09127",
      archivePrefix = "arXiv",
      primaryClass = "hep-ph",
}

@article{Winter:2003tt,
      author         = "Winter, Jan-Christopher and Krauss, Frank and Soff,
                        Gerhard",
      title          = "{A modified cluster-hadronisation model}",
      journal        = "Eur. Phys. J. C",
      volume         = "36",
      year           = "2004",
      pages          = "381-395",
      doi            = "10.1140/epjc/s2004-01960-8",
      eprint         = "hep-ph/0311085",
      archivePrefix  = "arXiv",
      XprimaryClass   = "hep-ph",
      reportNumber   = "CERN-TH-2003-272",
      SLACcitation   = "%%CITATION = HEP-PH/0311085;%%"
}

@article{Hoeche:2011fd,
      author         = "H{\"o}che, Stefan and Krauss, Frank and Sch{\"o}nherr, Marek and
                        Siegert, Frank",
      title          = "{A critical appraisal of NLO+PS matching methods}",
      journal        = "JHEP",
      volume         = "09",
      year           = "2012",
      pages          = "049",
      doi            = "10.1007/JHEP09(2012)049",
      eprint         = "1111.1220",
      archivePrefix  = "arXiv",
      primaryClass   = "hep-ph",
      reportNumber   = "SLAC-PUB-14661, IPPP-11-67, DCPT-11-134, LPN11-58,
                        FR-PHENO-2011-019, MCNET-11-24",
      SLACcitation   = "%%CITATION = ARXIV:1111.1220;%%"
}

@article{Catani:2001cc,
      author         = "Catani, S. and Krauss, F. and Webber, B. R. and Kuhn, R.",
      title          = "{QCD Matrix Elements + Parton Showers}",
      journal        = "JHEP",
      volume         = "11",
      year           = "2001",
      pages          = "063",
      doi            = "10.1088/1126-6708/2001/11/063",
      eprint         = "hep-ph/0109231",
      archivePrefix  = "arXiv",
      reportNumber   = "CERN-TH-2000-367, CAVENDISH-HEP-00-03",
      SLACcitation   = "%%CITATION = HEP-PH/0109231;%%"
}

@article{Siegert:2016bre,
      author         = "Siegert, Frank",
      title          = "{A practical guide to event generation for prompt photon
                        production with Sherpa}",
      journal        = "J. Phys. G",
      volume         = "44",
      year           = "2017",
      number         = "4",
      pages          = "044007",
      doi            = "10.1088/1361-6471/aa5f29",
      eprint         = "1611.07226",
      archivePrefix  = "arXiv",
      primaryClass   = "hep-ph",
      SLACcitation   = "%%CITATION = ARXIV:1611.07226;%%"
}

@article{Cascioli:2011va,
      author         = "Cascioli, Fabio and Maierh{\"o}fer, Philipp and Pozzorini, Stefano",
      title          = "{Scattering Amplitudes with Open Loops}",
      journal        = "Phys. Rev. Lett.",
      volume         = "108",
      year           = "2012",
      pages          = "111601",
      doi            = "10.1103/PhysRevLett.108.111601",
      eprint         = "1111.5206",
      archivePrefix  = "arXiv",
      primaryClass   = "hep-ph",
      reportNumber   = "ZU-TH-23-11, LPN11-66",
      SLACcitation   = "%%CITATION = ARXIV:1111.5206;%%"
}

@article{Buccioni:2017yxi,
    author = "Buccioni, Federico and Pozzorini, Stefano and Zoller, Max",
    title = "{On-the-fly reduction of open loops}",
    eprint = "1710.11452",
    archivePrefix = "arXiv",
    primaryClass = "hep-ph",
    reportNumber = "ZU-TH-29-17",
    doi = "10.1140/epjc/s10052-018-5562-1",
    journal = "Eur. Phys. J. C",
    volume = "78",
    number = "1",
    pages = "70",
    year = "2018"
}

@article{Buccioni:2019sur,
    author = "Buccioni, Federico and Lang, Jean-Nicolas and Lindert, Jonas M. and Maierh{\"o}fer, Philipp and Pozzorini, Stefano and Zhang, Hantian and Zoller, Max F.",
    title = "{OpenLoops 2}",
    eprint = "1907.13071",
    archivePrefix = "arXiv",
    primaryClass = "hep-ph",
    reportNumber = "IPPP/19/62, FR-PHENO-2019-12, PSI-PR-19-15, ZU-TH 37/19",
    doi = "10.1140/epjc/s10052-019-7306-2",
    journal = "Eur. Phys. J. C",
    volume = "79",
    number = "10",
    pages = "866",
    year = "2019"
}

@article{Denner:2016kdg,
      author         = "Denner, Ansgar and Dittmaier, Stefan and Hofer, Lars",
      title          = "{\textsc{Collier}: A fortran-based complex one-loop library in
                        extended regularizations}",
      journal        = "Comput. Phys. Commun.",
      volume         = "212",
      year           = "2017",
      pages          = "220-238",
      doi            = "10.1016/j.cpc.2016.10.013",
      eprint         = "1604.06792",
      archivePrefix  = "arXiv",
      primaryClass   = "hep-ph",
      reportNumber   = "FR-PHENO-2016-003, ICCUB-16-016",
      SLACcitation   = "%%CITATION = ARXIV:1604.06792;%%"
}

@article{Kallweit:2015dum,
    author = "Kallweit, Stefan and Lindert, Jonas M. and Maierh{\"o}fer, Philipp and Pozzorini, Stefano and Sch{\"o}nherr, Marek",
    title = "{NLO QCD+EW predictions for V + jets including off-shell vector-boson decays and multijet merging}",
    eprint = "1511.08692",
    archivePrefix = "arXiv",
    primaryClass = "hep-ph",
    reportNumber = "DCPT-15-140, FR-PHENO-2015-014, IPPP-15-70, MCNET-15-23, ZU-TH-41-15, MITP-15-108",
    doi = "10.1007/JHEP04(2016)021",
    journal = "JHEP",
    volume = "04",
    pages = "021",
    year = "2016"
}

@article{Czakon:2011xx,
  author         = "Czakon, Michal and Mitov, Alexander",
  title          = "{Top++: A program for the calculation of the top-pair cross-section at hadron colliders}",
  journal        = "Comput. Phys. Commun.",
  volume         = "185",
  year           = "2014",
  pages          = "2930",
  doi            = "10.1016/j.cpc.2014.06.021",
  eprint         = "1112.5675",
  archivePrefix  = "arXiv",
  primaryClass   = "hep-ph",
  reportNumber   = "CERN-PH-TH-2011-303, TTK-11-58",
}

@article{Agostinelli:2002hh,
      author         = "Agostinelli, S. and others",
      title          = "{\textsc{Geant4} -- a simulation toolkit}",
      journal        = "Nucl. Instrum. Meth. A",
      volume         = "506",
      year           = "2003",
      pages          = "250",
      doi            = "10.1016/S0168-9002(03)01368-8",
      reportNumber   = "SLAC-PUB-9350, FERMILAB-PUB-03-339",
      SLACcitation   = "%%CITATION = NUIMA,A506,250;%%"
}

@Article{PERF-2007-01,
    author         = "{ATLAS Collaboration}",
    title          = "{The ATLAS Experiment at the CERN Large Hadron Collider}",
    journal        = "JINST",
    volume         = "3",
    year           = "2008",
    pages          = "S08003",
    doi            = "10.1088/1748-0221/3/08/S08003",
    primaryClass   = "hep-ex",
}

@Article{SOFT-2010-01,
    author         = "{ATLAS Collaboration}",
    title          = "{The ATLAS Simulation Infrastructure}",
    journal        = "Eur. Phys. J. C",
    volume         = "70",
    year           = "2010",
    pages          = "823",
    doi            = "10.1140/epjc/s10052-010-1429-9",
    eprint         = "1005.4568",
    archivePrefix  = "arXiv",
    primaryClass   = "physics.ins-det",
}

@Article{SUSY-2011-20,
    author         = "{ATLAS Collaboration}",
    title          = "{Search for squarks and gluinos with the ATLAS detector in final states with jets and missing transverse momentum using \(4.7\,\text{fb}^{-1}\) of \(\sqrt{s} = 7\,\text{TeV}\) proton--proton collision data}",
    journal        = "Phys. Rev. D",
    volume         = "87",
    year           = "2013",
    pages          = "012008",
    doi            = "10.1103/PhysRevD.87.012008",
    reportNumber   = "CERN-PH-EP-2012-195",
    eprint         = "1208.0949",
    archivePrefix  = "arXiv",
    primaryClass   = "hep-ex",
}

@Article{PERF-2014-03,
    author         = "{ATLAS Collaboration}",
    title          = "{Performance of pile-up mitigation techniques for jets in \(pp\) collisions at \(\sqrt{s} = 8\,\text{TeV}\) using the ATLAS detector}",
    journal        = "Eur. Phys. J. C",
    volume         = "76",
    year           = "2016",
    pages          = "581",
    doi            = "10.1140/epjc/s10052-016-4395-z",
    reportNumber   = "CERN-PH-EP-2015-206",
    eprint         = "1510.03823",
    archivePrefix  = "arXiv",
    primaryClass   = "hep-ex",
}

@Article{PERF-2014-04,
    author         = "{ATLAS Collaboration}",
    title          = "{Performance of algorithms that reconstruct missing transverse momentum in \(\sqrt{s} = 8\,\text{TeV}\) proton--proton collisions in the ATLAS detector}",
    journal        = "Eur. Phys. J. C",
    volume         = "77",
    year           = "2017",
    pages          = "241",
    doi            = "10.1140/epjc/s10052-017-4780-2",
    reportNumber   = "CERN-EP-2016-134",
    eprint         = "1609.09324",
    archivePrefix  = "arXiv",
    primaryClass   = "hep-ex",
}

@Article{PERF-2014-07,
    author         = "{ATLAS Collaboration}",
    title          = "{Topological cell clustering in the ATLAS calorimeters and its performance in LHC Run~1}",
    journal        = "Eur. Phys. J. C",
    volume         = "77",
    year           = "2017",
    pages          = "490",
    doi            = "10.1140/epjc/s10052-017-5004-5",
    reportNumber   = "CERN-PH-EP-2015-304",
    eprint         = "1603.02934",
    archivePrefix  = "arXiv",
    primaryClass   = "hep-ex",
}

@Article{STDM-2014-09,
    author         = "{ATLAS Collaboration}",
    title          = "{Measurement of the inclusive isolated prompt photon cross section in \(pp\) collisions at \(\sqrt{s} = 8\,\text{TeV}\) with the ATLAS detector}",
    journal        = "JHEP",
    volume         = "08",
    year           = "2016",
    pages          = "005",
    doi            = "10.1007/JHEP08(2016)005",
    reportNumber   = "CERN-EP-2016-035",
    eprint         = "1605.03495",
    archivePrefix  = "arXiv",
    primaryClass   = "hep-ex",
}

@Article{PERF-2015-09,
    author         = "{ATLAS Collaboration}",
    title          = "{Jet reconstruction and performance using particle flow with the ATLAS Detector}",
    journal        = "Eur. Phys. J. C",
    volume         = "77",
    year           = "2017",
    pages          = "466",
    doi            = "10.1140/epjc/s10052-017-5031-2",
    reportNumber   = "CERN-EP-2017-024",
    eprint         = "1703.10485",
    archivePrefix  = "arXiv",
    primaryClass   = "hep-ex",
}

@Article{PERF-2015-10,
    author         = "{ATLAS Collaboration}",
    title          = "{Muon reconstruction performance of the ATLAS detector in proton--proton collision data at \(\sqrt{s} = 13\,\text{TeV}\)}",
    journal        = "Eur. Phys. J. C",
    volume         = "76",
    year           = "2016",
    pages          = "292",
    doi            = "10.1140/epjc/s10052-016-4120-y",
    reportNumber   = "CERN-EP-2016-033",
    eprint         = "1603.05598",
    archivePrefix  = "arXiv",
    primaryClass   = "hep-ex",
}

@Article{EXOT-2016-03,
    author         = "{ATLAS Collaboration}",
    title          = "{Measurement of detector-corrected observables sensitive to the anomalous production of events with jets and large missing transverse momentum in \(pp\) collisions at \(\sqrt{s} = 13\,\text{TeV}\) using the ATLAS detector}",
    journal        = "Eur. Phys. J. C",
    volume         = "77",
    year           = "2017",
    pages          = "765",
    doi            = "10.1140/epjc/s10052-017-5315-6",
    reportNumber   = "CERN-EP-2017-116",
    eprint         = "1707.03263",
    archivePrefix  = "arXiv",
    primaryClass   = "hep-ex",
}

@Article{PERF-2016-07,
    author         = "{ATLAS Collaboration}",
    title          = "{Performance of missing transverse momentum reconstruction with the ATLAS detector using proton--proton collisions at \(\sqrt{s} = 13\,\text{TeV}\)}",
    journal        = "Eur. Phys. J. C",
    volume         = "78",
    year           = "2018",
    pages          = "903",
    doi            = "10.1140/epjc/s10052-018-6288-9",
    reportNumber   = "CERN-EP-2017-274",
    eprint         = "1802.08168",
    archivePrefix  = "arXiv",
    primaryClass   = "hep-ex",
}

@Article{STDM-2016-01,
    author         = "{ATLAS Collaboration}",
    title          = "{Measurements of the production cross section of a \(Z\) boson in association with jets in \(pp\) collisions at \(\sqrt{s} = 13\,\text{TeV}\) with the ATLAS detector}",
    journal        = "Eur. Phys. J. C",
    volume         = "77",
    year           = "2017",
    pages          = "361",
    doi            = "10.1140/epjc/s10052-017-4900-z",
    reportNumber   = "CERN-EP-2016-297",
    eprint         = "1702.05725",
    archivePrefix  = "arXiv",
    primaryClass   = "hep-ex",
}

@Article{TRIG-2016-01,
    author         = "{ATLAS Collaboration}",
    title          = "{Performance of the ATLAS trigger system in 2015}",
    journal        = "Eur. Phys. J. C",
    volume         = "77",
    year           = "2017",
    pages          = "317",
    doi            = "10.1140/epjc/s10052-017-4852-3",
    reportNumber   = "CERN-EP-2016-241",
    eprint         = "1611.09661",
    archivePrefix  = "arXiv",
    primaryClass   = "hep-ex",
}

@Article{PERF-2017-03,
    author         = "{ATLAS Collaboration}",
    title          = "{Electron and photon energy calibration with the ATLAS detector using 2015--2016 LHC proton--proton collision data}",
    journal        = "JINST",
    volume         = "14",
    year           = "2019",
    pages          = "P03017",
    doi            = "10.1088/1748-0221/14/03/P03017",
    reportNumber   = "CERN-EP-2018-296",
    eprint         = "1812.03848",
    archivePrefix  = "arXiv",
    primaryClass   = "hep-ex",
}

@Article{STDM-2017-01,
    author         = "{ATLAS Collaboration}",
    title          = "{Measurement of the cross section for isolated-photon plus jet production in \(pp\) collisions at \(\sqrt{s}=13\,\text{TeV}\) using the ATLAS detector}",
    journal        = "Phys. Lett. B",
    volume         = "780",
    year           = "2018",
    pages          = "578",
    doi            = "10.1016/j.physletb.2018.03.035",
    reportNumber   = "CERN-EP-2017-265",
    eprint         = "1801.00112",
    archivePrefix  = "arXiv",
    primaryClass   = "hep-ex",
}

@Article{EGAM-2018-01,
    author         = "{ATLAS Collaboration}",
    title          = "{Electron and photon performance measurements with the ATLAS detector using the 2015--2017 LHC proton--proton collision data}",
    journal        = "JINST",
    volume         = "14",
    year           = "2019",
    pages          = "P12006",
    doi            = "10.1088/1748-0221/14/12/P12006",
    reportNumber   = "CERN-EP-2019-145",
    eprint         = "1908.00005",
    archivePrefix  = "arXiv",
    primaryClass   = "hep-ex",
}

@Article{JETM-2018-05,
    author         = "{ATLAS Collaboration}",
    title          = "{Jet energy scale and resolution measured in proton--proton collisions at \(\sqrt{s} = 13\,\text{TeV}\) with the ATLAS detector}",
    journal        = "Eur. Phys. J. C",
    volume         = "81",
    year           = "2021",
    pages          = "689",
    doi            = "10.1140/epjc/s10052-021-09402-3",
    reportNumber   = "CERN-EP-2020-083",
    eprint         = "2007.02645",
    archivePrefix  = "arXiv",
    primaryClass   = "hep-ex",
}

@Article{MUON-2018-03,
    author         = "{ATLAS Collaboration}",
    title          = "{Muon reconstruction and identification efficiency in ATLAS using the full Run~2 \(pp\) collision data set at \(\sqrt{s} = 13\,\text{TeV}\)}",
    journal        = "Eur. Phys. J. C",
    volume         = "81",
    year           = "2021",
    pages          = "578",
    doi            = "10.1140/epjc/s10052-021-09233-2",
    reportNumber   = "CERN-EP-2020-199",
    eprint         = "2012.00578",
    archivePrefix  = "arXiv",
    primaryClass   = "hep-ex",
}

@Article{STDM-2018-55,
    author         = "{ATLAS Collaboration}",
    title          = "{Differential cross-sections for events with missing transverse momentum and jets measured with the ATLAS detector in \(13\,\text{TeV}\) proton--proton collisions}",
    year           = "2024",
    reportNumber   = "CERN-EP-2024-034",
    eprint         = "2403.02793",
    archivePrefix  = "arXiv",
    primaryClass   = "hep-ex",
}

@Article{TRIG-2018-05,
    author         = "{ATLAS Collaboration}",
    title          = "{Performance of electron and photon triggers in ATLAS during LHC Run~2}",
    journal        = "Eur. Phys. J. C",
    volume         = "80",
    year           = "2020",
    pages          = "47",
    doi            = "10.1140/epjc/s10052-019-7500-2",
    reportNumber   = "CERN-EP-2019-169",
    eprint         = "1909.00761",
    archivePrefix  = "arXiv",
    primaryClass   = "hep-ex",
}

@Article{EGAM-2019-01,
    author         = "{ATLAS Collaboration}",
    title          = "{Tools for estimating fake/non-prompt lepton backgrounds with the ATLAS detector at the LHC}",
    journal        = "JINST",
    volume         = "18",
    year           = "2023",
    pages          = "T11004",
    doi            = "10.1088/1748-0221/18/11/T11004",
    reportNumber   = "CERN-EP-2022-214",
    eprint         = "2211.16178",
    archivePrefix  = "arXiv",
    primaryClass   = "hep-ex",
}

@Article{TRIG-2019-01,
    author         = "{ATLAS Collaboration}",
    title          = "{Performance of the missing transverse momentum triggers for the ATLAS detector during Run-2 data taking}",
    journal        = "JHEP",
    volume         = "08",
    year           = "2020",
    pages          = "080",
    doi            = "10.1007/JHEP08(2020)080",
    reportNumber   = "CERN-EP-2020-050",
    eprint         = "2005.09554",
    archivePrefix  = "arXiv",
    primaryClass   = "hep-ex",
}

@Article{DAPR-2021-01,
    author         = "{ATLAS Collaboration}",
    title          = "{Luminosity determination in \(pp\) collisions at \(\sqrt{s} = 13\,\text{TeV}\) using the ATLAS detector at the LHC}",
    journal        = "Eur. Phys. J. C",
    volume         = "83",
    year           = "2023",
    pages          = "982",
    doi            = "10.1140/epjc/s10052-023-11747-w",
    reportNumber   = "CERN-EP-2022-281",
    eprint         = "2212.09379",
    archivePrefix  = "arXiv",
    primaryClass   = "hep-ex",
}

@Booklet{ATL-SOFT-PUB-2021-001,
    author         = "{ATLAS Collaboration}",
    title          = "{The ATLAS Collaboration Software and Firmware}",
    howpublished   = "{ATL-SOFT-PUB-2021-001}",
    url            = "https://cds.cern.ch/record/2767187",
    year           = "2021",
}

@Article{CMS-SMP-18-003,
    author         = "{CMS Collaboration}",
    title          = "{Measurement of the \(Z\) boson differential production cross section using its invisible decay mode (\(Z \to \nu\bar{\nu}\)) in proton--proton collisions at \(\sqrt{s} = 13\,\text{TeV}\)}",
    journal        = "JHEP",
    volume         = "05",
    year           = "2021",
    pages          = "205",
    doi            = "10.1007/JHEP05(2021)205",
    reportNumber   = "CERN-EP-2020-223",
    eprint         = "2012.09254",
    archivePrefix  = "arXiv",
    primaryClass   = "hep-ex",
}

@Booklet{ATLAS-CONF-2015-029,
    author         = "{ATLAS Collaboration}",
    title          = "{Selection of jets produced in \(13~\text{TeV}\) proton--proton collisions with the ATLAS detector}",
    howpublished   = "{ATLAS-CONF-2015-029}",
    url            = "https://cds.cern.ch/record/2037702",
    year           = "2015",
}

@Booklet{ATLAS-CONF-2017-029,
    author         = "{ATLAS Collaboration}",
    title          = "{Measurement of the tau lepton reconstruction and identification performance in the ATLAS experiment using \(pp\) collisions at \(\sqrt{s} = 13~\text{TeV}\)}",
    howpublished   = "{ATLAS-CONF-2017-029}",
    url            = "https://cds.cern.ch/record/2261772",
    year           = "2017",
}

@Booklet{ATL-PHYS-PUB-2014-021,
    author         = "{ATLAS Collaboration}",
    title          = "{ATLAS Pythia~8 tunes to \(7~\text{TeV}\) data}",
    howpublished   = "{ATL-PHYS-PUB-2014-021}",
    url            = "https://cds.cern.ch/record/1966419",
    year           = "2014",
}

@Booklet{ATL-PHYS-PUB-2015-026,
    author         = "{ATLAS Collaboration}",
    title          = "{Vertex Reconstruction Performance of the ATLAS Detector at \(\sqrt{s} = 13~\text{TeV}\)}",
    howpublished   = "{ATL-PHYS-PUB-2015-026}",
    url            = "https://cds.cern.ch/record/2037717",
    year           = "2015",
}

@Booklet{ATL-PHYS-PUB-2015-034,
    author         = "{ATLAS Collaboration}",
    title          = "{Forward Jet Vertex Tagging: A new technique for the identification and rejection of forward pileup jets}",
    howpublished   = "{ATL-PHYS-PUB-2015-034}",
    url            = "https://cds.cern.ch/record/2042098",
    year           = "2015",
}

@Booklet{ATL-PHYS-PUB-2015-045,
    author         = "{ATLAS Collaboration}",
    title          = "{Reconstruction, Energy Calibration, and Identification of Hadronically Decaying Tau Leptons in the ATLAS Experiment for Run-2 of the LHC}",
    howpublished   = "{ATL-PHYS-PUB-2015-045}",
    url            = "https://cds.cern.ch/record/2064383",
    year           = "2015",
}

@Booklet{ATL-PHYS-PUB-2016-017,
    author         = "{ATLAS Collaboration}",
    title          = "{The Pythia~8 A3 tune description of ATLAS minimum bias and inelastic measurements incorporating the Donnachie--Landshoff diffractive model}",
    howpublished   = "{ATL-PHYS-PUB-2016-017}",
    url            = "https://cds.cern.ch/record/2206965",
    year           = "2016",
}

@Booklet{ATL-PHYS-PUB-2016-020,
    author         = "{ATLAS Collaboration}",
    title          = "{Studies on top-quark Monte Carlo modelling for Top2016}",
    howpublished   = "{ATL-PHYS-PUB-2016-020}",
    url            = "https://cds.cern.ch/record/2216168",
    year           = "2016",
}

@Booklet{ATL-PHYS-PUB-2017-005,
    author         = "{ATLAS Collaboration}",
    title          = "{Multi-Boson Simulation for \(13~\text{TeV}\) ATLAS Analyses}",
    howpublished   = "{ATL-PHYS-PUB-2017-005}",
    url            = "https://cds.cern.ch/record/2261933",
    year           = "2017",
}

@Booklet{ATL-PHYS-PUB-2018-027,
    author         = "{ATLAS Collaboration}",
    title          = "{ATLAS sensitivity to Two-Higgs-Doublet models with an additional pseudoscalar exploiting four top quark signatures with 3ab\(^{-1}\) of \(\sqrt{s} = 14~\text{TeV}\) proton--proton collisions}",
    howpublished   = "{ATL-PHYS-PUB-2018-027}",
    url            = "https://cds.cern.ch/record/2645845",
    year           = "2018",
}

@Booklet{ATL-PHYS-PUB-2019-033,
    author         = "{ATLAS Collaboration}",
    title          = "{Identification of hadronic tau lepton decays using neural networks in the ATLAS experiment}",
    howpublished   = "{ATL-PHYS-PUB-2019-033}",
    url            = "https://cds.cern.ch/record/2688062",
    year           = "2019",
}

@Booklet{ATL-PHYS-PUB-2021-011,
    author         = "{ATLAS Collaboration}",
    title          = "{Evaluating statistical uncertainties and correlations using the bootstrap method}",
    howpublished   = "{ATL-PHYS-PUB-2021-011}",
    url            = "https://cds.cern.ch/record/2759945",
    year           = "2021",
}
 
\clearpage
 
\begin{flushleft}
\hypersetup{urlcolor=black}
{\Large The ATLAS Collaboration}

\bigskip

\AtlasOrcid[0000-0002-6665-4934]{G.~Aad}$^\textrm{\scriptsize 103}$,
\AtlasOrcid[0000-0001-7616-1554]{E.~Aakvaag}$^\textrm{\scriptsize 16}$,
\AtlasOrcid[0000-0002-5888-2734]{B.~Abbott}$^\textrm{\scriptsize 121}$,
\AtlasOrcid[0000-0002-1002-1652]{K.~Abeling}$^\textrm{\scriptsize 55}$,
\AtlasOrcid[0000-0001-5763-2760]{N.J.~Abicht}$^\textrm{\scriptsize 49}$,
\AtlasOrcid[0000-0002-8496-9294]{S.H.~Abidi}$^\textrm{\scriptsize 29}$,
\AtlasOrcid[0009-0003-6578-220X]{M.~Aboelela}$^\textrm{\scriptsize 44}$,
\AtlasOrcid[0000-0002-9987-2292]{A.~Aboulhorma}$^\textrm{\scriptsize 35e}$,
\AtlasOrcid[0000-0001-5329-6640]{H.~Abramowicz}$^\textrm{\scriptsize 152}$,
\AtlasOrcid[0000-0002-1599-2896]{H.~Abreu}$^\textrm{\scriptsize 151}$
% ...
\bigskip
\\

% ...
$^{16}$Department for Physics and Technology, University of Bergen, Bergen; Norway.\\
% ...
$^{29}$Physics Department, Brookhaven National Laboratory, Upton NY; United States of America.\\
% ...
$^{35}$$^{(a)}$Facult\'e des Sciences Ain Chock, R\'eseau Universitaire de Physique des Hautes Energies - Universit\'e Hassan II, Casablanca;$^{(b)}$Facult\'{e} des Sciences, Universit\'{e} Ibn-Tofail, K\'{e}nitra;$^{(c)}$Facult\'e des Sciences Semlalia, Universit\'e Cadi Ayyad, LPHEA-Marrakech;$^{(d)}$LPMR, Facult\'e des Sciences, Universit\'e Mohamed Premier, Oujda;$^{(e)}$Facult\'e des sciences, Universit\'e Mohammed V, Rabat;$^{(f)}$Institute of Applied Physics, Mohammed VI Polytechnic University, Ben Guerir; Morocco.\\
% ...
$^{44}$Physics Department, Southern Methodist University, Dallas TX; United States of America.\\
% ...
$^{49}$Fakult\"{a}t Physik , Technische Universit{\"a}t Dortmund, Dortmund; Germany.\\
% ...
$^{55}$II. Physikalisches Institut, Georg-August-Universit\"{a}t G\"ottingen, G\"ottingen; Germany.\\
% ...
$^{103}$CPPM, Aix-Marseille Universit\'e, CNRS/IN2P3, Marseille; France.\\
% ...
$^{121}$Homer L. Dodge Department of Physics and Astronomy, University of Oklahoma, Norman OK; United States of America.\\
% ...
$^{151}$Department of Physics, Technion, Israel Institute of Technology, Haifa; Israel.\\
$^{152}$Raymond and Beverly Sackler School of Physics and Astronomy, Tel Aviv University, Tel Aviv; Israel.\\
% ...
$^{a}$ Also Affiliated with an institute covered by a cooperation agreement with CERN.\\
$^{b}$ Also at An-Najah National University, Nablus; Palestine.\\
$^{c}$ Also at Borough of Manhattan Community College, City University of New York, New York NY; United States of America.\\
$^{d}$ Also at Center for High Energy Physics, Peking University; China.\\
$^{e}$ Also at Center for Interdisciplinary Research and Innovation (CIRI-AUTH), Thessaloniki; Greece.\\
$^{f}$ Also at Centro Studi e Ricerche Enrico Fermi; Italy.\\
% ...

\end{flushleft}


\end{document}